\documentclass[pra,reprint,letterpaper,superscriptaddress,amsfonts,amsmath,amssymb,nofootinbib,,longbibliography]{revtex4-2}
\usepackage[T1]{fontenc}
\usepackage{calligra}
\usepackage{braket}
\usepackage{graphicx}
\usepackage{color}
\usepackage{ulem}
\usepackage{MnSymbol}
\usepackage{enumerate}
\usepackage{bbold}
\usepackage{subfigure}
\usepackage{subfiles}
\usepackage{units}
\usepackage[usenames,dvipsnames]{xcolor}
\usepackage{mleftright}
\usepackage[T1]{fontenc}

\usepackage{amsmath}
\usepackage{paralist}
\usepackage{psfrag}
\usepackage{setspace,tabularx,fancyhdr}
\usepackage{mathtools}
\usepackage[top=2cm,bottom=2cm,left=2cm,right=2cm]{geometry}
\usepackage{enumitem}
\setlist[itemize]{leftmargin=*}
\usepackage{wrapfig}
\usepackage{type1cm}
\usepackage{yfonts}
\usepackage{relsize}
\usepackage{multirow}
\usepackage{tabularx}
\usepackage{array}
\usepackage{rotating}
\usepackage{bigints}

\PassOptionsToPackage{breaklinks}{hyperref}
\usepackage{xurl} 

\usepackage{hyperref}
\hypersetup{colorlinks=true,linkcolor=MidnightBlue,citecolor=blue,filecolor=green,urlcolor=Violet}

\def\beq{\begin{equation}}
\def\eeq{\end{equation}}
\def\bsp{\begin{split}}
\def\esp{\end{split}}
\def\bea{\begin{eqnarray}}
\def\eea{\end{eqnarray}}
\def\ergo{\Rightarrow}
\def\TenD{\mathcal{D}}
\newcommand{\neweqline}{\nonumber\\}
\def\NM{M}
\def\Avx{\bar{x}^j}

\def\Tigam{\Tilde{\Gamma}}
\definecolor{mygreen}{rgb}{0.2, 0.8, 0.2}
\def\multi{\mathbb{M}}

\newcommand{\white}[1]{{\color{white} #1}}

\newcommand{\IGNORE}[1]{}

\newcommand{\al}{\alpha}

\newcommand{\eps}{\epsilon}
\newcommand{\kp}{\kappa}

\newcommand{\lp}{\ell_p}

\newcommand{\Christoffel}[2]{\ensuremath{{\mathlarger\Gamma}}^{#1}\!\!\!\!\!_{#2}}

\def\nablaST{\overset{\star}{\nabla}}
\def\otimesST{\overset{\star}{\otimes}}
\def\partialST{\overset{\star}{\partial}}
\def\commaST{\overset{\star}{,}}

\def\uST{\overset{\star}{u}}
\def\zST{\overset{\star}{z}}

\def\GeomR{\textgoth{R}}
\def\OprR{\mathcal{R}}

\def\lie{\mathcal{L}}
\def\bphi{\bar{\phi}}
\def\buphi{\bar{\phi}^{\{\alpha\}}}
\def\bdphi{\bar{\phi}_{\{\alpha\}}}
\def\invG{\overset{\star}{g}}

\graphicspath{{Images/}}

\begin{document}

\title{\texorpdfstring{$\kappa$-General-Relativity}{kappa-General-Relativity} I:
a Non-Commutative GR Theory \\ 
with the \texorpdfstring{$\kappa$-Minkowski}{kappa-Minkowski} Spacetime as its Flat Limit}

\author{Daniel Rozental}
\thanks{Author to whom any correspondence should be addressed. Email: \href{mailto:daniel.rozental@biu.ac.il}{daniel.rozental@biu.ac.il}.}

\affiliation{Department of Physics, Bar Ilan University, Ramat Gan 5290002, Israel}

\author{Ofek Birnholtz}
\affiliation{Department of Physics, Bar Ilan University, Ramat Gan 5290002, Israel}

\begin{abstract}
We employ a twist deformation of infinitesimal diffeomorphisms to construct a modification of General Relativity on a non-commutative spacetime extending the local $\kappa$-Minkowski geometry.
This spacetime arises in Deformed Special Relativity (DSR) models, where a fundamental length scale is incorporated into Special Relativity as an effective description of quantum gravitational effects.
To avoid the mathematical and physical inconsistencies associated with twisting the Poincaré group, we instead deform the dilatation-enlarged IGL(3,1) group, constructing a covariant and explicitly consistent gravitational theory (distinct from Weyl gravity).
The relativistic consistency of the twisted $\kappa$-Minkowski spacetime is demonstrated, including deformed transformations and differential structures.
A physically motivated Inönü–Wigner (IW) contraction procedure is suggested to enable a well-defined classical limit, addressing the correspondence issue.
This framework provides a consistent foundation for a dynamical sector of DSR and allows, in future treatment, explicit computations that could advance phenomenological predictions.


\end{abstract} 
\maketitle

 
\section{Introduction}\label{section: Introduction}
\subsection{Background and Motivation}
One of the phenomenological approaches to formulating a quantum gravity (QG) theory involves introducing deformations to the Poincaré symmetries underlying Special Relativity (SR). This deformation is aimed at ensuring that the resulting theory maintains an additional invariant length scale, which is considered fundamental in QG theories \cite{Multy_Mass, Relativity_with_length1, Relativity_with_length2, minimal_3, V_Objection_2}. These theories, often referred to as Deformed Special Relativity (DSR) or Deformed Relativistic Kinematics (DRK), represent a phenomenological ("bottom-up") approach to addressing the QG problem \cite{Multy_Mass}.

Phenomenological theories aim to capture essential properties of various QG theories and incorporate them as deformations into the classical theory. This enables the prediction of deviations from classical gravity in astrophysical systems, which could be probed using technologies within reach \cite{Relative_Locality, Relative_And_kappa_1, Relative_And_kappa_2}. In contrast, first-principle theories ("top-down"), such as String Theory and Loop Quantum Gravity, seek to address the problem from a more fundamental level but often yield physical predictions that are difficult or impossible to calculate and are typically applicable only to energies far beyond current technology \cite{TopDownQG, Multy_Mass}.

Fundamentally, DSR theories involve deforming the symmetries of SR in a manner that preserves the invariance of a QG-length scale, often identified as the Planck length $\lp$ \cite{Born, minimal_1, minimal_2, minimal_3}:
\begin{equation}
\label{Planck Length}
    \lp = \sqrt{\frac{\hbar G}{c^3}} \sim 10^{-35} \, \mathrm{m}~.
\end{equation}
We also define the Planck energy $\kp = E_p=\hbar c/\lp$, so that in Quantum Field Theory (QFT) units ($\hbar=c=1$), $\kp=\lp^{-1} \sim 10^{28} \, \mathrm{eV} $. It represents the energy scale at which General Relativity (GR), when treated as a QFT, is expected to diverge \cite{Effective_QG}. Correspondingly, one may introduce the Planck frequency:
\beq\label{Planck Frequency}
\omega_p = \frac{c}{l_p} = \frac{\kappa}{\hbar} \sim 10^{42} \,\mathrm{Hz}\,,
\eeq
a quantity that will play a role in the discussion of Sec.~\ref {ssection: IW-contraction our case}.

Various methods exist for implementing these deformations, each leading to different phenomenological consequences.
For example, in the seminal papers  \cite{Relativity_with_length1, Relativity_with_length2}, one finds two different approaches relying on slightly different basic arguments (whether the starting point is the invariant Planck length or the invariant Planck energy), which lead to different deformations of how Lorentz boosts act on momenta/energy. These differences affect \cite{Comparison_of_relativity_theories, Does_a_deformation_of} whether one expects a momenta-dependent speed of light, saturation of energy and momenta, etc. (see also \cite{Modified_Velocity_1, Modified_Velocity_2}).

Moreover, ensuring length scale invariance in causally connected states requires a deformed composition of momenta. From relativistic compatibility, a deformed action of transformations on multi-particle states is needed, an effect also known as backreaction on rapidity for the case of boost transformations \cite{Majid:1994cy, Boost_1, Boost_2}. These different possibilities can be understood as different bases of quantum deformations of the Poincaré group \cite{q_deformations, Majid:1994cy, Different_Realization}, establishing connections \cite{Relative_And_kappa_1, Relative_And_kappa_2} with related theories such as curved momentum space and Relative Locality \cite{Relative_Locality, RL_Test_1, RL_Test_2}.

Regardless of the specific model, DSR theories share two key properties.
The first is the promotion of spacetime variables into \textit{operators} with a non-vanishing commutation relation, known as $\kp$-Minkowski spacetime \cite{Majid:1994cy, QuantumGroup_GaugeTheory, Hopf_Book, noncommutative_spaces_of_worldlines, The_n_dimensional_kappa}, which we mark as $\mathcal{M}_\kp$:
\beq
\label{kappa minkowski} [\hat{x}^0,\hat{x}^{i}] = i \lp\hat{x}^{i}, \quad [\hat{x}^{i},\hat{x}^{j}] = 0, \quad \mathrm{for} \quad i,j = 1,2,3.
 \eeq
The second is that there are deformations in the Hopf algebra sector of the Poincaré symmetry (ISO(3,1)) \cite{Majid:1994cy, Poisson_bivector_map, Hopf_Book, Quantum_Poincare_group_related}, ensuring the relativistic invariance of relation \eqref{kappa minkowski} and the deformed symmetries.

Given a system's symmetry group $G$, a Hopf algebra is a structure embedded in the Universal Enveloping Algebra $\mathbf{U}(G)$ (the freely generated tensor algebra with the same representations/modules as $G$'s Lie algebra). It determines how observables (symmetry generators) act on multi-particle states represented as the tensor product of irreducible representations (this viewpoint on Hopf algebras as acting on the algebra of observables was mainly developed in \cite{BraidedAlgebra, CrossProductQuantization_AndTwisting, QuantumGroup_GaugeTheory}). Classically, this structure is trivial and often uncharacterized.
For example, the momentum generator (translation) with the one-particle (irreducible) representation $D(P)$ acts on one-particle and two-particle states as follows,
\bea
&&D (\mathbf{P})\ket{K} = {\bf k} \ket{K}~~,
~~D (\mathbf{P})\ket{L} = {\bf l} \ket{L}, \\
&&\ket{KL} \rightarrow(D\otimes D) (\Delta\mathbf{P})(\ket{K}\otimes \ket{L}) = (\mathbf{k}+\mathbf{l})\ket{KL},~~
\label{Bosonic Associative}
\eea
with the trivial co-product in the Hopf algebra defined as
\beq\label{co-product Poincare}
    \Delta\mathbf{P} \equiv \mathbf{P} \otimes 1 + 1 \otimes \mathbf{P}.
\eeq

In a DSR theory, the form of \eqref{co-product Poincare} is generally modified, along with the remaining Hopf algebra structure on $\mathbf{U}$(ISO(3,1)). A symmetry group $G$ with such non-trivial Hopf algebra is called a Quantum Group (in the sense presented in \cite{CrossProductQuantization_AndTwisting}). Any DSR model can be associated with a specific Quantum Group structure \cite{kappa_to_DSR} and the spacetime \eqref{kappa minkowski} as its invariant space.

Therefore, investigating these Quantum Group and  NC\footnote{From here on, we shall use the shorthand NC for ``non-commutative''.} space deformations in theories with a dynamical sector, such as the theory of gravity, will impose constraints on any DSR model.

Dynamical theories can often be deformed in their Hopf-algebra sector using deformation quantization tools \cite{QuantumGroup_GaugeTheory, DeformationQuantization1, DeformationQuantization2, DeformationQuantization3}, where definitive deformation operations on the classical structures derive the non-commutativity and deformed Hopf algebras. Twisted algebras \cite{CrossProductQuantization_AndTwisting, TwistingQuantumDiff_PlanckScaleHopf, Drinfeld, Diff_calculus_3, General_notion_of_Twist_2} or Seiberg-Witten (SW) maps \cite{SWmaps} are such operations schemes, with the former being applied to the Lie algebra of the gauge theory's symmetry. In contrast, the latter directly applies to the gauge variables.

Regarding the  NC deformation of GR, foundations of its twist-deformed version were already outlined in the seminal works \cite{TwistingQuantumDiff_PlanckScaleHopf, QuantumGroup_GaugeTheory} considering the construction of differential calculus and gauge theories on  NC space (NCS), and further developed in \cite{NC_Geometry2} to consider the specific case of GR, with consistency proofs provided in \cite{NC_Geometry3}. The general view is to formulate an invariant theory under deformed infinitesimal diffeomorphisms (viewed as the Lie algebra of vector fields deformed in its Hopf sector). One should view the  NC GR as describing the gravitational interaction on a  NC manifold through  NC differential geometry. The manifold is locally a  NC spacetime (NCST) characterized by the type of deformation describing the \textit{flat} sector of the NCST.
In DSR theories, one considered this flat limit to be the $\kappa$-Minkowski spacetime, described in \eqref{kappa minkowski}.

However, definitive results were only obtained for constant non-commutativity \cite{Twist_general_2,  Gravity_Non}, a result of a deformed symmetry group that lacks apparent physical meaning \cite{Twistgeneral1, Realization_of_Hopf}.
In contrast, no construction has been made explicitly for the physically attractive  NC scenario as defined by \eqref{kappa minkowski} (only special cases were treated, such as assuming specific symmetries in spacetime, see \cite{Metric_Perturbations_In_NC, NC_Geometry3} and references therein). This deficiency can be traced to a technical ambiguity: twisting classical algebras results only in deformations in the Hopf-algebra sector of the symmetry group, leaving the algebra sector classical. However, quantum deformations applied to the Poincaré algebra ($\kp$-Poincaré) that lead to the $\kp$-Minkowski spacetime are only consistent with simultaneous deformation of the algebra and the Hopf-algebra. 
Indeed, it was proven in \cite{NO_KAPPA_TWIST} that one cannot derive $\kp$-Poincaré via a Twist (see Sec.\ref{section: Building Blocks For Construction} for an extended discussion), an obstacle for the construction of GR (or any other gauge theory) on $\kp$-Minkowski spacetime by use of the twist formalism.

This is a problem because any reasonable attempt to assign a quantum nature to spacetime should admit a gravitational theory, which should coincide with the GR predictions at scales near the quantum-classical intersection.
Thus, for the DSR-type theories, we expect that, near Planckian scales, one should be able to construct a theory of gravity on the NCS of DSR.

\subsection{Overview}\label{section: Overview of Introduction}
Our aim in the current paper is to construct a \textit{physically consistent} theory of deformed GR, which we call $\kp$-GR, such that it both has the $\kappa$-Minkowski spacetime as its flat limit, and enables physical predictions that can be tested.
To achieve a predictable $\kp$-GR, we follow the construction of geometry in NC spaces using the established formalism of twist deformations \cite{CrossProductQuantization_AndTwisting, TwistingQuantumDiff_PlanckScaleHopf, QuantumGroup_GaugeTheory}. As discussed in previous sections, this cannot be accomplished with a twist belonging to the Poincaré group.
Thus, by following studies like \cite{kappa_minkowki_Twist, Twistgeneral1, IGL_TWIST_1}, we enlarge the underlying symmetry from the Poincaré group to the IGL(3,1) group (the Inhomogeneous General Linear group with Lorentzian signature, which can be viewed as a dilatation-enlarged Poincaré group - see Sec.\ref{section: Building Blocks For Construction} for more details).
We then construct a twist for this enlarged group such that its covariant spacetime corresponds to $\kp$-Minkowski, following studies like in \cite{A_Non_Commutative_From_Ads_Algebra, Quantum_AntiDeSitter}.

To asses what it means to pursue a \textit{consistent} $\kp$-GR within the twist formalism, let us outline some ambiguities that must be addressed before a $\kp$-GR can be called consistent.
The first key conceptual subtlety lies already when considering the \textit{flat sector} of the spacetime \eqref{kappa minkowski} constructed with twisted IGL(3,1) symmetry: the relations \eqref{kappa minkowski}
single out the time direction and may suggest a preferred frame.
This concern is sharpened by observing that while two observers agree on the spatial commutation sector $[x^i, x^j] = 0$, they will not agree regarding the time structure since the commutator $[x^0, x^j]$ is position-dependent, so observers at different spatial locations experience different \textit{local times}.
This subtlety is well understood when considering the symmetries to be $\kp$-Poincare (see Sec.\ref{section: Introduction} and e.g., \cite{NonCommutative_Spacetime_Interpretation1}), and we follow along the same principle lines and show that the NC structure remains invariant under the twisted symmetry transformations.
Specifically, for all generators $\xi \in igl(3,1)$ the deformed transformations, denoted for now as $\hat{\xi}$, we find
\beq
[\hat{\xi}(\hat{x}^\mu), \hat{x}^\nu] \in \mathcal{M}_{\kp},
\eeq
where $\mathcal{M}_{\kp}$
is defined by \eqref{kappa minkowski}, and therefore the deformed transformations $\hat{\xi}^\mu$ preserve the $\kappa$-Minkowski relations across all frames.
This confirms that although the deformation vector always aligns with the time direction, it transforms covariantly with the observer, enabling to interpret \eqref{kappa minkowski} as in the interpretation done in Relative Locality theories and DSR theories based on the $\kp$-Poincar\'e symmetries discussed in the previous section.

In addition to showing that the $\kappa$-Minkowski algebra is closed under the deformed igl(3,1) transformations, one must ensure that the framework admits a well-defined differential geometry and a self-consistent deformed Leibniz rule.
These properties are verified explicitly in Sec.\ref{section: Consistency of The NC-Relations} (especially see Sec.\ref{subsec: consistency transl} and Sec.\ref{subsec: consistency igl}), including for nontrivial transformations such as boosts and dilatation.

The next ambiguity to be addressed is that of the \textit{Correspondence Principle:} The classical limit of the deformed symmetry must recover standard GR with local Poincaré symmetry. In the paper we address this by proposing a contraction scheme of IGL(3,1) to ISO(3,1) using a dimensionless parameter $\epsilon_{RL} = \frac{\omega_{\text{local}}}{\omega_p}$ to quantify the scale of the NC effects on a system with given frequency $\omega_{\text{local}}$. A possible additional scaling factor of the form $\left(\frac{x}{l_{\text{sys}}}\right)^n$ may arise from Relative Locality (RL) considerations, but is left to future work \cite{Our_Next_Paper}. The resulting contracted algebra IGL'(3,1) ensures that in the classical limit $\epsilon_{RL} \to 0$, no residual dilatation symmetry remains, and the theory flows to standard GR.
This contraction is reflected in the twist through the rescaled generator $D' = \frac{1}{\alpha} D=\frac{1}{\omega_p}$.

The final consistency regards the \textit{deformed gravitational theory:} The theory must admit a well-defined deformation of GR, including a consistent notion of deformed diffeomorphisms, variation principles, and conservation laws. We thus show how deformed general coordinate transformations are deformed appropriately, derive the form of the deformed Einstein equation and the Einstein–Hilbert action, and discuss the possibility for coupling matter via a deformed energy–momentum tensor, outlining some of the persisting ambiguities (especially in the definition of local conserved quantities) and approaches for resolutions and future work.

Importantly, our construction respects twisted general covariance but is \textit{not} invariant under local scale transformations.
Although the dilatation generator $D$ appears in the twist, it does not correspond to a dynamical scale symmetry as in Weyl gravity \cite{WeylGrav1, WeylGrav2}. In Weyl gravity, dilatations preserve the action under local rescalings of the metric, whereas in our case, $D$ generates only kinematic symmetry of the flat sector, and the resulting theory is not Weyl-invariant.

All in all, when we refer to a \textit{consistent theory} in this study, we refer to one of the following:
\begin{itemize}
    \item[$\mathcal{A:}$] \textbf{Relativistic Consistency} of the deformed symmetry with the NC spacetime.
    \item[$\mathcal{B:}$] The \textbf{Correspondence Principle} of the theory, especially regarding its symmetry. 
    \item[$\mathcal{C:}$] \textbf{Well-defined Gravitational} theory; has a meaningful notion of deformed diffeomorphisms, variations, and conserved quantities.
\end{itemize}

The rest of the paper is organized as follows:
Sec.~\ref{section: NC Introduction} reviews the well-known general twist formalism in NCST and the construction of deformed GR. Sec.~\ref{section: Building Blocks For Construction} sets the building blocks of our construction with local $\kappa$-Minkowski with IGL(3,1) symmetry.
In Sec.~\ref{section: Consistency of The NC-Relations}, we establish the relativistic consistency of the twisted igl(3,1), including the deformed transformations and the rest of the twisted structure of the symmetry (consistency $\mathcal{A}$).
Sec.~\ref{Contraction of the Symmetry} introduces the IW-contraction ensuring classical correspondence (consistency $\mathcal{B}$). Sec.~\ref{section: The Construction of Gravity} develops the deformed gravitational theory, including the deformed diffeomorphisms,  Einstein–Hilbert action, variation, and energy–momentum tensor, discussing the advances made and what is left ambiguous (consistency $\mathcal{C}$).
Sec.~\ref{section: Final Words} concludes with remarks and future work.

\section{Basic Formulation: non-commutative geometry and gravity}\label{section: NC Introduction}

\subsection{Syntax and notation conventions}
This section gradually outlines the formulae and the mathematical constructions necessary for deformed general relativity extending the $\kappa$-Minkowski spacetime, following closely \cite{NC_Geometry2, NC_Geometry_Simplified}, which relies on the work in \cite{QuantumGroup_GaugeTheory, TwistingQuantumDiff_PlanckScaleHopf}, see \cite{FoundationsOfQuantumGroups} for a comprehensive treatment. In Sec.\ref{section: background A}, we describe the general formalism of  NC spaces by twisting the Lie algebra of vector fields. In particular, it includes the full derivations and definitions of the following operators and operands, enumerated here in brief:
\begin{enumerate}
    \item $\star$-product, a  NC product used as an isomorphism between NCST variables and their classical counterparts. 
    \item $\Phi$, the twist operator that generates the $\star$-product; it is an invertible rank two tensor composed of vector fields belonging to the symmetry we want to deform. Its inverse is $\bar{\Phi}$.
    \item $\OprR$, an invertible rank two tensor; the universal $\OprR$-matrix (as usually defined in Hopf algebras) which, for the current discussion, is used to flip the order of $\star$-multiplications, demonstrating the non-commutativity. Its inverse is $\bar{\OprR}$.
    \item $\mathbb{M}$, a bidifferential operator, acting as a linear map on products of vector fields ($X, Y$). We denote its action on some expression of ($X, Y$) with a ``$\circ$". 
    \item $\mathcal{L}^{\star},\  \langle \cdot\commaST \cdot \rangle,\ [\cdot\commaST \cdot],\ \otimesST ... $, the deformed versions of the classical operations (in this case: Lie derivative, the pairing map, vector fields' commutator, and the tensor product) required in a description of differential geometry. Generally, every operator/operand denoted with a $\star$-subscript is the deformed version consistent with NCST variables. 
\end{enumerate}

In Sec.\ref{section: background B}, we describe the required formalism for a GR theory on NCS, utilizing the formalism established in Sec.\ref{section: background A}.

Concerning conventions: we use late Greek alphabet letters for spacetime indices and mid-Latin letters for purely spatial indices (e.g., $\mu,\nu=0,i$, with $i,j=1,2,3$); early Latin letters ($a$,$b$,$c$) enumerate generators of algebras; early greek letters ($\al$) enumerate orders of expansion in a parameter $\lambda$; the Gothic \textgoth{R} is used for geometrical quantities, such as the Riemann tensor, the Ricci tensor, and the Ricci scalar.
Additionally, throughout the paper, we employ standard notation for symmetrization and anti-symmetrization of indices: Square brackets $[]$ denote anti-symmetrization, and parentheses $( )$ denote symmetrization.
For example, the notation in Equation \eqref{dependent: comparison to general non-commutativity} demonstrates the anti-symmetrization of the $0$ and $\rho$ indices.

\subsection{Non-commutative Spaces \& Twisting}\label{section: background A}

Here, we outline the known formalism of how one can deform a commutative space to obtain a  NC one in a manner that allows for defining similar concepts as in the classical (commutative) scenario, specifically concerning differential geometry. We note that everything listed in this section is \textit{known} and brought for completeness and for setting up the notations. The familiar reader can safely skip it and move directly to Sec.~\ref{section: background B} or to Sec.~\ref{section: Building Blocks For Construction}.

We start (see \cite{Hopf_Book}, and e.g., \cite{CrossProductQuantization_AndTwisting} for the view of spacetime as an algebra of functions) by considering the algebra $\hat{\mathcal{A}}_{\hat{x}}$ \textit{freely generated} by the  NC variables $\hat{x}^{\mu}$, i.e., consider the relation, 
\begin{equation}\label{non-commutativity}
[\hat{x}^{\mu},\hat{x}^{\nu}]=C^{\mu \nu }(\hat{x}),
\end{equation}
where $C^{\mu\nu}(\hat{x})$ is an expansion in $\hat{x}^{\mu}$ with constant coefficients, 
\bea\label{General non commutation}
 C^{\mu \nu }(\hat{x}) = i\theta^{\mu \nu }+ ic^{\mu \nu }_{\rho}\hat{x}^{\rho}+
 \mathcal{O}(\hat{x}^{\alpha}\hat{x}^{\beta}).
\eea
It turns out that the space of  NC polynomials generated by $\hat{\mathcal{A}}_{x}$ is isomorphic to polynomials in $\mathcal{A}_{x}$ equipped with a $\star$-product \cite{PBW_Condition, SWmaps,Hopf_Book} (a $\star$-isomorphism),
\beq\label{algebra isomorphism}
\hat{\mathcal{A}}_{\hat{x}} \sim (\mathcal{A}_{x} , \star),
\eeq implying  that polynomials $\hat{P}(\hat{x})\in \hat{\mathcal{A}}_{\hat{x}}$ and $P(x)\in \mathcal{A}_{x}$ are related through the mapping:
\begin{equation}\label{function isomorphism}
\hat{P}_{1}(\hat{x})\hat{P}_{2}(\hat{x}) \mapsto P_{1}(x)\star P_{2}(x).
\end{equation}
The specific form of this mapping arises from deformation in the Hopf algebra sector of the symmetry group on the space under consideration $\mathcal{M}$ (a smooth manifold in general), where spaces with the structures described in equation \eqref{General non commutation} become the deformed symmetry module algebras.  

This Hopf deformation is achieved by first considering the algebra of vector fields $\Xi$ (infinite dimensional, which is the algebra of infinitesimal diffeomorphisms) on $\mathcal{M}$. This structure possesses a Lie algebra structure, with elements generated by $\{v^{1},..,v^{N}\}$ (N is the group order), and has a Hopf algebra extension on its universal envelope $\mathbf{U}(\Xi)$ (the algebra of tensor fields modulo the ideal of the Lie algebra $\Xi$).
Denoting with Latin indices the generator number, the Lie algebra and the Hopf algebra are, 
\bea
[v^{a},v^{b}]&=&if^{abc}v^{c}, \label{undeformed Lie algebra structure}
\\
\Delta(v^{a}) &=&v^{a}\otimes1+1\otimes v^{a}\equiv \Sigma_{i=1}^{2} v^{a}_{(1)i}\otimes v^{a}_{(2)i}\nonumber
\\
&\equiv&v^{a}_{(1)}\otimes v^{a}_{(2)},\label{undeformed Hopf structure}
\eea
where we used the Sweedler notation for $v^{a}_{(1)},v^{a}_{(2)}$, with an understood summation, e.g., for \eqref{undeformed Hopf structure} we have  $v^a_{(1)2}=v^a_{(2)1}=1$, $v^{a}_{(1)1}=v^{a}_{(2)2}=v^{a}$. 

A possible deformation of this Hopf structure is through a \textbf{Twist Deformation} \cite{Drinfeld, CrossProductQuantization_AndTwisting, FoundationsOfQuantumGroups}: a general twist, denoted as the operator element $\Phi\in \mathbf{U}(\Xi)\otimes \mathbf{U}(\Xi)$, is an invertible operator with a  deformation nature in its expansion parameter $\Lambda$, with the order being denoted by the positive integer $\alpha$ ($\{\alpha\in\mathbb{Z}|\alpha\geq0\}$);
\bea\label{perturbative representation of the twist} 
\Phi &=& \phi^{\{\alpha\}}\otimes \phi_{\{\alpha\}} = 1\otimes 1 +\phi^{\{1\}}\otimes \phi_{\{1\}}+ \mathcal{O}(\Lambda^{2}),~~~
\\
\label{perturbative representation of the inverse twist}
\Phi^{-1}&\equiv&\bar{\Phi}=\bar{\phi}^{\{\alpha\}}\otimes \bar{\phi}_{\{\alpha\}}, \ \Leftrightarrow \Phi\bar{\Phi}=1.
\eea

A twist element $\Phi$ defines a consistent deformation quantization\footnote{I.e., a deformation of the co-product in \eqref{undeformed Lie algebra structure} via a twist $\Phi$, such that the resulting Hopf algebra induces a NC (i.e., non-abelian) product on the dual algebra of coordinate functions—leading to a NC spacetime.} if it satisfies two key properties with respect to the co-unit $\epsilon$ of $U(\Xi)$: the \textit{2-cocycle condition} and the \textit{normalization condition}, given by: 
\bea 
\Phi_{12}(\Delta\otimes\text{id})\Phi&=&\Phi_{23}(\text{id}\otimes \Delta)\Phi, 
\neweqline
(\epsilon \otimes \text{id})\Phi &=& (\text{id} \otimes \epsilon)\Phi,
\eea 
with the notation, 
\beq\label{reversed twist}
\Phi_{21}=\Phi\otimes 1, \quad \Phi_{23}=1\otimes \Phi.
\eeq

Given these conditions are fulfilled, when the twist acts on \eqref{undeformed Hopf structure}, it gives rise to the \textit{twisted symmetry} structure on the co-algebra, leaving the Lie algebra \eqref{undeformed Lie algebra structure} classical, 
\bea\label{twisted Hopf}
   \Delta^{\Phi}(v^a ):= \Phi\Delta(v^a) \Phi^{-1}=v^{a}_{(1)\Phi}\otimes v^{a}_{(2)\Phi}~.
\eea
As in \eqref{undeformed Hopf structure}, we have used the Sweedler notation in $v^{a}_{(i)\Phi}$, with subscript $\Phi$ denoting that this is the twisted quantity. 
This leads to the deformed Hopf algebra denoted as $(\mathbf{U}(\Xi)^{\Phi}, \cdot, \Delta^{\Phi})$, 
which shares the $"\cdot"$-product when $(\xi,\eta)\in\mathbf{U}(\Xi)$ act on some quantity:  $\mathcal{L}_{\xi}\mathcal{L}_{\eta}=\mathcal{L}_{\xi\eta}$, but differs from the classical one in its co-product \eqref{twisted Hopf} and therefore in its representations.
Thus, a deformation appears in the Leibniz rule of symmetry transformations when they act on field products. This fact allows the derivation of a quantum Hopf algebra from classical structure; the underlying symmetry of the NCS \eqref{algebra isomorphism} can be derived through a twist. Operationally, consider, for example, the algebra of commutative functions $Fun(\mathcal{M})\equiv\mathcal{A}$. 
The deformation of the usual product $\mu(f\otimes g):=fg$ with $(f,g)\in \mathcal{A}$ is constructed from the twist \eqref{perturbative representation of the twist} by 
\bea \label{function star product}
fg &\mapsto& \multi(\Phi^{-1}(f\otimes g)) 
\neweqline
&\equiv& f\star g  =\bar{\phi}^{\{\alpha\}}(f)\bar{\phi}_{\{\alpha\}}(g).
\eea
This deformed multiplication is associative and is defined for the twists we are considering since every $(\xi,\eta) \in\mathbf{U}(\Xi)$ has a defined action on $f\in \mathcal{A}$ via the Lie derivative of vector fields on functions \cite{NC_Geometry_Simplified}. Therefore, the twist deformation consistently defines an NC algebra of functions $Fun_{\star}(\mathcal{M})\equiv\mathcal{A}_{\star}$ defined by \eqref{function star product}. To observe the non-commutativity of this $\star$-product, it is useful to present an invertable element $\OprR\in U\Xi\otimes U\Xi$, named the \textit{universal} $\OprR$-\textit{matrix} \cite{Drinfeld:1986, NC_Geometry_Simplified}, which satisfies the following\footnote{These conditions are quasi-cocommutativity, co-cyclicty, and the Yang Baxter equations respectively. }
\bea 
v^{a}_{(2)\Phi}\otimes v^{a}_{(1)\Phi}&=&\OprR(v^{a}_{(1)\Phi}\otimes v^{a}_{(2)\Phi})\OprR^{-1},
\neweqline
(\Delta\otimes \text{id})\OprR&=&\OprR_{13}\OprR_{23}, \ (\text{id}\otimes \Delta)\OprR=\OprR_{13}\OprR_{12}, 
\neweqline
\OprR_{12}\OprR_{13}\OprR_{23}&=&\OprR_{23}\OprR_{13}\OprR_{12},  
\eea
with 
\bea
\OprR_{12}=\OprR\otimes 1, \ \OprR_{23} = 1\otimes\OprR, \ \OprR_{13}=\OprR^{\{\alpha\}}\otimes1\otimes \OprR_{\{\alpha\}}. ~~~~~
\eea
And the $\OprR$ could be expressed by 
\bea\label{Universal R-matrix}
\OprR  &:=& \Phi_{21}\Phi^{-1} = \text{R}^{\{\alpha\}}\otimes \text{R}_{\{\alpha\}} =\phi_{\{\alpha\}}\bar{\phi}^{\{\beta\}}\otimes \phi^{\{\alpha\}}\bar{\phi}_{\{\beta\}},
\neweqline
\bar{\OprR}&:=& \OprR^{-1}=\Phi\Phi_{21}^{-1}=\bar{\text{R}}^{\{\alpha\}}\otimes\bar{\text{R}}_{\{\alpha\}},
\eea

The non-commutativity of the product
\eqref{function star product} can now be expressed by
\beq\label{permutation definition}
f\star g = \bar{\text{R}}^{\{\alpha\}}(g) \star \bar{\text{R}}_{\{\alpha\}}(f) \neq g\star f.
\eeq 
The deformed algebra $\mathcal{A}_{\star}$ with the product \eqref{function star product} is a good representation of the deformed Hopf algebra $\mathbf{U}(\Xi)^{\Phi}$. In other words, $\mathcal{A}_{\star}$ is an algebra module of the deformed Hopf algebra; the algebra structure of $\mathcal{A}_{\star}$ is compatible with the action of $\mathbf{U}(\Xi)^{\Phi}$,
\beq\label{UEA on functions}
\xi(a\star b) = \xi_{(1)\Phi}(a)\star \xi_{(2)\Phi}(b), 
\eeq
where we have used Sweedler's notation for the Leibniz rule of $\xi$, similarly to \eqref{undeformed Hopf structure} and \eqref{twisted Hopf}, defined here by
\beq
\xi_{(1)\Phi}\otimes \xi_{(2)\Phi} \equiv \Delta^{\Phi}(\xi),
\eeq
for all $\xi\in \mathbf{U}(\Xi)\subset\mathbf{U}(\Xi)^{\Phi}$.

The same holds for the rest of the universal envelope; for any module algebra $A$ of $\mathbf{U}(\Xi)$, there exists a module $A_{\star}$ of $\mathbf{U}(\Xi)^{\Phi}$. Consequently, the algebras of vector fields and tensorfields have a meaningful definition of their $\star$-product by using the correct form of the Lie derivative (commutator for vector fields and adjoint action for tensor fields).
For example, in addition to $\mathcal{A}_{\star}$ the deformed algebra $\mathbf{U}(\Xi_{\star})$ can be defined since any $(\xi,\eta)\in \mathbf{U}(\Xi)$ has a definitive action of $\phi \in \mathbf{U}(\Xi)$ on them,
\beq\label{tensor star product}
\xi\star\eta:=\bar{\phi}^{\{\alpha\}}(\xi)\bar{\phi}_{\{\alpha\}}(\eta) =\mathcal{L}_{\bar{\phi}^{\{\alpha\}}}(\xi)\mathcal{L}_{\bar{\phi}_{\{\alpha\}}}(\eta),
\eeq
and, the algebra is also a $\mathbf{U}(\Xi)^{\Phi}$-module, as appears from, 
\beq
\zeta(\xi\star\eta)=\zeta_{(1)\Phi}(\xi)\star\zeta_{(2)\Phi}(\eta),
\eeq
for all $\zeta \in \mathbf{U}(\Xi)$.
These properties lead to a deformed commutator that closes in $\Xi$,
\bea \label{star lie brackets}
[u\commaST v] &=& u \star v - \bar{\text{R}}^{\{\alpha\}}(v)\star\bar{\text{R}}_{\{\alpha\}}(u)
\neweqline
&=&[\bar{\phi}^{\{\alpha\}}(u),\bar{\phi}_{\{\alpha\}}(v)].
\eea
Therefore, also $\Xi_{\star}$ is a deformed Lie algebra (a \textit{$\star$-Lie algebra}). However, any product of vector fields $(u,v)\in\Xi$ can be re-written as a product in $\mathbf{U}(\Xi_{\star})$ as $uv=\phi^{\{\alpha\}}(u)\star \phi_{\{\alpha\}}(v)$. Then, the $\star$-Lie algebra \eqref{star lie brackets} can also be regarded as being generated by $\star$-vector fields, and consequently, $\mathbf{U}(\Xi_{\star})$ as the universal envelope of $\Xi_{\star}$. 
As a universal envelope structure, $\mathbf{U}(\Xi_{\star})$ is a Hopf algebra $(\mathbf{U}(\Xi_{\star}), \star, \Delta_{\star})$, with the $\star$-co-product $\Delta_\star$ is defined by \cite{NC_Geometry_Simplified} 
\bea\label{isomorphism map}
     \Delta_{\star}=(D^{-1}\otimes D^{-1})\circ\Delta^{\Phi}\circ D,
\eea
with a corresponding universal $\OprR_{\star}$-matrix,
\bea
\OprR_{\star}:= (D^{-1}\otimes D^{-1})(\OprR), 
\eea
where the operator $D$ is defined by, 
\beq\label{isomorphism side1}
 D(\xi):= \bar{\phi}^{\{\alpha\}}(\xi)\bar{\phi}_{\{\alpha\}}, \quad \text{s.t. } \ D(\xi\star\eta)=D(\xi)D(\eta).
\eeq
For the formulation of a gravity theory, where a deformed infinitesimal diffeomorphism is central, one mostly uses the $\mathbf{U}(\Xi_{\star})$ structure \cite{Star_Hopf_Algebra, NC_Geometry_Simplified}. 
To use $\mathbf{U}(\Xi_{\star})$, its module algebras must be constructed; they coincide with the modules of $\mathbf{U}(\Xi)^{\Phi}$. Thus, $\mathcal{A}_{\star}$ is a $\mathbf{U}(\Xi_{\star})$-module with the action given by the \textit{$\star$-Lie derivative} (Sweedler notation is understood):
\bea
\label{star lie derivative functions}
\mathcal{L}^{\star}_{\xi}(f) &:=& \mathcal{L}_{D(\xi)}(f)  ,~ \\ \mathcal{L}^{\star}_{\xi}(f\star g)
&=&\mathcal{L}^{\star}_{\xi_{(1)\star}}(f)\star\mathcal{L}^{\star}_{\xi_{(2)\star}}(g), \nonumber
\eea
and similarly for the module 
$\mathbf{U}(\Xi_{\star})$,
\bea\label{star lie derivative tensors} 
\mathcal{L}^{\star}_{\xi}(\tau)
&:=& \mathcal{L}_{D(\xi)}(\tau) ,~ \\ \mathcal{L}^{\star}_{\xi}(\tau\star \tau')
&=&\mathcal{L}^{\star}_{\xi_{(1)\star}}(\tau)\star\mathcal{L}^{\star}_{\xi_{(2)\star}}(\tau'). \nonumber
\eea
Note that here the product is a $"\star"$-one, hence the product of actions of $(\xi,\eta)\in\mathbf{U}(\Xi_{\star})$ on some quantity will be $\mathcal{L}^{\star}_{\xi}\mathcal{L}^{\star}_{\eta}=\mathcal{L}^{\star}_{\xi\star\eta}$. Then, the action of the $\star$-Lie derivatives of $u\in\Xi_{\star}$ on another $v\in\Xi_{\star}$ or that of $\xi\in\mathbf{U}(\Xi_{\star})$ on $\zeta\in\mathbf{U}(\Xi_{\star})$ is the $\star$-Lie commutator or the $\star$-adjoint action respectively:
\beq\label{star lie derivative vectors}
\mathcal{L}^{\star}_{u}(v)=[u\commaST v]  ,\quad \mathcal{L}_{\xi}^{\star}(\zeta) = adj_{\xi}^{\star}(\zeta) \equiv adj_{D(\xi)}(\zeta).
\eeq

The generalization to the deformed algebra of tensorfields $\mathcal{T}_{\star}\in \Omega_{\star}\otimes... \otimes\Omega_{\star}\otimes \Xi_{\star}...\otimes\Xi_{\star}$ (with the $\star$-1-forms denoted as $\Omega_{\star}$) is now achieved through the $\star$-tensor-product,
\beq\label{tensor product star}
\forall
(\tau,\eta) \in \mathcal{T}_{\star}  : \quad  \tau \otimesST  \eta = \bar{\phi}^{\{\alpha\}}(\tau)\otimes\bar{\phi}_{\{\alpha\}}(\eta), 
\eeq
with the following non-commutativity,
\bea\label{tensor star product Lie derivative}
\tau \star h &=& \mathcal{L}^{\star}_{\bar{\phi}^{\{\alpha\}}}(\tau)\mathcal{L}^{\star}_{\bar{\phi}_{\{\alpha\}}}(h)=\mathcal{L}^{\star}_{\bar{\phi}_{\{\alpha\}}}(h)\mathcal{L}^{\star}_{\bar{\phi}^{\{\alpha\}}}(\tau)
\neweqline
&=&\mathcal{L}^{\star}_{\bar{\text{R}}^{\{\alpha\}}}(h)\star \mathcal{L}^{\star}_{\bar{\text{R}}_{\{\alpha\}}}(\tau) =\bar{\text{R}}^{\{\alpha\}}(h)\star\bar{\text{R}}_{\{\alpha\}}(\tau).\quad
\eea

The action of $\mathbf{U}(\Xi_{\star})$ on a general tensor is determined by the $\star$-Lie derivative, as in \eqref{star lie derivative tensors}. The action on each component is derived using \eqref{star lie derivative vectors}. Consequently, the algebra of tensorfields $\mathcal{T}_{\star}$ is also a $\mathbf{U}(\Xi_{\star})$-module algebra. Specifically, the $\star$-Lie derivative along $u\in\Xi_{\star}$ possesses a deformed Leibniz rule:
\beq\label{deformed Leibniz rule}
\mathcal{L}^{\star}_{u}(h\star g) = \mathcal{L}^{\star}_{u}(h)\star g +\bar{\text{R}}^{\{\alpha\}}(h)\star\mathcal{L}^{\star}_{\bar{\text{R}}_{\{\alpha\}}(u)}(g),
\eeq
which agrees with the $\star$-co-product \eqref{isomorphism map}.

With the above definitions of $\star$-vector and $\star$-tensor fields, it is apparent that the notion of commutative \textit{pairing} of $\star$-vector fields and $\star$-1-forms must be deformed (to maintain compatibility with the $\star$-Lie derivative and $\mathcal{A}_{\star}$-Linearity, see \cite{NC_Geometry_Simplified}). Thus a $\star$-pairing is introduced, 
\beq\label{star pairing definition}
\begin{split}
&\forall (\xi \in \Xi_{\star} , w\in\Omega_{\star}): 
\\
&(\xi,w) \mapsto \langle \xi \commaST w\rangle:=\langle \bar{\phi}^{\{\alpha\}}(\xi),\bar{\phi}_{\{\alpha\}}(w)\rangle.
\end{split}
\eeq
A proper basis of $\star$-vector fields and $\star$-1-forms must satisfy the duality condition dictated from \eqref{star pairing definition}. Choosing the $\star$-1-form basis as in the commutative scenario leads to,
\bea\label{basis condition}
\langle \partialST_{\mu}\commaST dx^{\nu} \rangle =\delta_{\mu}^{\nu} \Rightarrow \partialST_{\mu} &=& N_{\mu}^{-1_{\star}\nu}\star\partial_{\nu},
\neweqline
\text{where,}\ \
N^{\nu}_{\mu}&=& \langle \partial_{\mu}\commaST dx^{\nu} \rangle,
\eea
but this product does not generally have to be equal to $\delta_{\mu}^{\nu}$ as if it were classical.
This definition concludes the introduction of the theory of  NC geometry necessary for the present work.
\\

{\bf Let us now summarize the key ideas} presented in this subsection. Formula (\ref{non-commutativity}) defines the algebra of functions on the NCS, and (\ref{algebra isomorphism}, \ref{function isomorphism}) define the form of the isomorphism to the commutative scenario. In \eqref{perturbative representation of the twist}, the twist element was defined, with the twist-deformed Hopf algebra of vector fields in \eqref{twisted Hopf}. Using these definitions, we have shown the shape of the $\star$-product isomorphism map for functions in \eqref{function star product} and the representations of the deformed algebra that are consistent with a quantum differential calculus in (\ref{star lie derivative functions}-\ref{star lie derivative vectors}). Additional noteworthy formulae include the deformed Leibniz rule \eqref{deformed Leibniz rule}, the deformed definition of basis vector fields \eqref{basis condition}, and each of the  NC products (e.g., \eqref{tensor product star} for the tensor field representation).

\subsection{General Relativity on Non-Commutative Spaces}\label{section: background B}
Here, we demonstrate how the formulation from the previous section can be used to construct a deformed version of Einstein's equation on NCST given its flat limit\footnote{This section covers \textit{only known} material and the familiar reader can safely skip it.}; we again follow \cite{NC_Geometry_Simplified}, which relied on the studies in \cite{Majid:1994cy, QuantumGroup_GaugeTheory, TwistingQuantumDiff_PlanckScaleHopf}.
The underlying philosophy is that GR is a theory with an action invariant under infinitesimal diffeomorphisms, represented by vector fields and their Lie algebra structure. The previous section showed that the structure of infinitesimal diffeomorphism can be adapted to the NCS scenario, with consistent operations defining a differential geometry. Consequently, the construction will follow the classical approach to define the appropriate deformed geometrical quantities that are $\star$-isomorphic to the classical quantities in GR. Thus, the Einstein equation can be composed similarly to the classical scenario, leading to an equation whose associated action is invariant under \textit{deformed} infinitesimal diffeomorphisms.
Note that in contracts to the usual construction of GR, geometrical definitions, like the covariant derivative, curvature tensor, and so on, come before introducing a compatible metric. 
We begin with the  NC version of the covariant derivative. 

\textbf{$\star$-Covariant Derivative:} The map $\nablaST_{u}:\Xi_{\star}\rightarrow\Xi_{\star}$ is defined such that the following holds for all $(u,v,z)\in \Xi_{\star}$ and $h\in \mathcal{A}_{\star}$ \cite{NC_Geometry_Simplified}:
\begin{equation}\label{star covariant derivative}
    \begin{split}
        &\nablaST_{u+v}z =\nablaST_{u}z+\nablaST_{v}z ,\ \ \nablaST_{h\star u}v =h\star\nablaST_{u}v,\\
        &\nablaST_{u}(h\star v) = \mathcal{L}^{\star}_{u}(h)\star v + \bar{\text{R}}^{\{\alpha\}}(h)\star \nablaST_{\bar{\text{R}}_{\{\alpha\}}(u)}(v).
    \end{split}
\end{equation}
In a local coordinate neighborhood $\mathbf{U}$ with coordinate set $\{x^{\mu}\}$ and a suitable, $\star$-dual basis according to \eqref{basis condition}, the  NC connection coefficients are determined by\footnote{The classical counterpart of this relation is identical, just drop the $\star$-products and use the classical $\partial_{\mu}$'s.}:
\beq\label{connection formula}
\nablaST_{\partialST_{\mu}}\partialST_{\nu} = \Gamma_{\mu \nu }^{\sigma}\star\partialST_{\sigma} , \ \ \nablaST_{\partialST_{\mu}}dx^{\nu} = -\Gamma_{\mu \sigma }^{\nu}\star dx^{\sigma},
\eeq
where the $\star$-covariant derivative of a function $h\in \mathcal{A}_{\star}$ is the $\star$-Lie derivative:
\beq\label{star covariant derivative of function} 
\nablaST_{u}(h)=\mathcal{L}^{\star}_{u}(h)=\mathcal{L}_{D(u)}(h).
\eeq
For a general vector field $u$ and a 1-form $\omega$, one has the following rules 
\bea 
\nablaST_\mu u &=&\partialST_\nu\star(\mathbf{d}u^\nu+\Gamma^\nu_{\mu\rho}\star u^\rho), 
\neweqline
\nablaST_\mu \omega &=&(\mathbf{d}\omega_\nu-\omega_\rho\star\Gamma^\rho_{\nu\mu})\star dx^\nu,
\eea
respectively. 
The $\star$-covariant derivative of a general $\star$-tensorfield (of a $\star$-bi-vector field, of a $\star$-1-form, etc.) is defined using the deformed Leibniz rules:
\beq\label{star covariant derivative on tensor products}
  \nablaST_{u}(\tau \otimesST  \tau') = \nablaST_{u}(\tau)\otimesST  \tau' + \bar{\text{R}}^{\{\alpha\}}(\tau)\otimesST  \nablaST_{\bar{\text{R}}_{\{\alpha\}}(u)}(\tau').
\eeq
We Now define a \textbf{$\star$-Curvature} for all $(u,v,z) \in \Xi_{\star}$:
\begin{equation}\label{curvature definition}
     \textgoth{R}(u,v,z):=\nablaST_{u}\nablaST_{v}z
-\nablaST_{\bar{\text{R}}^{\{\alpha\}}(v)}\nablaST_{\bar{\text{R}}_{\{\alpha\}}(u)}z-\nablaST_{[u\commaST v]}z.
\end{equation}
In a local coordinate neighborhood, the expression \eqref{curvature definition} can be written using its components, and a \textbf{$\star$-Riemann tensor} and a \textbf{$\star$-Ricci tensor} are then defined,
\bea\label{coefficient frame curvature}
\textgoth{R}_{\mu \nu \rho}^{\sigma}\star\partialST_{\sigma}&=&\textgoth{R}(\partialST_{\mu},\partialST_{\nu},\partialST_{\rho})  ,
\\
\textgoth{R}_{\mu \nu }&=&\textgoth{R}(\partialST_{\mu},\partialST_{\nu}) := \langle dx^{\sigma}\commaST\textgoth{R}(\partialST_{\sigma},\partialST_{\mu},\partialST_{\nu})\rangle'.
\nonumber
\eea
The "\textbf{$'$}" in the contraction emphasizes that the contraction is between a form on the left and a vector field on the right---in NC geometry, the $\star$-Ricci tensor $\textgoth{R}_{\mu\nu}$ is not necessarily symmetric under the exchange of $\mu$ and $\nu$.
The \textbf{$\star$-Metric tensor} which we call $\mathbf{g}$ can be constructed by considering a symmetric rank-two $\star$-tensor in $\mathcal{T}_{\star}$, 
\beq\label{fundamental metric}
\mathbf{g}=\mathbf{g}^{a}\otimesST \mathbf{g}_{a} \in \Omega_{\star}\otimesST \Omega_{\star},
\eeq
the left $\mathcal{A}_{\star}$-linear map is well defined for $(u,v)\in \Xi_{\star}$,
\beq\label{metric consistency stared}
\begin{split}
    & \mathbf{g}: \Xi_{\star}\otimesST \Xi_{\star} \rightarrow \mathcal{A}_{\star},\\
    &(u,v)\mapsto \mathbf{g}(u,v) =\langle u \otimes v\commaST \mathbf{g}\rangle := \langle u\commaST\langle v\commaST\mathbf{g}\rangle\rangle.
\end{split}
\eeq
In a local neighborhood, with a suitable $\star$-basis, one can write (see \cite{NC_Geometry_Simplified}) for the $\star$-metric component, and for its associated $\star$-inverse metric\footnote{From the component expression one can understand the $"\star"$ in the metric inverse; we demand that $\mathbf{g}_{\mu\nu}\star\overset{\star}{\mathbf{g}}^{\nu\rho}=\delta_{\mu}^{\rho}$} $\mathbf{g}^{-1^\star}$,
\bea\label{non-commutative metric component}
\mathbf{g}=dx^{\nu}\otimesST dx^{\mu}\star\mathbf{g}_{\mu \nu },
\quad\mathbf{g}^{-1^{\star}}=\invG^{ \mu\nu}\star \partialST_{\nu}\otimesST \partialST_{\mu}.
\eea

The condition of the metric being compatible as a manifold (assuming a torsion-free metric) is, 
\beq \label{metric compatibility}
\nabla^{\star
}_{u}\mathbf{g}=0.
\eeq

Then, the connection can be determined in terms of the metric, and the
\textbf{$\star$-Ricci scalar} can thus be constructed.
We here consider the
\textit{left $A_\star$-linear} contraction rule: the inverse metric always
acts from the left (see
\cite{Gravity_Non,Twist_general_2,NC_Geometry2}).
Therefore,
\bea\label{star definition of the Ricci scalar} 
\textgoth{R}&:=&\textgoth{R}((g^a)^{-1},g_a^{-1})=\textgoth{R}(\overset{\star}{\mathbf{g}}^{ \mu\nu}\star\partialST_\nu,\partialST_{\mu})
\neweqline
&=&\overset{\star}{\mathbf{g}}^{ \mu\nu}\star \textgoth{R}_{\nu \mu},
\eea
where the order of indices reflects the left $A_\star$-linearity of the contraction.
Because the $\star$-product is non-commutative the contraction in
\eqref{star definition of the Ricci scalar} is not unique: a right-acting definition would
differ by a total $\star$-commutator. However, when considering the \textit{action formulation} (see \eqref{deformed Einstein Hilbert}) and $\star$-products constructed by an Abelian twist, such terms integrate to zero \cite{Gravity_Non}.

With these definitions for the geometrical objects, it is now
possible to write a deformation of GR on NCST (see the discussion in
\cite{NC_Geometry3}) through the deformed Einstein equation which is
supposed to be, by definition, invariant under deformed infinitesimal
diffeomorphisms.
Thus, one can write the Einstein tensor and the
Einstein equation in vacuum,
\beq\label{deformed EE}
\textgoth{G}_{\mu \nu }=\textgoth{R}_{\mu \nu }-\frac{1}{2}\mathbf{g}_{
\mu \nu }\star\textgoth{R}=0,
\eeq
where we again use the left-multiplication convention in
$\mathbf g_{\mu\nu}\star\textgoth{R}$ (see
\cite{Gravity_Non}).
This equation can be perturbatively expanded in
the deformation parameter $\lambda$, which appears in the twist
\eqref{perturbative representation of the twist}.
However, two notes are in order.  
\textbf{First}, the $\star$-Ricci scalar defined in~\eqref{star definition of the Ricci scalar} is generally complex.  
One way to handle this is to work in an \textit{action formulation} and add the appropriate complex-conjugate term~\cite{Gravity_Non}.  
\textbf{Second}, in \textit{vacuum} Eq.~\eqref{deformed EE} can be rewritten as $\textgoth{R}_{\mu\nu}=0$.  
This form, however, is likewise not fully bona fide~\cite{Metric_Perturbations_In_NC}.  
Several prescriptions exist to interpret it consistently, and we discuss both points in Sec.~\ref{section: The Construction of Gravity} around ~\eqref{expanded defomed EE} and~\eqref{equation from action!}.

In summary, we now have all the formulations necessary for constructing a GR theory on a NCST. The main features to take from this formulation are the deformed actions of the various Lie derivatives $\mathcal{L}^{\star}$, the $\star$-product appearing in all definitions, and the deformed Leibniz rules (such as \eqref{star covariant derivative on tensor products}). Finally, we are left with equation \eqref{deformed EE}, describing the dynamics of a NCS.

\section{Building blocks of geometry in $\kp$-Minkowski (position-dependent) non-commutativity}\label{section: Building Blocks For Construction} 
Here, we shall outline known and new results that will pave the road for our desired construction of a  NC gravity theory (of the Einstein Tensor) that extends a local $\kp$-Minkowski spacetime, in which the non-commutativity is position-dependent. For the first time, we give the explicit action of the twist on the basis of vector fields and one-forms. These yield a clear and tractable characterization of the deformed basis and Leibniz rule, which we verify using the previously obtained results in \cite{kappa_minkowki_Twist, IGL_TWIST_1}. Again, the detailed derivations of various expressions are found in Appendix \ref{Appendix: Building Blocks} to keep the discussion compact.   

The $\kp$-Minkowski spacetime's non-commutativity (see Sec.\ref{section: Introduction}), can be written in the \textit{operator} representation as 
\beq \label{Kappa minkowsky}
[\hat{x}^0,\hat{x}^{i}]= \frac{i}{\kp}\hat{x}^{i} \  , \ \ [\hat{x}^{i},\hat{x}^{j}] = 0\ , \ \ \ i,j =1,2,3.
\eeq
this is equivalent to the second term in \eqref{General non commutation} with the anti-symmetric constant tensor $C_{\rho}^{\mu\nu}$, while setting the first term ($\theta^{\mu\nu}$) to zero (we use here $\lambda:=1/\kp=l_{p}$ as the deformation parameter): 
\beq\label{dependent: comparison to general non-commutativity}
C_{\rho}^{\mu\nu}=\frac{1}{\kp}(\delta_{0}^{\mu}\delta_{\rho}^{\nu}-\delta_{0}^{\nu}\delta_{\rho}^{\mu}):= \lambda\delta_{[0}^{\mu}\delta_{\rho]}^{\nu}.
\eeq

We would now like to utilize the $\star$-isomorphism \eqref{algebra isomorphism} for this non-commutativity. In principle, it is possible to construct a $\star$-product suitable for this non-commutativity that will be a module algebra of the $\kp$-Poincaré algebra (see Sec.\ref{section: Introduction} and \cite{Hopf_Book} with references therein) and to derive a deformation quantization procedure from establishing gauge theories on that space.
At first sight, this seems a preferable approach; it has a $\kp$-Minkowski spacetime structure, preserves SR symmetries in the classical limit, and adds a fundamental length scale as an additional invariant.
However, when one computes the suitable differential calculus \cite{Star_Hopf_Algebra, Differential_Kappa_Poincare} and the associated $\star$-product \cite{star_product_kappa1,star_product_kappa2}, a profound ambiguity arises: the integral on $\star$-products of fields is not cyclic.
Hence, no gauge theory can be built out with this $\star$-product \cite{Twist_general_2, Hopf_Book}.
Several approaches exist to bypass this ambiguity: one is to consider a measure function $\mu(x)$ that will make the integral cyclic \cite{measure_function1,measure_function2}; however, such a measure function will spoil the classical limit $\lambda\rightarrow0$ since it is not identity there. Others are to consider different notions of integration \cite{measure_integral3}, etc., see \cite{Twist_general_2,ongoing_integral1,ongoing_integral2} for ongoing research on the subject.

Here, we consider approaching the spacetime \eqref{Kappa minkowsky} as a result of an abelian twist deformation as outlined in previous sections. This way, we are guaranteed to have a meaningful definition of the integral. However, there is no twist element $\Phi \in \mathbf{U}(ISO) \otimes \mathbf{U}(ISO)$ such that \eqref{kappa minkowski} is a module of the twisted algebra $\mathbf{U}(ISO)_{\star}$. This non-existence was demonstrated in \cite{NO_KAPPA_TWIST}, where a condition for an Abelian twist to produce $\kappa$-Minkowski was derived:
\beq\label{condition twisting}
\exists X\in\Xi \ \text{s.t.} \ [X,P_{0}]=P_{0}. 
\eeq
This condition is not fulfilled for $\Xi = ISO(3,1)$. Another way to show the non-existence is to note that while the algebraic sector remains classical in the twist formalism, a bona fide $\kappa$-Poincaré must be deformed in its algebra. To see why, we examine (part of) the co-algebra in the \textit{classical basis} of the $\kappa$-Poincaré (classical algebraic sector) as considered in \cite{Classical_basis}:
\bea\label{classical basis co-product}
\Delta_{\kappa}(P_{0}) &=& P_{0} \otimes \Pi_{0} + \Pi_{0}^{-1} \otimes P_{0} + \frac{1}{\kappa} P_{m}\Pi_{0}^{-1} \otimes P^{m},
\neweqline
    \Delta_{\kappa}(P_{i}) &=& P_{i} \otimes \Pi_{0} + 1 \otimes P_{i},
\eea
with 
\beq\label{Definition for co-product} 
\Pi_{0} \equiv \frac{1}{\kappa}P_{0} +\sqrt{1-\frac{P^2}{\kappa^2}} \quad \text{and} \quad \Pi_{0}^{-1}\equiv \frac{\sqrt{1-\frac{P^2}{\kappa^2}}-\frac{P_{0}}{\kappa}}{1-\frac{P^2}{\kappa^2}}
.
\eeq
The square roots are understood as a formal power series. Thus, the deformation parameter $\kappa$ must remain formal, which makes the physical meaning of the basis ambiguous.

Overall, to get the $\kappa$-Minkowski spacetime as a module of a twisted algebra, we must enlarge the Poincaré group so that condition \eqref{condition twisting} is fulfilled. The minimal enlargement \cite{A_Non_Commutative_From_Ads_Algebra, IGL_TWIST_1, Quantum_AntiDeSitter} is to consider the $\mathcal{WP}$-algebra (Weyl-Poincaré), which is the Poincaré algebra enlarged with a dilatation generator $D=x^{\mu}\partial_{\mu}$ and is a subalgebra of so(4,2). Here, for the benefit of using an abelian twist, we consider (as in \cite{Twistgeneral1, kappa_minkowki_Twist}) enlarging the Poincaré algebra to the $\text{igl(3,1)}=\text{gl(3,1)}\rtimes t^{4}$
; the semi-direct product of the general linear algebra gl(3,1) with the 4-translations algebra $t^4$. 
In the vector field representation and the Schwinger realization\footnote{Note that in \cite{IGL_TWIST_1}, they used slightly different notation. Namely, $L^{\nu}_{\mu},P_{\mu}$ were used as the basis for igl(3,1), with the representation $L^{\nu}_{\mu}=x^{\nu}\partial_{\mu}$. The representation \eqref{Generators of IGL} is equivalent through lowering indices with the metric.}, we write for the generators of igl(3,1), 
\beq\label{Generators of IGL}
P_{\mu}=\partial_{\mu}, \quad L_{\mu\nu}=x_{\mu}\partial_{\nu},
\eeq
with the associated Lie algebra,
\bea\label{Lie algebra of IGL}
&&[P_{\mu},P_{\nu}]=0,\quad [L_{\mu\nu},P_{\rho}]=\eta_{\mu\rho}P_{\nu}, \neweqline
&&[L_{\mu\nu},L_{\rho\sigma}]=\eta_{\nu \rho}L_{\mu\sigma}-\eta_{\mu\sigma}L_{\rho\nu}.
\eea
In comparison, the Poincar\'e algebra $\text{iso(3,1)}=\text{so(3,1)}\rtimes t^4$, the algebra that generates the symmetry of SR, and the one that $\kp$-Poinca\'e is built upon is generated by the following vector fields:
\beq\label{Poincare generators}
P_{\mu}=\partial_{\mu},\quad M_{\mu\nu}= x_\mu\partial_\nu-x_\nu\partial_\mu.
\eeq
Hence, igl(3,1) can be thought of as the Poincar\'e algebra enlarged by the elements $\TenD_{\mu\nu}=\eta_{\mu\nu}\ \text{tr}\left[L_{\mu\nu}\right]$, as discussed in Sec.\ref{ssection: IW-contraction our case}, where we deal with the physical consistency of the enlargement.
\\\\
By enlarging the symmetry, we can now consider the general form of \textit{Abelian Twists}, defined with vector fields $(X_{a}, X_{b})\in \Xi$ \cite{Abelian_Twist1, Realization_of_Hopf}:
\bea\label{abelian twist1}
\Phi&=&\exp[-\frac{i}{2}\Lambda\Theta^{ab} X_{a}\otimes X_{b}]=\phi^{\{\alpha\}}\otimes \phi_{\{\alpha\}},
\neweqline
\Phi^{-1}\equiv\bar{\Phi}&=&\exp[+\frac{i}{2}\Lambda\Theta^{ab} X_{a}\otimes X_{b}]\equiv\bar{\phi}^{\{\alpha\}}\otimes \bar{\phi}_{\{\alpha\}}.
\eea
In this general form, the vector fields $X_a$ and $X_b$ mutually commute (hence the name \textit{Abelian}), and the matrix $\Theta^{ab}$ is anti-symmetric with constant entries.

For our specific case, the twist \eqref{abelian twist1} generating the desired twisted algebra is given by \cite{A_Non_Commutative_From_Ads_Algebra, Quantum_AntiDeSitter, kappa_minkowki_Twist}:
\beq\label{twist vector fields}
\begin{split}
&\Theta^{00}=0,\ \Theta^{01}= 1,\
\Theta^{10}=-1,\ \Theta^{11}=0,\\
&X_{0}=\partial_{0},\ X_{1}=x^{j}\partial_{j} \in \text{IGL(3,1)}.
\end{split}
\eeq
Writing the twist explicitly, we get:
\beq\label{dependent: kappa twist}
\Phi = \exp\left[-i\frac{\lambda}{2}\left(\partial_{0}\otimes x^{j} \partial_{j}-x^{j}\partial_{j}\otimes \partial_{0}\right)\right].
\eeq
The associated permutation operator \eqref{Universal R-matrix} is, 
\beq \label{dependent: permutation kappa}
\OprR=\Phi_{21}\Phi^{-1}=\exp\left[+i\lambda\left(\partial_{0}\otimes  x^{j}\partial_{j}-x^{j}\partial_{j}\otimes\partial_{0}\right)\right].
\eeq

We first want to show that \eqref{dependent: kappa twist} implies \eqref{Kappa minkowsky} (which is a known result \cite{kappa_minkowki_Twist}, quoted here for completeness). Thus, we calculate the $\star$-product to second order, 
\bea\label{dependent: star product kappa}
f\star g &:=&\multi_\star(f\times g)=fg
+i\frac{\lambda}{2}x^{j}f_{,[0} g_{,j]}
\\
&&+\frac{1}{8}(i\lambda)^2x^jx^l\left(f_{,0[0}\cdot g_{,l]j}
+ f_{,j[l}\cdot g_{,0]0}
\right)
+\mathcal{O}(\lambda^3),
\nonumber
\eea 
where $\multi_\star$ is the $\star$-multiplication map $\multi_\star(a\otimes b):=a\star b$. 

The form of $\bar{\phi}_{\{\alpha\}}\otimes\bar{\phi}^{\{\alpha\}}$ for an Abelian twist is, 
\bea\label{dependent: side 1}
\bar{\phi}_{\{\alpha\}}\otimes\bar{\phi}^{\{\alpha\}}&=&\exp\left[i\frac{\lambda}{2}\Theta^{ba}X_{b}\otimes X_{a}\right].
\eea
Using \eqref{twist vector fields}, we can explicitly calculate \eqref{star lie brackets} to derive,
\bea\label{dependent: computation of brackets}
[x^{\mu}\commaST x^{\nu}]&=&\bar{\phi}^{\{\alpha\}}(x^{\mu})\bar{\phi}_{\{\alpha\}}(x^{\nu})-\bar{\phi}_{\{\alpha\}}(x^{\nu})\bar{\phi}^{\{\alpha\}}(x^{\mu})
\neweqline 
&=&
x^{\mu}x^{\nu}+i\frac{\lambda}{2}x^{j}\delta_{[0}^{\mu}\delta_{j]}^{\nu}- x^{\nu}x^{\mu}-i\frac{\lambda}{2}x^{j}\delta_{[0}^{\nu}\delta_{j]}^{\mu}
\neweqline
&=&i\lambda x^{j}\delta_{[0}^{\mu}\delta_{j]}^{\nu}
\\
\ergo [x^{0}\commaST x^{j}]&=& i\lambda x^{j}=-[x^{j}\commaST x^{0}],
\nonumber
\eea
in accordance with \eqref{kappa minkowski}, confirming that \eqref{dependent: kappa twist} is a suitable twist.
The next step is constructing a suitable $\star$-dual basis of vector fields using the condition \eqref{basis condition}. Thus, we must compute the action of the twist on 1-forms and vectors, i.e., to calculate the Lie derivative \eqref{tensor product star} along the twists' vector fields \eqref{twist vector fields} of the basis 1-forms and basis vector fields. Note that in \cite{General_notion_of_Twist_2}, the basis was found by different means. However, we calculate with the $\star$-dual basis construction; by comparing the results with those in \cite{General_notion_of_Twist_2}, we can verify our side calculations of the twist's actions on the bases. Then, we can use them for calculations in the next section \ref{section: The Construction of Gravity}.

The vector fields act on basis 1-forms as (see \eqref{App. action on 1forms}),
\bea\label{action on 1forms}
X_{0}(dx^{\mu})=0,\quad X_{1}(dx^{\mu})=\delta_{j}^{\mu}dx^{j}\neq 0.
\eea
And, on the basis vectors as (see \eqref{App. action on basis vectors}),
\bea \label{action on basis vectors assumed}
X_{0}(\partial_{\mu})=0,  
\quad
X_{1}(\partial_{\mu})=-\delta^{j}_{\mu}\partial_{j}.
\eea
Thus, we are not dealing with a \textit{nice} basis (as defined in \cite{Metric_Perturbations_In_NC}); the following calculations will not coincide with theirs. The formulae (\eqref{action on 1forms},\eqref{action on basis vectors assumed}) dictate the action of the twist on the classical basis of vector fields and of 1-forms, which we now outline. However, we first define a new notation for brevity:
\beq\label{Notation Definition O[l,n]}
O^{\mu'}_\mu[\lambda,n]\equiv e^{in\frac{\lambda}{2}\partial_{0}}\delta^{j}_{\mu}\delta_{j}^{\mu'}+\delta^{0}_{\mu}\delta_{0}^{\mu'}.
\eeq
Using this notation, we now write the (explicit) twist's action:
\begin{subequations}\label{partial twist action on classical derivatives}
\begin{align}
\label{partial twist action on classical derivatives a}\phi^{\{\alpha\}}(\partial_{\mu})\phi_{\{\alpha\}}&= O^{\mu'}_{\mu}[\lambda,-1]\partial_{\mu'}
\\
\label{partial twist action on classical derivatives b}
\bar{\phi}^{\{\alpha\}}(\partial_{\mu})\bar{\phi}_{\{\alpha\}}&=O^{\mu'}_{\mu}[\lambda,+1]\partial_{\mu'}
\\
\label{partial twist action on classical derivatives c}
\phi^{\{\alpha\}}(dx^{\mu})\phi_{\{\alpha\}}&= O^{\mu}_{\mu'}[\lambda,+1]dx^{\mu'}
\\
\label{partial twist action on classical derivatives d}
\bar{\phi}^{\{\alpha\}}(dx^{\mu})\bar{\phi}_{\{\alpha\}}&=O^\mu_{\mu'}[\lambda,-1]dx^{\mu'}
\end{align}
\end{subequations}
By use of \eqref{basis condition}, we now explicitly derive the $\star$-pairing of the classical basis (see \eqref{App. basis proof}),
\bea\label{basis proof}
\langle\partial_{\mu}\commaST dx^{\nu}\rangle&=&
\buphi(\partial_{\mu})\bdphi(x^\nu)
=O^{\mu'}_{\mu}[\lambda,1]\partial_{\mu'}(x^\nu)
\neweqline
&=&O^{\mu'}_{\mu}[\lambda,1]\delta_{\mu'}^\nu
=O^\nu_\mu[\lambda,1].
\eea
The $\star$-dual basis is therefore (denoted with a $"\star"$),
\beq \label{kappa basis differentials} 
\partialST_{0}=\partial_{0}   ,  \ \partialST_{j}=e^{-i\frac{\lambda}{2}\partial_{0}}\partial_{j} \ \Leftrightarrow \ \langle \partialST_{\mu}\commaST dx^{\nu} \rangle =\delta_{\mu}^{\nu}.
\eeq

In \cite{General_notion_of_Twist_2}, various results were derived through a different method from ours. We now recalculate them using our method and results to mutually validate them.

\textbf{The non-commutativity of basis 1-forms:} by definition we have, 
\bea\label{Deriving 1-form non-commutativity_1}
f\star dx^{\mu} &=& \bar{\text{R}}^{\{\alpha\}}(dx^{\mu})\star\bar{\text{R}}_{\{\alpha\}}(f),
\eea
then, utilizing \eqref{action on 1forms} and \eqref{partial twist action on classical derivatives}, we derive (see \eqref{App. Deriving 1-form non-commutativity_1}),
\bea 
f\star dx^{\mu}=dx^{\mu'}\star O_{\mu'}^{\mu}[\lambda,2](f),
\eea
in agreement with the known results in \cite{Twistgeneral1}. 

\textbf{The deformed Leibniz rule of $\star$-basis vectors:} in \cite{Twistgeneral1} they derived the appropriate Leibniz rule by first considering:
\bea\label{1-form differential}
  \mathbf{d}f  &=& (\partial_{\mu}f)dx^{\mu} = (\partialST_{\mu}f)\star dx^{\mu},
 \\
  \mathbf{d}(f\star g)  &=& \mathbf{d}f\star g + f\star \mathbf{d}g = (\partialST_{\mu}(f\star g))\star dx^{\mu}.
  \nonumber
\eea
Then, by equating between the first and the second equalities in the second line, they derived the following Leibniz rule (see \eqref{1-form differential Appendix}), 
\bea\label{deformed Leibniz rule on star derivatives}
\partialST_{\mu}(f\star g)=\partialST_{\mu'}(f)\star O^{\mu'}_{\mu}[\lambda,-2](g)+f\star(\partialST_{\mu}g).\quad
\eea

The validity of \eqref{partial twist action on classical derivatives} can now be confirmed by recalculating the Leibniz rule, this time with \eqref{star lie derivative functions} and by calculating the $\star$-co-product of the $\star$-derivative (same as the one for $\partial$ by simple replacement) from \eqref{isomorphism map}; after calculations, we derive the $\star$-co-product (see \eqref{App. Star co-product}),
\bea 
\Delta_\star(\partialST_{\mu})=\partialST_{\mu'}\otimes O^{\mu'}_{\mu}[\lambda,-2]\,+\,1\otimes \partialST_{\mu},
\eea
which produces the deformed Leibniz rule (see \eqref{App. Leibniz rule final}),
\bea 
\partialST_{\mu}(f\star g)=\partialST_{\mu'}(f)\star O^{\mu'}_{\mu}[\lambda,-2](g)+f\star(\partialST_{\mu}g),\ \ 
\eea
in agreement with \eqref{deformed Leibniz rule on star derivatives}, supporting the result in \eqref{partial twist action on classical derivatives}.

This concludes our analysis of the foundations of geometry in the NCST \eqref{Kappa minkowsky}. We note a significant difference compared to the constant non-commutativity (see e.g., \cite{Gravity_Non}); no 'trivial' twist actions appear here; the differential calculus is not classical, etc. Consequently, we anticipate differences in the deformed GR, which we will develop and analyze in Sec.\ref{section: The Construction of Gravity}, after making sure that the construction will yield a \textit{consistent} physical theory, which is the subject of the following two sections.

Consequently, we expect deviations in the deformed GR, which will be developed and analyzed in Sec.\ref{section: The Construction of Gravity}. However, before proceeding, we will ensure the construction results in a \textit{consistent} physical theory in the next two sections.

\section{Relativistic Consistency of the Non-Commutative Relations }\label{section: Consistency of The NC-Relations}

In this section, our goal is to establish that the NCST \eqref{kappa minkowski}, along with the twisted symmetry framework we employed in Sec.~\ref{section: Building Blocks For Construction}, maintains a consistent physical meaning \textit{in the context} of its \textbf{relativistic invariance}; the first consistency condition $\mathcal{A}$  outlined in Sec.\ref{section: Overview of Introduction}. To achieve this, with regard to the translational part of deformed symmetries, we parallel the study in \cite{NonCommutative_Spacetime_Interpretation1}, which considered kinematical Hilbert space realizations for NCST variables and their associated deformed transformations. This study drew on earlier works \cite{A_No_Pure_Uncertainty, Generalizing_The_Noether} that established the non-commutativity of transformation parameters, such as the non-commutativity of translations derived from the requirement that translated points remain within the $\kappa$-Minkowski spacetime.

However, the studies mentioned above focused on the $\kappa$-Minkowski spacetime with $\kappa$-Poincar\'e symmetry in the bicrossproduct basis (see Sec.~\ref{section: Introduction}).
In contrast, our approach uses the twisted IGL(3,1) symmetry, a broader structure incorporating dilatations in addition to Lorentz transformations and translations (see Sec.\ref{section: Building Blocks For Construction}). While these frameworks are not algebraically equivalent -- the twisted IGL(3,1) is a twist-deformed version of a dilatation-enlarged Poincar\'e group, and $\kappa$-Poincar\'e is a quantum deformation of the Poincar\'e algebra without dilatations -- they share a common translational sector. Thus, to apply the insights of \cite{NonCommutative_Spacetime_Interpretation1} to our setting, we must verify the consistency of translations, and more generally, of the full symmetry transformations under the twist we use here.

In Sec.~\ref{subsec: Consistency till now}, we review how the non-commutativity of translation transformations was determined in the bicrossproduct basis, providing context and additional insights. Then, in Sec.~\ref{subsec: consistency transl}, we derive the structure of the deformed translations in our twisted-IGL(3,1) case and show their consistency. We find that translations remain undeformed, thereby enabling the definition of exterior differential calculus and confirming that the Leibniz rule remains intact under the translational twisted symmetry.

However, to fully ensure that our framework admits a physically consistent realization of $\kappa$-Minkowski spacetime, it is necessary to go beyond translations and examine the consistency of the remaining generators of the symmetry algebra: Lorentz transformations and dilatations. This extended consistency analysis is performed in Sec.~\ref{subsec: consistency igl}, where we derive explicit expressions for the deformed Lie derivatives $\lie^\star_\xi$ corresponding to all generators $\xi \in gl(3,1)$ (the homogeneous part of igl(3,1)), and compute their associated $\star$-coproducts $\Delta_\star(\xi)$. With these tools, we evaluate the following three consistency conditions\footnote{In the context of $\kappa$-Poincaré, the term “consistent differential calculus” often refers to a bicovariant differential structure, which necessitates a five-dimensional basis including an extra differential element (see \cite{Diff_calculus_1, 5d_DIff}). This is distinct from the four-dimensional constructions used in the twist approach, where consistency is defined via compatibility with the star-product and twisted coproducts. Such 4D frameworks are employed, for example, in \cite{Generalizing_The_Noether, Differential_Kappa_Poincare, Hamiltonian_Considirations}, where differential structures are constructed directly from the twist element without invoking a bicovariant extension.}:
\begin{itemize}
    \item $\mathcal{A.}1$ Differential compatibility: $\lie^\star_\xi(\mathbf{d}f) = \mathbf{d}(\lie^\star_\xi(f))$
    \item $\mathcal{A.}2$ Invariance of $\kappa$-Minkowski under deformed transformations: $[\lie^\star_\xi(x^\mu)\commaST x^\nu] \in \mathcal{M}_\kappa$
    \item $\mathcal{A.}3$ Deformed Leibniz rule consistency: $\lie^\star_\xi([x^\mu\commaST  x^\nu]) = \mu_\star(\Delta_\star(\xi)(x^\mu \otimes x^\nu - x^\nu \otimes x^\mu))$
\end{itemize}

Each of these conditions is verified explicitly in Sec.~\ref{subsec: consistency igl}, with particular attention paid to the non-trivial case of boost transformations, where the use of the undeformed Leibniz rule is apparent. Overall, setting the structure of the deformed Lie actions and coproducts reveals how the twisted framework preserves the covariance of $\kappa$-Minkowski spacetime and supports the physical validity of our generalized symmetry construction. Some of the results in this section are summarized in Table.~\ref{summary-table}.

\subsection{Consistency: Translations in $\kappa$-Poincar\'e}\label{subsec: Consistency till now}

In the framework of q-deformations, fields such as spacetime variables are inherently non-commutative. Adopting the \textit{time to the right} convention\footnote{in addition to the \textit{metric to the left} convention}\cite{A_No_Pure_Uncertainty, Generalizing_The_Noether}, we represent functions as
\beq\label{Fourier time to right}
f(\hat{x}) = \int d^4k e^{i\vec{k} \cdot \vec{\hat{x}}} e^{-ik_0\hat{x}^0} \Tilde{f(k)},
\eeq
with the $\hat{x}$-coordinates exhibiting the \eqref{kappa minkowski}  NC, while $k$-parameters are considered classical (commutative), following e.g. \cite{SWmaps}.
Considering  \cite{NonCommutative_Spacetime_Interpretation1}, the infinitesimal translation $T = 1 + \mathbf{d}$ of a function along a vector field $\epsilon \in T(\mathcal{M}_\kappa)$ in the $\kappa$-Minkowski manifold $\mathcal{M}_\kappa$ is expressed with the following differential one-form (a function's exterior derivative, see eq.(3.4) in \cite{Generalizing_The_Noether}):
\beq\label{classical diff one-form}
\mathbf{d}f(\hat{x}) = \epsilon^\mu \left(P_\mu f(\hat{x})\right),
\eeq
with $P_\mu$ as the real-defined translations written in \eqref{Poincare generators}.\footnote{Here, $\mathbf{d}f$ is understood as a 1-form acting on the vector field $\epsilon$ via the dual pairing $\langle \mathbf{d}f, \epsilon \rangle = \epsilon^\mu P_\mu f$, yielding the directional derivative along $\epsilon$.}
To determine the non-commutativity of $\epsilon^{\mu}$, one translates the point  $\hat{x}^{\mu}$ in the $\kappa$-Minkowski spacetime:
\beq\label{translation of operator coordinates}
\hat{x}^{\mu} \mapsto \hat{\Tilde{x}}^{\mu} = \hat{x}^{\mu} + \epsilon^\nu(\partial_\nu)\hat{x}^\mu = \hat{x}^{\mu} + \epsilon^{\mu},
\eeq
and require that the translated point remains within $\kappa$-Minkowsky spacetime. This leads to the commutation relation:
\beq\label{commutation of operator coordinate}
[\epsilon^{\mu}, \hat{x}^{0}] = i\lambda \epsilon^{j} \delta_{j}^{\mu}.
\eeq

To validate the consistency with $\kappa$-Poincar\'e, one starts with the Leibniz rule for differentials:
\beq\label{Leibniz rule classical}
\mathbf{d}(fg) = \mathbf{d}(f)g + f\mathbf{d}(g). 
\eeq
However, in the bicrossproduct basis, the co-product of the $P_{\mu}$ generator is given by \cite{Majid:1994cy}:
\beq\label{co-product bicross}
\Delta_{\kappa}(P_{\mu}) = P_{\mu} \otimes 1 + O^{\mu'}_{\mu}[\lambda,-2] \otimes P_{\mu'},
\eeq
which, by use of \eqref{classical diff one-form}, gives an alternative expression for the Leibniz rule: 
\bea\label{Leibniz co-product bicross}
\mathbf{d}(fg)  
&=&i\epsilon^{\sigma}P_{\sigma}(fg)
=\mu\{\Delta_{\kappa}(P_{\sigma})(f\otimes g)\}
\neweqline
&=&i\epsilon^{\sigma}[ (k_\sigma + O^{\sigma'}_{\sigma}[\lambda,-2] q_{\sigma'} ) fg ]
\neweqline
&=& (i\epsilon^{\sigma}P_{\sigma}f)g + i\epsilon^{\sigma}O^{\sigma'}_{\sigma}[\lambda,-2]f(\hat{x})(P_{\sigma'}g(\hat{x})) .\quad\quad
\eea
Equating this with \eqref{Leibniz rule classical} for consistency, one derives the condition (note that the $'$ in $\epsilon^{\mu'}$ stands for a dummy index):
\bea\label{equating with Leibniz classical}
0 &=& \left(f(\hat{x})\epsilon^{\mu'} - \epsilon^{\mu}O^{\mu'}_\mu[\lambda,-2]f(\hat{x})\right)P_{\mu'}g(\hat{x}) 
\neweqline
&\Rightarrow&
f(\hat{x})\epsilon^{\mu'} = \epsilon^{\mu}O^{\mu'}_\mu[\lambda,-2]f(\hat{x}). 
\eea
This condition is satisfied by \eqref{commutation of operator coordinate}, as shown in \cite{Generalizing_The_Noether}, thereby demonstrating that the $\kappa$-Poincaré translational symmetry and the $\kappa$-Minkowski spacetime are compatible, in the sense that \eqref{co-product bicross}, \eqref{commutation of operator coordinate}, and \eqref{classical diff one-form} are consistent with \eqref{Leibniz rule classical}. 
It is important to note that the translation parameters $\epsilon^{\mu}$ must be promoted to noncommutative quantities, as dictated by \eqref{commutation of operator coordinate}. This ensures consistency between the $\kappa$-Poincar\'e coproduct \eqref{co-product bicross}, the operator commutation relation \eqref{commutation of operator coordinate}, and the classical Leibniz rule \eqref{Leibniz rule classical} when applied to the differential form \eqref{classical diff one-form}.

\subsection{Consistency: Twisted Translational Symmetry}\label{subsec: consistency transl}
In this subsection, we consider the problem within our setup, i.e., when no NC functions are represented with the $\star$-isomorphism generated by the twisted-IGL(3,1) group outlined in Sec.\ref{section: Building Blocks For Construction}. Our goal is to demonstrate that translations in the twisted symmetry preserve the invariance of \eqref{kappa minkowski} and are consistent with the structure of differential calculus (with the Leibniz rule as done previously, cf. \eqref{equating with Leibniz classical}). 

We consider the differential one-form structure in noncommutative geometry within the twisted Hopf algebra approach (see, e.g., \cite{Twistgeneral1}). The differential of a function $f$ is given by
\bea\label{differential form star}
\mathbf{d}f = (\partial_{\mu}f)\,dx^{\mu}
= (\partialST_{\mu}f) \star dx^{\star\mu}
= (\partialST_{\mu}f) \star dx^{\mu},
\eea
where the last equality follows from using a suitable $\star$-basis as constructed in \eqref{basis proof} (see also eq.(2.22) in \cite{Twistgeneral1}). In this basis, the sets $\{\partial^\star_\mu\}$ and $\{dx^\mu\}$ satisfy the deformed dual pairing $\langle \partial^\star_\mu, dx^\nu \rangle_\star = \delta^\nu_\mu$, as in \eqref{kappa basis differentials}. 

The Leibniz rule in  NC geometry for the exterior derivative $\mathbf{d}$ (and specifically for $\mathbf{d}$, see \cite{pedadogical_twits_formalism} and Sec.\ref{subsec: consistency igl} where such a Leibniz rule does not hold) is given by the
\textit{undeformed Leibniz rule}:
\bea\label{Leibniz rule using star non-commutativity}
\mathbf{d}(f \star g) = \mathbf{d}(f) \star g + f \star \mathbf{d}(g),
\eea
which ensures consistency with the deformed symmetry action. 
On the other hand, the twisted co-product of the derivative operator (written here for $\partial_\mu$, and equivalently valid for $\partialST_\mu$ by simple replacement) is given by:
\bea\label{twisted co-product}
\Delta^{\star}(\partial_{\mu}) = \partial_{\mu'} \otimes O^{\mu'}_\mu[\lambda,-2] + 1 \otimes \partial_\mu.
\eea
Consequently, by use of \eqref{UEA on functions} and the identification \eqref{classical diff one-form}, the associated Leibniz rule is:
\bea\label{Leibniz co-product ours}
\mathbf{d}(f \!\star\! g) &=& \partial_{\mu}(f \star g) dx^{\mu} = \mu_{\star} \circ \{\Delta_{\star}(\partial_{\mu})(f \otimes g)\}
\neweqline
&=& ((\partial_{\mu'} f) \!\star\! O^{\mu'}_\mu[\lambda,-2] (g) + f \!\star\! (\partial_{\mu} g)) dx^{\mu}\,. ~~~~~~
\eea

To verify that \eqref{Leibniz rule using star non-commutativity} is identical to \eqref{Leibniz co-product ours}, we proceed as follows:
\bea\label{Leibniz identity ours}
\mathbf{d}\left(f\!\star\! g\right)&=& (\partial_{\mu}(f)dx^{\mu})\!\star\! g+f\!\star\!\left(\partial_{\mu}(g)dx^{\mu}\right)
\neweqline
&=& \partialST_{\mu} f \!\star\! \left(\bar{\text{R}}^{\{\alpha\}} (g) \!\star\! \bar{\text{R}}_{\{\alpha\}} (dx^{\mu})\right) + f \star (\partialST_{\mu} g \!\star\! dx^{\mu})
\neweqline
&=& \left[(\partialST_{\mu} f) \!\star\! O^{\mu'}_\mu[\lambda,-2](g) + f \!\star\! (\partialST_{\mu'} g) \right]\!\star\! dx^{\mu'}
\neweqline
&=& \left[\left(\partial_{\mu}f\right)\!\star\! O^{\mu'}_\mu[\lambda,-2](g)+f\!\star\!(\partial_{\mu'}g)\right]dx^{\mu'}.
\eea

We applied the non-commutativity relation to the one-form basis $dx^{\mu}$ utilizing the $\text{R}$-matrix, yielding: 
\bea\label{non-commutativity of form basis}
dx^{\mu}\star g &=& \bar{\text{R}}^{\{\alpha\}}(g)\star\bar{\text{R}}_{\{\alpha\}}(dx^{\mu}) \nonumber \\
&=& O^{\mu'}_\mu[\lambda,-2](g)\star dx^{\mu'}.
\eea
Thus, consistency -- namely, that the operator $\mathbf{d}(\cdot) = \partial_\mu(\cdot) dx^\mu$ satisfies the twisted Leibniz rule \eqref{Leibniz rule using star non-commutativity} -- follows from the non-commutativity of the product with a one-form basis.

Regarding the translation differential, we must use the definition of the $\star$-Lie derivative in NCST.
Using \eqref{star lie derivative functions} and \eqref{partial twist action on classical derivatives b}, we calculate the \textbf{$\star$-Lie derivative} for a constant vector field $\xi$, 
\bea \label{kappa Lie function}
\lie_{\xi^\mu\partialST_{\mu}}^{\star}(h) &=& \xi^\mu\mathcal{L}_{\bar{\phi}^{\{\alpha\}}(\partialST_{\mu})\bar{\phi}_{\{\alpha\}}}(h)
\\
&=& \xi^\mu O^{\mu'}_{\mu}[\lambda,1] \partial_{\mu'}^{\star}(h) 
=\xi^\mu\partial_{\mu}(h),
\nonumber
\eea
which can be applied to the generator of infinitesimal translations because of its constant nature: 
\beq\label{Lie derivative translation star}
\mathcal{L}^{\star}_{\epsilon^{\mu}\partial_{\mu}^{\star}}(f(x))=\epsilon^{\mu}\partial_{\mu}(f(x)):= \mathbf{d}(f).
\eeq
We can simplify the $\star$-Lie derivative since $\epsilon^{\mu}$ is constant. If $\epsilon^{\mu}$ were not constant, the $\star$-Lie derivative would be more complex, see the next section. 

Comparing \eqref{Lie derivative translation star} with \eqref{differential form star}, we assess that $dx^{\mu}$ and $\epsilon^{\mu}$ are on equal footing in the manner the $\star$ product acts on them.
Thus, the relation \eqref{non-commutativity of form basis} also holds for the translation parameters:
\beq
\epsilon^{\mu}\star g= \bar{\text{R}}^{\{\alpha\}}(g)\star\bar{\text{R}}_{\{\alpha\}}(\epsilon^{\mu})=O^{\mu}_{\mu'}[\lambda,-2](g)\star \epsilon^{\mu'},
\eeq
where summation over the dummy index $\mu'$ is implied. Substituting $g = x^{\mu}$, we obtain:
\bea\label{epsilon star NC}
\epsilon^{\mu}\!\star\! x^{\nu}&=&\bar{\text{R}}^{\{\alpha\}}(x^{\nu})\!\star\! \bar{\text{R}}_{\{\alpha\}}(\epsilon^{\mu})
=O^{\mu}_{\mu'}[\lambda,-2](x^{\nu})\!\star\! \epsilon^{\mu'}
\neweqline
&=&x^{\nu}\star \epsilon^{\mu}-i\delta_{\mu'}^j\delta_{j}^{\mu}\lambda\partial_{0}(x^{\nu})\star \epsilon^{\mu'}
\neweqline \ergo\text{[}x^\nu,\epsilon^\mu\text{]}_{\star}&=& i\lambda\delta_{[0}^{\mu}\delta_{j]}^{\nu}\epsilon^{j}.
\eea
This validates that the spacetime \eqref{kappa minkowski} is invariant under \eqref{Lie derivative translation star}. From here we turn to show that the translation differential \eqref{Lie derivative translation star} exhibits consistency as a differential form by exploiting \eqref{epsilon star NC} and exhibiting equality between the $\star$-Leibniz rule \eqref{Leibniz rule using star non-commutativity} and the Leibniz rule derived from \eqref{twisted co-product}.  
Following a procedure similar to that used in equations (\eqref{Leibniz rule classical}-\eqref{equating with Leibniz classical}), we obtain the consistency condition: 
\beq\label{consistency star epsilon}
f(x)\star\epsilon^{\mu}=\epsilon^{\mu'}\star O^{\mu}_{\mu'}[\lambda,-2]f(x).
\eeq
To solve this, we first apply \eqref{Fourier time to right} (note that now there are no \textit{hats} on the co-ordinates, and the  NC is expressed via the $\star$-product) to obtain:
\bea\label{fourier decomposition classical}
e^{-\lambda  P_{0}} f(x) &=& e^{-\lambda k_{0}} f(x)
\neweqline 
\Rightarrow
e^{-ik_{0}x^{0}}\star\epsilon^{\mu} &=& \epsilon^{\mu'}\star e^{-ik_{0}x^{0}}\left(e^{-\lambda k_{0}}\delta_{\mu'}^j\delta_{j}^{\mu}+\delta_{\mu'}^0\delta_{0}^{\mu}\right)\quad\quad
\eea
Implying the requirement (for the non-trivial spatial part):
\beq\label{commutator condition classical} 
\left[\epsilon^{j},
        e^{-ik_{0}x^{0}}
    \right]_{\star}
= \epsilon^{j}\star
    e^{-ik_{0}x^{0}}
    \left(1-e^{-\lambda k_{0}}\right).
\eeq
 
On the other hand, by use of \eqref{epsilon star NC}, we find (see \eqref{appendix: epsilon non-commutativity}):
\beq
\left[\epsilon^{j}, (x^0)^n             \right]_{\star}
= \epsilon^{j}
    \star
    \left[ \left(x^0\right)^n - \left(x^0 - A\right)^n \right], 
\eeq
where $ A = i\lambda $. This leads to:
\bea\label{exponential commutator result}
\left[\epsilon^{j}, e^{-ik_{0}x^0}\right]_{\star} &=& \sum_{n=0}^{\infty} \frac{1}{n!} (-ik_{0})^{n} [\epsilon^{j}, (x^0)^n]_{\star}
\neweqline
&=& \epsilon^{j} \star\left(e^{-ik_{0}x^0} - e^{ik_{0}(A-x^0)}\right)
\neweqline
&=& \epsilon^{j}\star e^{-ik_{0}x^0}\left(1-e^{-\lambda k_{0}}\right).
\eea
Therefore, the fact that the translation differential carries the non-commutativity \eqref{epsilon star NC}\footnote{Note that one could go the other way around, i.e., by insisting that for consistency \eqref{commutator condition classical} hold, \eqref{epsilon star NC} must be true.} ensures that the non-commutativity relation \eqref{commutator condition classical}, derived from the Leibniz rule \eqref{Leibniz rule using star non-commutativity} and from \eqref{twisted co-product}, is satisfied, as demonstrated in \eqref{exponential commutator result}. Consequently, as shown for \eqref{classical diff one-form}, the differential \eqref{Lie derivative translation star} remains consistent with the $\star$-Leibniz rule and can be interpreted as a differential form of the  NC translation parameters.

\subsection{Consistency: Twisted General Linear Symmetries}\label{subsec: consistency igl}

Having established the consistency of the deformed translations which turned out to remain undeformed, we now turn to analyze the behavior of the remaining generators in $ \xi \in igl(3,1) $: namely, the Lorentz transformations and dilatations. In doing so, we will have an additional and non-trivial test of the internal consistency of the deformed symmetry structure, deriving several general expressions for deformed quantities, for later, broader use.

In the case of translations, we focused on checking the following schematic relations (with $ T := \epsilon^\mu \partialST_\mu $ for brevity):
\begin{align}
\mathbf{d}(f \star g) &= \multi_\star\left\{ \Delta_\star(\mathbf{d})(f \otimes g) \right\}\nonumber
\\
&= \mathbf{d}(f) \star g + f \star \mathbf{d}(g),\nonumber
\\
[\lie^\star_T(x^\mu)\commaST x^\nu] &\overset{?}{=} -i \lambda \epsilon^{j} (\delta_j^\mu\delta_0^\nu-\delta_j^\nu\delta_0^\mu), 
\\
[\epsilon^\mu\commaST x^\nu] &= i \lambda \epsilon^{j} \delta_{[0}^{\mu}\delta_{j]}^{\nu}.
\end{align}
Here, $ \mathbf{d}(\cdot) $ is the exterior derivative constructed from the deformed translations, and $ \multi_\star(a \otimes b) := a \star b $ denotes the deformed multiplication map. The validity of these relations clarified the fact that translations allowed us to construct a differential satisfying the classical Leibniz rule. However, this property is special to translations.

We can no longer assume such undeformed behavior for
a generic transformation $\xi\in gl(3,1)$. Specifically, acting with $ \lie^\star_\xi $ on a $ \star $-product of functions requires using the deformed Leibniz rule (see Eq.~\eqref{star lie derivative functions}):
\beq\label{deformed lie leibniz}
\lie^\star_\xi(f\star g) := \xi^\star(f\star g) = \multi_\star\{\Delta_\star(\xi^\star)(f \otimes g)\}.
\eeq

To assess the consistency of the deformed $ gl(3,1) $ transformations, we outline the following steps:
\begin{enumerate}
    \item Compute the action of $ \lie^\star_\xi $ for each $ \xi \in gl(3,1) $.
    \item Compute the deformed coproduct $ \Delta_\star(\xi) $ for all $ \xi \in gl(3,1) $.
    \item \textbf{Consistency $\mathcal{A.}1$:}
    Verify the compatibility with the differential: $ \lie^\star_\xi(\mathbf{d}(f)) = \mathbf{d}(\lie^\star_\xi(f))$.
    \item \textbf{Consistency $\mathcal{A.}2$:} Check the algebraic closure: $ [\lie^\star_\xi(x^\mu)\commaST x^\nu] = i\lambda x^{\mu'} \delta_{\mu'}^j \delta^\mu_j \delta^\nu_0 $.
    \item \textbf{Consistency $\mathcal{A.}3$:} Validate that\\
    $ \lie^\star_\xi([x^\mu\commaST x^\nu]) = \multi_\star\left(\Delta_\star(\xi)(x^\mu \otimes x^\nu - x^\nu \otimes x^\mu)\right) $.
\end{enumerate}
Note that Consistency $\mathcal{A.}3$ addresses compatibility at first order, as it involves only the coordinates $ x $ and $ x' $, not general functions $ f $ and $ g $. This should be contrasted with a naive attempt to impose the undeformed coproduct by checking $ [\lie^\star_\xi(x^\mu)\commaST x^\nu] + [x^\mu\commaST \lie^\star_\xi(x^\nu)] $, which we will later see fails to hold. Once these steps are verified, we can conclude that the deformed symmetry transformations of $ gl(3,1) $ preserve consistency at the level of the algebraic structure.

\subsubsection*{Evaluation of Differential Quantities}
We begin with the first task: evaluating $ \lie^\star_\xi $ for non-constant vector fields. The derivation is deferred to Appendix~\eqref{appendix: Lie derivative non non constant}, and we summarize the result here:
\beq\label{Lie derivative non non constant}
\lie^{\star}_{\xi^{\mu}\partialST_{\mu}} \!\! =
\!\!\sum_{\substack{n=0\\n'=0}}^{\infty}\!\!\!
\frac{\left(i\lambda/2\right)^{n+n'}}{n!n'!}
[(-x^j\partial_{j})^{n'}[\partial_{0}^{n}(\xi^{\mu})\partialST_{\mu}]\partial_{0}^{n'}(x^{j}\partial_{j})^{n}],
\eeq
with the first order given by 
\bea\label{first order lie}
\lie^{\star}_{\xi}
&=&
\xi^{\mu}\partialST_{\mu}
+\frac{i\lambda}{2}\left[x^j\partial_{[0,}(\xi^\mu)\partial_{j]}+\xi^\mu\delta_j^\mu\partial_0\right]\partialST_{\mu}.
\eea
with anti-symmetrization notation $"[]"$, see the end of Subsec.\ref{section: Overview of Introduction}. 
Higher-order terms ($ \mathcal{O}(\lambda^2) $) are provided in Eq.~\eqref{appendix: Lie derivative non non constant expansion}.

To compute $\lie^\star_\xi$ for $\xi\in gl(3,1)$, let us first recall the homogeneous part of the differential representation of $gl(3,1)$ in \eqref{Generators of IGL}: 
\beq
L_{\mu\nu}:=x_\mu\partial_\nu,\quad M_{\mu\nu}:=x_\mu\partial_\nu-x_\mu\partial_\nu
\eeq

Let us then evaluate first the expression of $\lie^\star_\xi$ for $\xi=x_\nu\partialST_\mu$ to first non-trivial order. Applying Eq.~\eqref{first order lie}, we thus compute:
\bea\label{star lie for x^mu partial_mu}
\lie^\star_{x_\nu\partialST_\mu} &=&x_\nu\partialST_\mu
\neweqline
&&+\frac{i\lambda}{2}\left[x^j\left(\partial_0(x_\nu)\partial_j -\partial_j(x_\nu)\partial_0\right)+x_\nu\delta_j^{\mu'}\delta_\mu^j\partialST_{\mu'}\partial_0\right]
\neweqline
&=& x_\nu\partialST_\mu
+\frac{i\lambda}{2}\left(x_\nu\delta_j^{\mu'}\delta_{\mu}^j\partial_0\right)\partialST_{\mu'}+\frac{i\lambda}{2}x^j\delta_{\nu[0,}\partial_{j]}\partialST_\mu
\neweqline
&=&x_\nu\partial_\mu+\frac{i\lambda}{2}x_\nu\partial_{\mu'}\left(-(\partial_0\delta_j^{\mu'}\delta_{\mu}^j) + (\partial_0\delta_j^{\mu'}\delta_{\mu}^j) \right)
\neweqline
&&+\frac{i\lambda}{2}x^j\delta_{\nu[0,}\partial_{j]}\partial_\mu+\mathcal{O}(\lambda^2)
\neweqline
&=&\boxed{x_\nu\partial_\mu+\frac{i\lambda}{2}x^j\partial_\mu\delta_{\nu[0,}\partial_{j]}=\lie^\star_{x_\nu\partialST_\mu}}\, ,
\eea
We now extract the deformed transformations for specific generators (note that all the results ignore $O(\lambda^{n\geq2})$): 
\\

\textbf{Dilatation:}
\bea\label{star lie dilation}
\lie^\star_{x_\mu\partialST_\mu}&=&x_\mu\partial_\mu+\frac{i\lambda}{2}x^j\partial_{[0}\partial_{j]}
=\boxed{x_\mu\partial_\mu}=\lie_{x_\mu\partial_\mu}.
\eea
This is the classical version of the transformation, which should come as no surprise because the twist is composed of dilatation, just as happened for the translations in the previous section.

\textbf{Lorentz Transformations:}
For the generators $ M_{\nu\mu}=x_\nu \partial_\mu - x_\mu \partial_\nu:=x_{[\nu,}\partial_{{\mu]}} $:
\bea\label{star lie four rotations}
\lie^\star_{(x_\nu\partialST_\mu-x_\mu\partialST_\nu)}&=&
x_{[\nu,}\partial_{\mu]}
\\
&&+\frac{i\lambda}{2}x^j\left[\partial_{\mu}\delta_{\nu[0,}\partial_{j]}-\partial_\nu\delta_{\mu[0,}\partial_{j]}\right].\nonumber
\eea
As a result, the deformed \textbf{boost} transformations are 
\bea\label{star lie boost}
\lie^\star_{(x_0\partialST_i-x_i\partialST_0)}=\boxed{\lie_B+\frac{i\lambda}{2}x^j\Box_{E}}\,,
\eea
where $ B := x_0\partial_i - x_i\partial_0 $ and $ \Box_E := \partial_0^2 + \partial_j^2 $ is the Euclidean d'Alembertian. 
Finally, the deformed \textbf{spatial rotation} can also be calculated
\bea\label{star lie rotations}
\lie^\star_{(x_i\partialST_j-x_j\partialST_i)}= \boxed{\lie_R-\frac{i\lambda}{2}\partial_0\lie_R}\, , 
\eea
where $R=x_i\partial_j-x_j\partial_i$. 

In summary, we have obtained the explicit form of all deformed $ gl(3,1) $ generators to $ \mathcal{O}(\lambda) $. Generalizing to higher orders might introduce exponential corrections involving operators like $ e^{\lambda^n x^j \cdots \partial^n} $, but such effects are beyond the scope of the current treatment.

Next, we evaluate $ \Delta_\star(x_\nu \partial_\mu) $ via the isomorphism relation \eqref{isomorphism map}. This requires computing the twisted coproduct $ \Delta^\Phi(x_\nu \partial_\mu) $. While this result is available in the literature (e.g., \cite{IGL_TWIST_1}), we rederive it here in a manifestly covariant form, which simplifies the evaluation of $ \Delta_\star $.

We begin with its definition:
\bea\label{twisted co gl 1}
\Delta^\Phi(x_\nu\partial_\mu)&=&\Phi(x_\nu\partial_\mu\otimes1+1\otimes\ x_\nu\partial_\mu)\Phi^{-1} 
\\
&=&\left(\phi^{\{\alpha\}}(x_\nu\partial_\mu)\otimes\phi_{\{\alpha\}}\right.
\neweqline
&&+\left.\phi^{\{\alpha\}}\otimes\phi_{\{\alpha\}}(x_\nu\partial_\mu)\right)(\bphi^{\{\beta\}}\otimes\bphi_{\{\beta\}}).\nonumber
\eea
To evaluate such an expression, and for later use in $\Delta_\star$, we need the computation of $\phi^{\{\alpha\}}(x_\nu\partial_\mu)\phi_{\{\alpha\}}$. But before that, let us evaluate $\phi^{\{\alpha\}}(x_\nu)\phi_{\{\alpha\}}$, using the shorthand $\phi^{\{\alpha\}}(\partial_\mu)\phi_{\{\alpha\}}=O^{*\mu'}_{\mu}\partial_{\mu'}$ for the notation in \eqref{Notation Definition O[l,n]}: 
\bea\label{phi(x)} 
\phi^{\{\alpha\}}(x_\nu)\phi_{\{\alpha\}}&=&\phi^{\{\alpha\}}(x_\nu)\otimes \phi_{\{\alpha\}} =x_\nu\otimes1
\neweqline
&&- \frac{i\lambda}{2}\left[\partial_0(x_\nu)\otimes x^j\partial_j-x^j\partial_j(x_\nu)\otimes \partial_0\right]
\neweqline
&=&x_\nu\otimes1-\frac{i\lambda}{2}x^j\left[\delta_{0\nu}\otimes\partial_j-\delta_{j\nu}\otimes \partial_0\right] \Rightarrow
\neweqline
\phi^{\{\alpha\}}(x_\nu)\phi_{\{\alpha\}}&=&x_\nu-\frac{i\lambda}{2}x^j\left(\delta_{0\nu}\partial_j-\delta_{j\nu} \partial_0\right):=B^{*}_\nu\,. 
\eea
And now we can evaluate: 
\bea\label{phi(x partial)}
\phi^{\{\alpha\}}(x_\nu\partial_\mu)\!\otimes\!\phi_{\{\alpha\}}&=&\phi^{\{\alpha\}}(x_\nu)\partial_\mu\!\otimes\!\phi_{\{\alpha\}}
\!+\!x_\nu\phi^{\{\alpha\}}(\partial_\mu)\!\otimes\!\phi_{\{\alpha\}}
\neweqline
&=&\partial_\mu\otimes B^{*}_\nu+x_\nu\partial_{\mu'}\otimes O^{*\mu'}_{\mu}\Rightarrow 
\neweqline
\phi^{\{\alpha\}}\!\otimes\!\phi_{\{\alpha\}}(x_\nu\partial_\mu)&=&B_\nu\otimes\partial_\mu+x_\nu O^{\mu'}_{\mu}\otimes\partial_{\mu'},
\eea
where $(B_\nu, O^{\mu'}_\mu)$ are the complex conjucates of $(B^{*}_{\nu},O^{*\mu'}_{\mu})$. 
Applying these expressions to the last equality in \eqref{twisted co gl 1}, we derive
\bea\label{twisted co gl 2}
\Delta^\Phi(x_\nu\partial_\mu) &=& \partial_\mu\otimes B^{*}_{\nu}+x_\nu\partial_{\mu'}\otimes O^{*\mu'}_{\mu}
\neweqline
&&+B_\nu\otimes\partial_\mu+x_\nu O^{\mu'}_{\mu}\otimes\partial_{\mu'}
\neweqline
&=&\partial_\mu\otimes x_\nu+\frac{i\lambda}{2}\left(-\delta_{0\nu}\partial_\mu\otimes D +\delta_{j\nu}\partial_\mu\otimes x^j\partial_0 \right) 
\neweqline
&&+x_\nu\partial_{\mu'}\otimes O^{*\mu'}_\mu +x_\nu\otimes\partial_{\mu}
\neweqline
&&+\frac{i\lambda}{2}\left(\delta_{0\nu} D\otimes\partial_\mu-\delta_{j\nu} x^j\partial_0\otimes\partial_\mu\right)+x_\nu O^{\mu'}_{\mu}\otimes \partial_{\mu'} 
\neweqline
&=& \left[x_\nu\partial_{\mu'}\otimes e^{-\frac{i\lambda}{2}\partial_0\delta^{\mu'}_j\delta_\mu^j}+x_\nu e^{\frac{i\lambda}{2}\partial_0\delta^{\mu'}_j\delta_\mu^j}\otimes\partial_{\mu'}\right]
\neweqline
&&+\frac{i\lambda}{2}\left(\delta_{0\nu}[D\otimes\partial_\mu-\partial_\mu\otimes D]\right)
\\
&&+\frac{i\lambda}{2}\left(\delta_{j\nu}[x^j\partial_\mu\otimes\partial_0-x^j\partial_0\otimes \partial_\mu]\right)+\mathcal{O(\lambda^2)}, \nonumber
\eea
where we used $D=x^j\partial_j$. This result, when inserting the relevant $(\mu,\nu)$, matches the known result in \cite{IGL_TWIST_1}, and note that it matches it \textit{exactly}, exhibiting that expanding $B_\nu$ only to $\mathcal{O}(\lambda)$ gives the full result. 

We now have all the tools for computing $\Delta_\star$; specifically, we shall use the previous result, the action \eqref{phi(x partial)}, and the fact that, to first order in $\lambda$, terms of the form $\lambda\phi(\cdot)\phi\sim\lambda(\cdot)+ \mathcal{O}(\lambda^2)$ are approximated to $\lambda(\cdot)$:
\bea
\Delta_\star(x_\nu\partial_\mu)&=&(\phi^{\{\alpha\}}\phi_{\{\alpha\}}\!\otimes\!\phi^{\alpha'}\phi_{\alpha'})\bigg[L_{\nu\mu'}\otimes O^{*\mu'}_{\mu}+O^{\mu'}_{\mu}\otimes L_{\nu\mu'}
\neweqline
&&+\frac{i\lambda}{2}\left(\delta_{j\nu}[x^j\partial_\mu\otimes\partial_0-x^j\partial_0\otimes \partial_\mu]\right)
\neweqline
&&+\frac{i\lambda}{2}\left(\delta_{0\nu}[D\otimes\partial_\mu-\partial_\mu\otimes D]\right) \bigg]
(\bphi^{\{\beta\}}\otimes\bphi_{\{\beta\}})
\neweqline
&=&L_{\nu\mu'}\otimes S^{*\mu'}_{\mu}+1\otimes L_{\mu\nu}
\neweqline
&&+\frac{i\lambda}{2}\left(\delta_{0\nu}[D\otimes\partial_\mu-\partial_\mu\otimes D]\right)
\neweqline
&&+\frac{i\lambda}{2}\left(\delta_{j\nu}[x^j\partial_\mu\otimes\partial_0-x^j\partial_0\otimes \partial_\mu]\right),
\eea
where we used $S^{*\mu'}_\mu$ as denoting $O^{\mu'}_{\mu}[\lambda,-2]$. Further simplifying, we finally derive:
\bea\label{star co gl 1}
\Delta_\star(x_\nu\partial_\mu)&=&\Delta(L_{\nu\mu})-i\lambda \delta^{\mu'}_j\delta_{\mu}^j L_{\nu\mu'}\otimes \partial_0 
\neweqline
&&+\frac{i\lambda}{2}\left(\delta_{0\nu}[D\otimes\partial_\mu-\partial_\mu\otimes D]\right)
\neweqline
&&+\frac{i\lambda}{2}\left(\delta_{j\nu}[x^j\partial_\mu\otimes\partial_0-x^j\partial_0\otimes \partial_\mu]\right).~~~~
\eea

This is the final result for $\Delta_\star(L_{\nu\mu})$, and one must use it when considering the action of $\star$-elements on $\star$-products, as in \eqref{Leibniz co-product ours}. Note that the second term in the last equality is what we searched for in evaluating $\Delta_\star$; it has the form of a $\star$-co-product like \eqref{twisted co-product}.

\subsubsection*{Evaluation of Consistency Conditions}
We are now equipped to verify the three consistency conditions for the deformed differentials we constructed. To do so, note that all of the deformed transformations in Eqs.~\eqref{star lie dilation}, \eqref{star lie boost}, and \eqref{star lie rotations} can be schematically expressed as
\bea\label{star lie schematic}
\lie^\star_\xi\sim \lie_\xi+G(\lambda,x)\partial^2,
\eea
where $G(\lambda,x)$ is a function of the indicated parameters and $\partial^2$ denotes second-order derivatives such as $\partial_0^2,\partial_i\partial_0$, etc.

\textbf{Consistency $\mathcal{A.}1$:} by use of \eqref{star lie schematic}, we can see that $\lie_\xi^\star$ commutes with the classical exterior derivative $\mathbf{d}\sim\partial_\mu(\cdot)dx^\mu$:
\bea\label{commutation lie exterior}
[\lie^\star_\xi,\mathbf{d}]\sim[\lie_\xi+G\partial^2,\mathbf{d}]=[\lie_\xi,\mathbf{d}]=0.
\eea
We used the commutativity of the classical $\mathbf{d}$ with the classical $\lie_\xi$, and the fact that $\partial^2$ also trivially commutes with $\mathbf{d}$ (write $\partial^2=\partial\cdot\partial$, and use the commutation of $\partial$ with $\partial$). 
Given the commutation we showed, it is immediate to see that we have consistency: 
\bea\label{gl consistency a}
\lie^\star_\xi(\mathbf{d}f)=\mathbf{d}(\lie^\star_\xi(f)). 
\eea
Note that this equality is not trivially satisfied for all conceivable deformations. In particular, if the deformation introduces terms of the form $ G(\lambda,x)$ without second derivatives (i.e., terms without $\partial^2$, as in \eqref{star lie schematic}), then the commutator in \eqref{commutation lie exterior} will generally no longer vanish.

\textbf{Consistency $\mathcal{A.}2$:} again by use of the schematic form \eqref{star lie schematic}, we observe the following
\bea\label{lie star on x}
\lie^\star_\xi(x^\mu)\sim\lie_\xi(x^\mu)+G\partial^2(x^\mu)\sim\lie_\xi(x^\mu)\equiv \tilde{x}^{\mu}. 
\eea
In the last equality, we used the fact that $x^\mu$ belongs to the (flat) Minkowski spacetime $\mathcal{M}$, and as such, for all Lorentz generators $\xi$ we have $\lie_\xi(x^\mu)\in \mathcal{M}$; this remains true also for dilatation: for all $\xi \in gl(3,1)$, we indeed have $\lie_\xi(x^\mu)\in \mathcal{M}$. Although the result $\tilde{x}^\mu$ is not generally $x^\mu$ itself, it remains within the Minkowski coordinate algebra, e.g., $\lie_B(x^\mu)=x_{[i,}\delta_{j]}^0\in \mathcal{M}$. Using \eqref{lie star on x}, we can then trivially verify that
\bea\label{consistency gl 2}
[\lie^\star_\xi(x^\mu)\commaST x^\nu]\sim [\tilde{x}^{\mu}\commaST x^\nu]\sim i\lambda x^j\delta^{\mu}_{[0}\delta_{j]}^\nu\in \mathcal{M}_\kappa. 
\eea
Therefore, $\lie^\star_\xi$ transformations leave the spacetime \eqref{kappa minkowski} invariant.

\textbf{Consistency $\mathcal{A.}3$:} 
We shall now turn to validate the consistency of $\lie^\star$ with respect to its $\star$-Leibniz rule. Let us consider the case of $\lie^\star_{B}$, with $B$ the boost generator. Such an example is not trivial, involving various results we have derived, and thus functions as a good overall consistency evaluation.
We consider boosts to be represented by $M_{j0}=x_j\partialST_0-x_0\partialST_j$, and note that from \eqref{star lie schematic} we know that $\lie^\star_B(x^\mu)=\lie_B(x^\mu)$ since $\partial^2(x^\mu)=0$. The $\star$-co-product $\Delta_\star(B)$ can be read off from the general expression \eqref{star co gl 1} to derive 
\bea\label{star co boost}
\Delta_\star(M_{0j})&=&\Delta(M_{j0})+i\lambda\left[M_{j0}\otimes\partial_0
+\frac{1}{2}(D\otimes\partial_j-\partial_j\otimes D)\right]
\neweqline
&=& \boxed{\Delta(M_{j0})-i\lambda M_{j0}\otimes\partial_0}.
\eea

And let us recall that we must find that the following holds:
\beq\label{consistency c condition}
\lie^\star_\xi([x^\mu\commaST x^\nu])=\multi_\star(\Delta_\star(M_{\rho\sigma})(x^\mu\otimes x^\nu-x^\nu\otimes x^\mu)).
\eeq
We begin with the LHS of the equality, focusing on the non-trivial situation of $(\rho=0,\sigma=j)$ \eqref{kappa minkowski} or interchanged\footnote{i.e, we are not considering the cases where \eqref{kappa minkowski} vanishes and everything is trivial}: 
\bea\label{LHS of consis c}
\lie^\star_{M_{j0}}([x^\mu\commaST x^\nu])&=&\lie^\star_{M_{j0}}([x^0\commaST x^k])
\\
&=&\lie_\xi^\star(i\lambda x^j)=i\lambda\lie_{M_{j0}}(x^k)=\boxed{-i\lambda x^0\delta_j^k}\,.\nonumber 
\eea

Now we shall compute the RHS in \eqref{consistency c condition}
\bea\label{RHS cons c}
&& \multi_\star\left[(\Delta(M_{j0})-i\lambda M_{j0}\otimes\partial_0)(x^0\otimes x^k-x^k\otimes x^0)\right]
\neweqline
&=& \multi_\star\left[\right. M_{j0}(x^0)\otimes x^k+x^0\otimes M_{j0}(x^k)
\neweqline
&&~~~~~-M_{j0}(x^k)\otimes x^0-x^k\otimes M_{j0}(x^0)\left.\right]
\neweqline
&&~~~~~+\multi_\star\left[-i\lambda M_{j0}(x^0)\otimes \partial_0(x^k)+i\lambda M_{j0}(x^k)\otimes\partial_0(x^0)\right]
\neweqline
&=&\multi_\star\left[x^j\otimes x^j-x^0\otimes x^0+x^0\otimes x^0-x^j\otimes x^j\right]
\neweqline
&&~~~~~+i\lambda\multi_\star\left[-x^j\otimes\partial_0(x^k)-\delta_j^kx^0\otimes1\right] =\boxed{-i\lambda x^0\delta_j^k},
\eea
where the last equality holds to all orders since $x^0\star1=x^0$, ensuring a match with the LHS \eqref{LHS of consis c}. This confirms the consistency condition \eqref{consistency c condition} for the non-trivial case of deformed boosts. Note, using the undeformed co-product would have resulted in an inconsistency, as the RHS would have vanished while the LHS would not have. For the remaining generators $M_{\nu\mu}$, including dilatations and spatial rotations, the structure ensures simpler consistency, and detailed verification is omitted here.

\subsection*{Summary: Twisted Symmetries}

To summarize, our goal in this section was to establish that the deformed symmetry transformations in $\kappa$-Minkowski spacetime, defined through the twist-deformed IGL(3,1) group, follow the $\mathcal{A}$ outlined in Subsec.\ref{section: Overview of Introduction}: the deformed symmetries are consistent from a differential-geometric perspective and preserve the noncommutative structure \eqref{kappa minkowski} without introducing a preferred frame.

Initially, we verified this explicitly for translations. We demonstrated that the twisted differential form \eqref{differential form star} obeys the appropriate Leibniz rule \eqref{Leibniz co-product ours}, consistent with its classical counterpart \eqref{Leibniz identity ours}. Subsequently, using the deformed Lie derivative for translations \eqref{Lie derivative translation star}, we showed that the translated point remains within the $\kappa$-Minkowski spacetime, and that consistency is maintained in the differential calculus sense (\eqref{consistency star epsilon}, \eqref{fourier decomposition classical}, \eqref{exponential commutator result}). Importantly, unlike \cite{NonCommutative_Spacetime_Interpretation1}, where translation parameters were postulated to be noncommutative, here this property arose naturally from the twist structure and the $\star$-product calculus.

Building on this, we extended the analysis to the full $ gl(3,1) $ algebra in Subsec.~\ref{subsec: consistency igl}. There, we computed the $\star$-Lie derivatives \eqref{star lie dilation}, \eqref{star lie boost}, and \eqref{star lie rotations}, verifying that they reduce to their classical counterparts up to corrections of order $\mathcal{O}(\lambda)$ involving second-order derivatives, as captured schematically in \eqref{star lie schematic}. We then performed the three explicit consistency checks of $\mathcal{A}$: $\mathcal{A.}1:$ The commutation of the deformed Lie derivative with the exterior derivative \eqref{commutation lie exterior}. $\mathcal{A.}2:$ The preservation of the $\kappa$-Minkowski commutation relations under $gl(3,1)$ transformations \eqref{consistency gl 2}. $\mathcal{A.}1:$ The compatibility with the deformed Leibniz rule via the coproduct \eqref{consistency c condition}, including a nontrivial demonstration for boost transformations (\eqref{LHS of consis c} vs.~\eqref{RHS cons c}).

Crucially, these results show that even the deformed boosts, which are often considered problematic in $\kappa$-spacetime contexts, are \textit{fully consistent} within our twisted framework. The coproducts of the generators, such as \eqref{star co gl 1} and the specific case of boosts \eqref{star co boost}, play a central role in ensuring this consistency.

Therefore, our construction proves that the full deformed igl(3,1) symmetry algebra, including translations, dilations, rotations, and boosts, can be consistently implemented on $\kappa$-Minkowski spacetime using twist deformation techniques, without violating the differential geometric structure or the invariance of the noncommutative spacetime algebra.

\section{Contraction of the Symmetry}\label{Contraction of the Symmetry}
Before developing the deformed gravity theory with $\kp$-Minkowski as its flat limit, a concern arises regarding the physical realism of such a theory. As outlined in Sec.\ref{section: Building Blocks For Construction}, we had to extend the local symmetry to IGL(3,1) before applying the twist. Consequently, after constructing GR within this twisted symmetry framework, the classical limit of the resulting theory would reproduce the classical GR equations, but with a local symmetry described by IGL(3,1) - instead of the expected local symmetry of classical GR, the Poincaré group ISO(3,1). This discrepancy implies that the classical limit of the constructed deformed GR would have an incorrect local symmetry, violating the correspondence principle, raising concerns about the physicality of the deformed GR.

In this section, in the pursuit of the \textit{second consistency} $\mathcal{B}$ mentioned in Sec.~\ref{section: Overview of Introduction}, we propose a solution for maintaining the correspondence principle in the context of deformed GR. The full deformed GR theory, incorporating this solution, will be derived in Sec.~\ref{section: The Construction of Gravity}. 
The proposal relies on employing the In\"on\"u-Wigner (IW) contraction procedure to address this correspondence issue. In Subsec.\ref{ssection: IW-contraction Intro}, we briefly overview the IW contraction. In Subsec.\ref{ssection: IW-contraction our case}, we apply this contraction scheme to our specific scenario.

\subsection{Introduction: In\"on\"u-Wigner Contractions}\label{ssection: IW-contraction Intro}
In this section, we provide a brief overview of the In\"on\"u-Wigner (IW) contractions, following closely \cite{gilmore} (for additional details, see \cite{IW_Contraction}). In short, the IW contraction is a mathematical and physical mechanism that introduces a parameter into a given Lie algebra, for which there exists a limiting case where a different Lie algebra emerges, not isomorphic to the old one. 

To begin, let us consider a Lie algebra $g$ of dimension $\mathcal{N}$ with a vector field basis $\{X_{a}\}_{1}^{\mathcal{N}}$. This basis satisfies the Lie bracket relations with the algebra's structure constants:
\beq\label{un-contracted commutation}
[X_{a},X_{b}]=C_{ab}^{c}X_{c}.
\eeq
Now, introduce a new basis $\{Y_{a}\}_{1}^\mathcal{N}$ for $g$, defined as a $\epsilon$-parameterized transformation of the original basis:
\beq \label{transformed basis}
Y_{a}=U(\epsilon)_{a}^{b}X_{b}.
\eeq
The new basis $\{Y_{a}\}_{1}^\mathcal{N}$ will satisfy a different Lie algebra, denoted $\Tilde{g}$, with commutation relations:
\beq\label{commutator of divergence} 
[Y_{a},Y_{b}]=C(\epsilon)_{ab}^{c}Y_{c}.
\eeq
The transformed structure constants are related to the original ones via:
\beq\label{transformed structures}
U(\epsilon)_{a}^{d}U(\epsilon)_{b}^{c}C_{dc}^{e}U^{-1}(\epsilon)_{e}^{f}=C_{ab}^{f}(\epsilon).
\eeq

To determine the nature of the transformation matrices $U(\epsilon)$, we ask that \eqref{commutator of divergence} produce a Lie algebra that is \textit{non-isomorphic} to \eqref{un-contracted commutation}. Therefore, we consider transformations that are singular in some limit. Specifically, for $\epsilon=1$, we have $U(1)_{i}^{j} = \delta_{i}^{j}$, which is the identity, and for $\epsilon=0$, $\det||U(0)|| = 0$, which is the singular part.

For concreteness, consider decomposing the vector space of $g$ into a direct sum of two subspaces $g = V_R \oplus V_N$, with associated bases $\{X_{\alpha}\}_{1}^{\text{dim}(V_{R})}$ and $\{X_{i}\}_{1}^{\text{dim}(V_{N})}$ (we use latin and greek alphabet for $V_R$ and $V_N$ respectively). We can then apply a unity transformation on $V_{R}$ and a singular transformation on $V_{N}$:
\bea\label{trnasformation form V_R}
U(\epsilon)_{i}^{j}&=&1 \quad \forall \ i=j, \quad \& \quad i,j\leq dim(V_R), 
\\\label{trnasformation form V_N}
U(\epsilon)_{i}^{j}&=&\epsilon \ \forall \ i=j, \ \&\  dim(V_R) < i,j\leq dim(V_N).\ \ \ \ 
\eea

In the language of the new Lie algebra \eqref{commutator of divergence}, the transformation reduces to 
\beq \label{transformed decomposed}
U(\epsilon)X_{\alpha}=X_{\alpha}\equiv Y_{\alpha}, \quad U(\epsilon)X_{i}=\epsilon X_{i}\equiv Y_{i}.
\eeq

We require that \eqref{commutator of divergence} defines a (new) genuine Lie algebra under the singular transformation \eqref{transformed decomposed}. Therefore, the new structure constants in \eqref{transformed structures} must have a well-defined limit as $\epsilon = 0$. The problem arises when the left-hand side of \eqref{transformed structures} includes terms proportional to $\epsilon^{-1}$; this can only happen, as seen from \eqref{transformed decomposed}, if the commutator of elements from $V_R$ is in $V_N$ before the transformation. Thus, to ensure consistency, we must assume that $C_{\alpha\beta}^{i}=0$. 
Once this consistency condition is satisfied, we can describe the new Lie algebra using the transformed structure constants $c' \equiv c(\epsilon)$. Taking the singular limit $\epsilon \rightarrow 0$, we obtain:
\bea\label{contracted structure constant}
c'^{\gamma}_{\alpha\beta}&=&c_{\alpha\beta}^{\gamma},\quad \quad\quad
c'^{i}_{\alpha\beta}=c_{\alpha\beta}^{i}=0,
\neweqline
c'^{\gamma}_{\alpha j}&=&\epsilon c_{\alpha j}^{\gamma}\rightarrow0,
\quad c'^{k}_{\alpha j}=c_{\alpha j}^{k},
\neweqline
c'^{\gamma}_{ij}&=&\epsilon^{2}c_{ij}^{\gamma}\rightarrow 0, \quad c'^{k}_{ij}=\epsilon c_{ij}^{k}\rightarrow0.
\eea

This yields the following conclusions:
\begin{itemize}
    \item[\textbf{1}] The algebras $\{X_{\alpha}\}$ and $\{Y_{\alpha}\}$ are closed and form sub-algebras in $g$ and $\Tilde{g}$, respectively.
    \item[\textbf{2}] The algebra $\{Y_{i}\}$ spans an invariant sub-algebra in $\Tilde{g}$.
    \item[\textbf{3}] The algebra $\{Y_{i}\}$ is an invariant sub-algebra of $\Tilde{g}$, and it is Abelian.
\end{itemize}

We state now the IW-theorem: 
\textit{Let $g$ be a Lie algebra that can be decomposed to $g=V_R\oplus V_N$. Let $U(\epsilon)$ be a singular transformation matrix when $\epsilon \rightarrow 0$, with the decomposed action $U(0)V_R=V_R,\ U(0)V_N=0$. Then $g$ can be contracted with respect to $V_R$ if and only if $V_R$ is closed. Then, $V_R$ forms a sub-algebra in both $g\text{ and }\Tilde{g}$, and $V_N$ is an abelian invariant sub-algebra of $\Tilde{g}$. The last property implies that $\Tilde{g}$ is non-semi-simple.}

This theorem allows us to consider the singular limits of non-isomorphic physical symmetry groups.
For example (see \cite{gilmore}), starting from the Poincar\'e group ISO(3,1), which is the symmetry group of relativistic mechanics, one can \textit{contract} it to obtain the Galilean group, the symmetry group of classical mechanics. The contraction in 
$\text{ISO(3,1)}\underset{\epsilon\rightarrow 0}{\longrightarrow}\text{GL(3)}$ 
uses the inverse of the speed of light, $\epsilon = c^{-1}$, as the contraction parameter, which is unity in the relativistic region and zero in the classical limit.

\subsection{IW-Contraction of IGL(3,1)}\label{ssection: IW-contraction our case}
In this subsection, we apply the IW-contraction described earlier to preserve the correspondence principle for our theory. The goal is to contract the IGL(1,3) group with respect to the Poincar\'e group, such that in the commutative (classical) limit, we recover the Poincar\'e group as the underlying symmetry.

To proceed with the IW-contraction, we first decompose the vector space of IGL(1,3) into two vector spaces. Using the representations \eqref{Generators of IGL} and the algebra \eqref{Lie algebra of IGL}, we see that the generator $L_{\mu\nu}$ can be decomposed into an anti-symmetric part and a trace part $\TenD_{\mu\nu}$: 
\bea\label{algebra decomposition}
M_{\mu\nu}:=L_{[\mu,\nu]},\quad \TenD_{\mu\nu}=\eta_{\mu\nu}\ \text{tr}\left[L_{\mu\nu}\right].
\eea
Whereas for the Schwinger realization \eqref{Generators of IGL}, we can compute: 
\bea
M_{[\mu\nu]}= x_{[\mu,}\partial_{\nu]},\quad \TenD_{\mu\nu}=\eta_{\mu\nu}(\delta^{\rho\sigma}x_{\rho}\partial_{\sigma}). 
\eea
It is apparent that $M_{\mu\nu}$ together with $\TenD_{\mu\nu}$ generate each of the generators in $L_{\mu\nu}$. 
We now identify the above composition of IGL(3,1) (together with the translations in \eqref{Generators of IGL}) as a composition into a 
the vector space of a Poincar\'e part and an enlarged part as follows:
\\

\textbf{The Poincar\'e part:}
\bea\label{Poincare part of New basis of IGL}
P_{\mu}=\partial_{\mu}, \quad M_{\mu\nu}=x_{[\mu,}\partial_{\nu]}.
\eea

\textbf{The enlarged part}
\bea\label{Dilatation part of New basis of IGL}
\TenD_{\mu\nu}=\eta_{\mu\nu}(\delta_{\rho\sigma}x^{\rho}\partial^{\sigma}), 
\eea
Note that the dilatation operator $D = x^j \partial_j$, which is used in the twist \eqref{dependent: kappa twist}, is defined in terms of $\TenD_{\mu\nu}$ through:
\beq\label{Dilatation}
\TenD=\eta^{\mu\nu}\TenD_{\mu\nu}=
\sum x^{\mu}\partial_{\mu}\Rightarrow D=\TenD-\TenD^0_0.  
\eeq
The enlargement of the Poincar\'e algebra by including only the $\TenD$ generator is known as the $\mathcal{WP}$-algebra, discussed previously. However, we work with the full IGL(3,1) algebra in our framework so that the generated twists will be Abelian.
A straightforward calculation shows that the redefinitions in \eqref{Poincare part of New basis of IGL} indeed represent the ISO(3,1) group:
\bea\label{Poincare algebra}
&&[P_{\mu},P_{\nu}]=0,\quad [M_{\mu\nu},P_{\rho}]=\eta_{\mu\rho}P_{\nu}-\eta_{\nu\rho}P_{\mu}
\\
&&\text{[}\NM_{\mu\nu},\NM_{\rho\sigma}\text{]}=\eta_{\mu\sigma}\NM_{\nu\rho}+\eta_{\nu\rho}\NM_{\mu\sigma}-\eta_{\mu\rho}\NM_{\nu\sigma}-\eta_{\nu\sigma}\NM_{\mu\rho}.
\nonumber
\eea
The Lie algebra of \eqref{Dilatation part of New basis of IGL} is trivially Abelian, as is the Lie algebra associated with $D$ in \eqref{Dilatation}. The commutation relations between the elements in \eqref{Poincare part of New basis of IGL} and those in \eqref{Dilatation part of New basis of IGL} are given by: 
\bea\label{Commutators of Dilatation and iso}
[\TenD^\rho_\sigma,\NM_{\mu\nu}]=0,~~ [\TenD^\rho_\sigma,P_{\mu}]=P_{\mu}.\quad  
\eea
We can now assign the vector space spanned by the elements in \eqref{Poincare part of New basis of IGL} as $V_{R}$, and the space spanned by the elements in \eqref{Dilatation part of New basis of IGL} as $V_{N}$. Then, we can summarize the commutation results in a convenient form: 
\bea\label{decomposite algebra of igl}
&&[X,Y]=0 ~~~~~ \forall X,Y\in V_{N}, 
\neweqline
&&[X,Y]\in V_{R} ~~~ \forall X,Y \in V_{R}, 
\neweqline
&&[X,Y]\in V_{R} ~~~ \forall X \in V_{R}\ \& \ Y\in V_{N}.
\eea

These lead to the following conclusions: 
\begin{itemize}
    \item The group $g=IGL(3,1)=GL(3,1)\rtimes T^4$ can be decomposed as a vector space to $g=V_{R}\oplus V_{N}$ defined above. 
    \item The vector space $V_{R}$ is \textbf{closed} under commutation in $g$, and is \textbf{invariant} in $g$.
    \item The vector space $V_{N}$ is Abelian but not invariant in $g$. 
\end{itemize}

Therefore, according to the IW-theorem, we can contract the group $g = IGL(3,1)$ with respect to $V_{R}$. Furthermore, regardless of the contraction parameter we choose, from the properties of $g$ stated above and from \eqref{contracted structure constant}, we conclude that the $\Tilde{g}$ structure is unique. Specifically, we have the following new Lie algebra for the contracted generators (in the sense of \eqref{transformed decomposed}):
\bea
\label{contracted Lie algebra}
&&[X', Y']=0, ~~~~~~~~~~~~~~~ \forall X,Y\in V_{N},
\neweqline
&&[X', Y']=[X,Y]\in V_{R}, ~~ \forall X,Y\in V_{R},
\neweqline
&&[X',Y']=0, ~~~~~~~~~~~~~~~ \forall X\in V_{R}\ \& \ Y\in V_{N}.~~
\eea

In particular, we find that $[D', P_{\mu}] = [\TenD', P_{\mu}] = 0$ after taking the limit $\epsilon = 0$. Then, regardless of the contraction parameter, the contracted group $\Tilde{g} $'s physical behavior is identical to that of the Poincar\'e group (or indistinguishable from it). As an example, consider the Casimir invariant $C = P_{\mu} P^{\mu}$ of the Poincar\'e group, which remains an invariant in $\Tilde{g}$. This can be expressed mathematically as $\text{IGL}' = \text{ISO} \oplus \TenD'^0_0\oplus \TenD'^1_1\oplus\TenD'^2_2\oplus \TenD'^3_3$ 

To explicitly describe the contraction procedure, we must identify the dimensionless parameter $ \epsilon $ used in \eqref{trnasformation form V_N}. This parameter must be defined so that the limiting cases $ \epsilon \rightarrow 0 $ and $ \epsilon \rightarrow 1 $ correspond to meaningful physical interpretations in relation to the dilatation generator and the $\kappa$-deformation scale.
To establish such a parameter, we draw inspiration from the contraction $ \text{ISO(3,1)} \rightarrow \text{GL(3)} $ (see \cite{gilmore}), where GL(3) represents the 3-dimensional Galilean group. In that context, the contraction parameter was chosen (see eqs. (1.44, 1.45) in Ch.~10 of \cite{gilmore}) by examining the representation of boosts. These boosts have a natural velocity scale $ v = c $, which determines whether a system with a given velocity $ v $ behaves relativistically ($ v \sim c $) or Galilean ($ v \ll c $). Consequently, the natural choice for a contraction parameter is $\epsilon_{\text{SR}} = v/c $, such that, the limiting cases $ \epsilon_{\text{SR}} = 0 $ and $ \epsilon_{\text{SR}} = 1 $ correspond $v\ll c$ and $v\sim c$ respectively. This identification of $\eps_{SR}$ provides a clear physical interpretation of the contraction, even though only $ 1/c $ appears in the structure constants. Hence, we can understand that a general, dimensionless \textit{small parameter} induced by a theory can be written in the following form:
\beq\label{small parameter general}
\epsilon:=\frac{\alpha_{\text{sys}}}{\alpha_{\text{fund}}}\leq1,
\eeq
where $\alpha_{\text{sys}}$ is some \textit{system-dependent} scale, and $\alpha_{\text{fund}}$ is some \textit{fundamental constant} parameter emerging from the theory. Thus, for the case of $\epsilon_{\text{SR}}$, we identify $\alpha_{\text{sys}} = v$ and $\alpha_{\text{fund}} = c$.

In our case, the situation is closely analogous. We therefore analyze the dilatation generator in the vector space representation:  
\bea
\eta_{\rho\sigma}(\delta^{\nu\mu}x_\nu \partial_\mu).
\eea
To better understand what scale is required here, let us consider the action of this operator on some eigenfunction\footnote{This is also closely analogous to the analysis of the boost generator in SR, which leads to the velocity scale.}, say on the free wave $\Psi := e^{-ik_\mu x^\mu}$: 
\bea \label{Dilatation action: free wave}
\mathcal{L}_{D_{00}} \Psi &=& t \partial_0 \Psi = t\omega_{\text{local}}\Psi,
\neweqline
\mathcal{L}_{D_{11}} \Psi &=& x \partial_x \Psi = x \frac{\omega_{\text{local}}}{c}\Psi,
\eea
where we used $\omega_{\text{local}}$ as the wave's frequency in \textit{its} own frame—a distinction that will become necessary in a moment. 

The appearance of the wave's frequency $\omega_{\text{local}}$ suggests that when considering the contraction of the IGL(3,1) symmetry, the relevant physical quantity—the $\alpha_{\text{fund}}$ in \eqref{small parameter general}—should be linked to some fundamental scale with units of frequency.
Importantly, we stress that $ \alpha_{\text{fund}} $ plays the role of the \textit{contraction parameter} in the algebraic sense (just as $ 1/c $ does in SR). At the same time, the terms $ x\omega_{\text{local}}/c,\ t\omega_{\text{local}} $ appearing in \eqref{Dilatation action: free wave} reflect the \textit{physical scaling} of a given system with respect to the action of the dilatation generator. 

Hence, following \eqref{small parameter general}, we can say that the parameter $\epsilon$ serves as a classifier of the system’s position within the classical or quantum-deformed regime. Still, it does not enter as a dynamical field, nor does it control the contraction procedure. The contraction is formally defined with the fixed scale $ \alpha_{\text{fund}} $, not the system-dependent ratio.

The question then arises: what physical limit can the theory of $\kappa$-Minkowski provide for $\alpha_{\text{fund}}$ that has a relation with frequency? To address this, we note that in DSR, the NC nature of \eqref{kappa minkowski} is known to lead to an uncertainty between space and time, which depends on the distance of the system from the observer \cite{Kappa_Deformed_Phase_Space, NonCommutative_Spacetime_Interpretation1}:
\beq\label{Uncertainty Relation} 
\sigma_{\hat{x}^{0}}\sigma_{\hat{x}^{j}} \geq \frac{1}{2\kappa}\Avx.
\eeq
For the effect of this uncertainty to be observable, two physical conditions must be simultaneously satisfied \cite{NonCommutative_Spacetime_Interpretation2, NonCommutative_Spacetime_Interpertation_3, Relative_Locality, NonCommutative_Spacetime_Interpretation1}:
\begin{enumerate}  
    \item The distance $x^j$ of the observer from the system must be nonzero and sufficiently large.  
    \item The frequency of the system $p^0$ must be high, approaching Planckian frequencies.
\end{enumerate} 
Now, let us consider a system with some associated frequency, and study the situation \textit{locally}—we observe the system with $\bar{x}^j \sim l_{\text{sys}}$, where $l_{\text{sys}}$ is the system's size, and thus measure $\omega_{\text{local}}$. In such a measurement, it is physically reasonable to claim (e.g., from local QFT arguments) that $\omega_{\text{local}} \leq \omega_p$, with $\omega_p$ as the Planck frequency \eqref{Planck Frequency}; put differently, the system's energy is lower than the Planck energy. Considering that if the system approaches the maximal value $\omega_{\text{local}} \sim \omega_p$, then $l_{\text{sys}} \sim l_p$, we realize that in such a case the uncertainty \eqref{Uncertainty Relation} reaches order unity (since $l_p \sim \mathcal{O}(1) \Rightarrow \kappa \sim \mathcal{O}(1)$). 
Now, we know that for \textit{non-local} measurements, \eqref{Uncertainty Relation} introduces a \textit{magnification} of the uncertainty effect, scaling with the measurement distance $\bar{x}^j$. It is also understood (consider this as an assumption we make here) that \eqref{Uncertainty Relation} is \textit{bounded} by an order-unity effect—it can reach $\mathcal{O}(1)$ but does not exceed it. Therefore, when $\bar{x}^j > l_{\text{sys}}$, we should require that the \textit{magnified} effect does not reach unity—the \textit{local} frequency $\omega_{\text{local}}$ is now bounded \textbf{below} its local maximal value $\omega_p$.

In light of this, one might expect that the frequency should be further suppressed when measured from afar, perhaps by a factor of the form $\left(x / l_{\text{sys}} \right)^n$ for some $n \geq 1$. However, since the precise form and justification of such a factor require a more detailed physical analysis, which we defer to future work \cite{Our_Next_Paper}, we do not include it in the current definition. 

Instead, we define the following \textbf{ansatz} for a dimensionless parameter that quantify the system’s deviation from the classical regime:
\beq\label{epsilon of omega}
\epsilon_{RL} := \frac{\omega_{\text{local}}}{\omega_p}.
\eeq
Hence, the relevant \textit{fundamental, system-independent} scale is $\alpha_{\text{fund}} = \omega_p$,
which sets the scale at which a system begins to exhibit $\kappa$-induced NC effects, analogous to how the speed of light $c$ separates relativistic and non-relativistic regimes. In particular, our ansatz \eqref{epsilon of omega} implies:
\begin{itemize}
    \item For $\omega_{\text{local}} \ll \omega_p$, the system resides in a classical regime where $\kappa$-deformations are negligible, so $\epsilon_{RL} \rightarrow 0$.
    \item For $\omega_{\text{local}} \sim \omega_p$, the system enters a quantum–Planckian regime where noncommutative effects become manifest. We name this region the \textit{Quantum Planckian Scale} (QPS), so $\epsilon_{RL} \rightarrow 1$.
\end{itemize}
We acknowledge that a more rigorous and quantitative analysis is required to determine the precise form of the parameter $\epsilon_{RL}$, particularly regarding a potential multiplicative factor of the form $(x/l_{\text{sys}})^n$. We intend to investigate this in future work \cite{Our_Next_Paper}, aiming to derive $\epsilon_{RL}$ from a more fundamental perspective within the framework of RL. In the present treatment, however, we make the simplifying assumption that 
$\frac{\omega_{\text{local}}}{\omega_p}$ is sufficiently small that even 
$\omega_{\text{local}}(x/l_{\text{sys}})^n \ll \omega_p$ for all systems under consideration.
Nevertheless, the usefulness of our ansatz can be demonstrated momentarily, following the definition of the contracted dilatation.

Following the transformations given in \eqref{trnasformation form V_R}–\eqref{transformed decomposed}, the contraction of $g = IGL(3,1)$ with respect to $V_R = ISO(3,1)$ can be expressed in terms of the generators in two parts:

\textbf{The $V_R$ part:} remains unmodified,
\bea\label{contracted vectorfields iso}
P'_{\mu}&=&\lim_{\epsilon\rightarrow0}P_{\mu}=P_{\mu},
\neweqline
\NM_{\mu\nu}'&=&\lim_{\epsilon\rightarrow0}\NM_{\mu\nu}=\NM_{\mu\nu}.
\eea

\textbf{The $V_N$ part:} is contracted, by first rescaling 
\beq\label{Dilatation Rescaled}
\mathcal{D}'_{\mu\nu} := \frac{1}{\alpha_{\text{fund}}} \mathcal{D}_{\mu\nu} = \frac{1}{\omega_p} D_{\mu\nu},
\eeq
which leads to the following classical limits: 
\bea\label{contracted vectorfields Dilation}
\TenD'_{\mu\nu} &=& \lim_{\alpha_{\text{fund}}\rightarrow\infty} \frac{1}{\alpha_{\text{fund}}} \TenD_{\mu\nu} = \lim_{\alpha_{\text{fund}} \rightarrow \infty} \frac{1}{\alpha_{\text{fund}}}\ \eta_{\mu\nu}(x_{\rho}\partial_{\sigma}\delta^{\rho\sigma})
\neweqline
&=& \eta_{\mu\nu} \frac{1}{\omega_p}(x_{\rho}\partial_{\sigma}\delta^{\rho\sigma})\Big|_{\epsilon \rightarrow 0}
\rightarrow 0.
\eea

Let us now pause for a moment and demonstrate how these limits directly follow from \eqref{Dilatation Rescaled} and \eqref{epsilon of omega}. To do so, we repeat the procedure from \eqref{Dilatation action: free wave}, now explicitly using $\omega_p$:
\bea \label{contraction Demonestrated}
\mathcal{L}_{D'_{00}} \Psi &=& \frac{1}{\omega_p} t \partial_0 \Psi = t \frac{\omega_{\text{local}}}{\omega_p} \Psi = t \epsilon_{RL}  \Psi,
\neweqline
\mathcal{L}_{D'_{11}} \Phi &=& \frac{1}{\omega_p} x \partial_x \Psi = x \frac{\omega_{\text{local}}}{c \omega_p} \Psi = x \epsilon_{RL}  \Psi,
\eea
as before, we assume $\omega_{\text{sys}}\ll\omega_p$ and so in the classical limit $\epsilon_{RL} \rightarrow 0$, the limits in \eqref{contracted vectorfields Dilation} directly follow from \eqref{epsilon of omega}\footnote{One could also regard this in the opposite way: the fact that the rescaling \eqref{Dilatation Rescaled} of the dilatation generator should admit a meaningful classical limit dictates the form of the ansatz \eqref{epsilon of omega}.}.

The contraction scheme \eqref{contracted vectorfields Dilation} confirms (in a trivial manner) the invariance of the Poincar\'e Casimir $C = P_{\mu} P^{\mu}$:
\beq\label{Casimir Contracted}
[C,X]=0 \ \forall X\in \text{ISO(3,1)},\quad \& \quad [D,C]=0. 
\eeq

The proposed contraction scheme leads exactly to the Poincaré group in the classical limit. This behavior can be represented pictorially, using the system’s spacetime coordinate distance $x$, and local frequency $\omega_{\text{local}}$ as:
\beq\label{contraction pictorial}
\text{IGL}'(3,1) \underset{ \omega_{\text{local}}\ll \omega_p}{\longrightarrow} \text{ISO(3,1)}. \nonumber
\eeq
We note here that this limit is not independent of the scale $\kappa$, since $l_p \rightarrow 0$ implies $\omega_p \rightarrow \infty$, as seen from \eqref{Planck Frequency}. Thus, when there are no noncommutative deformations, the algebra is necessarily contracted, i.e., $\alpha_{\text{fund}} \rightarrow \infty \Leftrightarrow \kappa \rightarrow \infty$.

Here, we use the notation IGL$'$(3,1) to denote the group composed of the generators in \eqref{contracted vectorfields iso} and \eqref{contracted vectorfields Dilation}, before taking the contraction limit $\epsilon_{RL} \rightarrow 0$, or, equivalently, before taking $\kp\rightarrow\infty$. The corresponding Lie algebra is given by:
\beq\label{generators of igl'}
P_\mu, \quad M_{\mu\nu}, \quad \mathcal{D}'_{\mu\nu} = \frac{1}{\alpha} \mathcal{D}_{\mu\nu}.
\eeq

Note that we still need to determine the multiplication scaling factor $\left(\frac{x}{l_{\text{sys}}}\right)^n$ in our ansatz \eqref{epsilon of omega}; its precise value should ultimately be derived either from experimental input or from theoretical considerations. We intend to pursue this in future work by analyzing specific physical systems \cite{Our_Next_Paper}.

We again stress that the $1/\alpha_{\text{fund}} = 1/\omega_p = l_p/c$ contraction is in direct analogy with the use of $1/c$ in the contraction from Lorentz to Galilean symmetry. The dimensionless ratio $\epsilon_{RL} = \frac{\omega_{\text{local}}}{\omega_p}$ serves only as a physical indicator of the system’s regime, not as the contraction parameter itself. This treatment is justified by the fact that $\alpha_{\text{fund}}$ arises physically from the NC structure of spacetime, as captured in the \textit{original} uncertainty relation \eqref{Uncertainty Relation}.

Our suggestion here is that the $\kappa$-Minkowski structure \textit{already} encodes a scale of the form $ \omega_{\text{local}} \sim \alpha_{\text{fund}}$, implying that the twist deformations should be reconstructed with the scale $\alpha_{\text{fund}}$ appearing in $\mathcal{D}'$. However, the NC relations \eqref{kappa minkowski} and \eqref{Uncertainty Relation} should remain intact under the contraction. Indeed, it is important to distinguish between these structures: while the twist deformation of the symmetry is contracted, the underlying NC relations of spacetime \textit{remain unaffected} throughout the contraction procedure.

\subsection*{Summary of Symmetry Contraction}
In this section, we proposed a way to address the correspondence issue arising from the use of the enlarged IGL(3,1) symmetry in the $\kappa$-deformed framework by employing an Inönü–Wigner (IW) contraction scheme. This procedure ensured that, in the classical limit, the symmetry group reduces cleanly to the Poincaré group $ISO(3,1)$, thereby preserving consistency with General Relativity.

The contraction was performed with respect to the vector space decomposition $g = V_R \oplus V_N$, where $V_R \cong \text{ISO(3,1)}$ is a closed and invariant subalgebra, and $V_N$ contains the affine generators. The contracted symmetry algebra, denoted IGL$'$(3,1), is characterized by the generators $P_\mu$, $M_{\mu\nu}$, and the rescaled dilatations $\mathcal{D}'_{\mu\nu} = \frac{1}{\alpha_{\text{fund}}} \mathcal{D}_{\mu\nu}$, with $\alpha_{\text{fund}} = \omega_p$ identified as the fundamental frequency scale.

We introduced a dimensionless parameter $\epsilon_{RL} = \frac{\omega_{\text{local}}}{\omega_p}$, inspired by RL, to quantify the deviation from the classical regime. We noted that RL may also imply a scaling factor of the form $\left( \frac{x}{l_{\text{sys}}} \right)^n$, whose precise role remains to be established. In the classical limit, $\epsilon_{RL} \to 0$, and based on RL arguments, we assume $\epsilon_{RL} \leq 1$.
While $\epsilon_{RL}$ arises from the physical representation of dilatation on wavefunctions (see \eqref{contraction Demonestrated}), the actual contraction parameter is $\alpha_{\text{fund}} = \omega_p$, which is physically related to $\kappa$ via the Planck scale. In the classical limit $\kappa \to \infty$, one also has $\alpha_{\text{fund}} \to \infty$, ensuring $\epsilon_{RL} \to 0$, i.e., the symmetry contracts as IGL'(3,1) $\to$ ISO(3,1).

At the quantum–Planckian scale ($\epsilon_{RL} \sim 1$), the full twisted IGL(3,1) symmetry is active. In contrast, the classical limit justifies using the contracted algebra IGL$'$(3,1) for $\kappa$-deformed gravity. Accordingly, the twist vector fields (Sec.~\ref{section: NC Introduction} and \ref{section: Building Blocks For Construction}) involve the rescaled generator $D' = \frac{1}{\alpha_{\text{fund}}} D$, encoding this contraction.

Finally, while this treatment ensures consistency at the level of the symmetry structure, further work is required to derive $\epsilon_{RL}$ from first principles within the noncommutative theory, and in particular, to determine the form and origin of the possible scaling factor $\left( \frac{x}{l_{\text{sys}}} \right)^n$ in the ansatz \eqref{epsilon of omega}.

\section{Gravity extension of the \texorpdfstring{$\kp$-Minkowski Spacetime}{kappa-Minkowski}}\label{section: The Construction of Gravity}
We now have the complete toolkit for constructing a gravitational theory on a flat (local) sector of the NC spacetime given in \eqref{kappa minkowski}. As outlined in Sec.~\ref{section: background B}, such a construction within the twist-algebraic framework requires setting up a suitable \textit{twist element} that generates the deformation from a commutative spacetime to the NC spacetime of \eqref{kappa minkowski}. As discussed in Sec.~\ref{section: Building Blocks For Construction}, a consistent twist requires enlarging the \textit{local} ISO(3,1) symmetry to IGL(3,1), applying the twist deformation then yields the \textit{local} symmetry group of twisted IGL(3,1).

In Sec.~\ref{section: Consistency of The NC-Relations}, we analyzed the physical features of this twisted symmetry, examining the associated deformed transformations, the $\star$-coproduct, and the deformed Leibniz rule, and showed their internal consistency with the NC structure in \eqref{kappa minkowski}. With this \textit{flat spacetime} setup, one can, in principle, compute the $\star$-geometrical quantities needed to formulate the deformed Einstein equations \eqref{deformed EE}.

However, formulating a gravitational theory involves more than just stating the equations. We aim to construct gravitation based on the twisted local IGL(3,1) symmetry, which raises some questions: What form do \textit{deformed general covariance} or \textit{deformed general coordinate transformations} take? How is the classical notion of field variation defined in this setting? What becomes of local symmetries and their associated conservation laws, such as $\nabla_\lambda G^{\lambda\mu} = 0$, under noncommutativity? And importantly, how is matter coupled to gravity, i.e., what is the structure of the deformed energy–momentum tensor in this framework?

In the pursuit of the \textit{third consistency} condition $\mathcal{C}$ outlined in Sec.\ref{section: Overview of Introduction}, we investigate these issues in what follows. Some of these questions we answer with new insights, and for others, the remaining ambiguities are highlighted.

We begin by deriving the deformed Einstein equation in vacuum via the deformed Einstein tensor on a NC spacetime exhibiting local $\kappa$-Minkowski behavior. The underlying symmetry is the \textit{twisted} group  IGL'(3,1), defined in \eqref{generators of igl'}, which is a rescaled version of the original twisted IGL(3,1). We thus compute the relevant $ \star $-geometrical objects using the formalism outlined in Secs.~\ref{section: NC Introduction} and~\ref{section: Building Blocks For Construction}. Detailed derivations are deferred to Appendix~\ref{Appendix. the construction}.

An important modification arises from the use of IGL'(3,1): the generator $ D $ is replaced by $ D' = \frac{1}{\omega_p} D $. Since $ D' $ appears in the twist definition \eqref{dependent: kappa twist}, this rescaling can be absorbed into a redefinition of the deformation parameter. We thus define the effective deformation scale as
\bea\label{rescaling lambda deformation}
\lambda' \equiv\frac{\lambda}{\omega_p}= \frac{\lambda}{\alpha_{\text{fund}}},
\eea
where $\lambda'$ is to be treated as the physically relevant deformation parameter. This redefinition arose from an Inönü–Wigner contraction of the dilatation generator aimed at retaining the classical limit of the theory, much like the $ v/c $ contraction from special to Galilean relativity. As noted in Sec.~\ref{ssection: IW-contraction our case}, the scale $\alpha_{\text{fund}}=\omega_p$ originates from the commutator $[x^0\commaST x^j] = i\lambda x^j$, and our proposal was that the twist alone cannot fully reproduce this structure without introducing the additional structure in $\lambda'$. Hence, in the NC relations \eqref{kappa minkowski}, the scale $\lambda = l_p$ remains fixed, while $\lambda'$ appears in the twist exponent to reflect the contraction.

It is crucial to emphasize that although the twist involves the dilatation generator $ D $, our theory does not exhibit dynamical invariance under scale transformations in the Weyl gravity sense. In Weyl gravity, dilatations are true dynamical symmetries \cite{WeylGrav1,WeylGrav2}, preserving the form of the action under local scalings of the metric and fields. In contrast, in our setup, $ D \in \text{IGL(3,1)} $ plays the role of a generator for the flat local sector and enters the twist to produce the $\kappa$-deformation, but does not represent a dynamical scale invariance.
Thus, although the appearance of $ D $ may superficially resemble Weyl-invariant formulations, our construction is fundamentally distinct. The resulting deformed gravitational theory respects general diffeomorphism invariance in its twisted form, but it is not invariant under local Weyl rescalings. Whether one can combine Weyl gravity with $\kappa$-deformed geometry into a scale-invariant NC gravitational theory is an interesting question that deserves more study.

We note that some of the results of this section, including comparison to the classical and constant NC case, can be found in the Table.\ref{summary-table}.

\subsubsection*{Deformed Diffeomorphisms}
We shall now account for deformed general coordinate transformations and their actions on geometrical quantities; we must define the deformed diffeomorphism.
To do so, we can utilize the $\star$-Lie derivative, developed in \eqref{Lie derivative non non constant} to construct the diffeomorphism algebra on the non-commutative space. Working similarly to \cite{Gravity_Non}, we first define the transformation law of a scalar field $f$ by a vector field $\xi$
\bea\label{scalar transformation}
f\hookrightarrow f' := \delta^\star_\xi(f):=\lie^\star_{\xi}(f),
\eea
which is simply the $\star$-Lie derivative \eqref{Lie derivative non non constant}. From \eqref{deformed Leibniz rule}, the multiplication of scalars $f\star g$ transforms also as a scalar and we show it explicitly to first non-trivial order in \eqref{appendix: scalar multiplied trans}. For the transformation rule of a vector $\delta^\star_{\xi}(V)$, we follow the definitions of the $\star$-Lie derivative \eqref{star lie derivative functions}, \eqref{star lie derivative vectors} and derive (cf. eq(7.2) in \cite{Gravity_Non})
\beq\label{trans rule vectors}
\delta^\star_{\xi}(V_\mu)=\lie^\star_{\xi}(V_{\mu})+\lie^\star_{(\partialST_\mu\xi^\nu)}(V_{\nu}),
\eeq
where $\lie^\star_\xi$ is \eqref{Lie derivative non non constant} and 
\bea
&\lie&^\star_{(\partialST_\mu\xi^\nu)}(V_{\rho})
\\
&=&
\sum_{\substack{n=0\\n'=0}}^{\infty}\!\!\!
\frac{\left(i\lambda'/2\right)^{n+n'}}{n!n'!}
[(-x^j\partial_{j})^{n'}[\partial_{0}^{n}\partialST_{\mu}(\xi^{\nu})]\partial_{0}^{n'}(x^{j}\partial_{j})^{n}].\nonumber
\eea
And, as in the scalar case, the multiplication of a vector and a scalar will transform as a vector, and a multiplication of two vectors will transform as a tensor (which is a simple generalization of \eqref{trans rule vectors}, see \cite{Gravity_Non}). This fact can be seen by inserting vectors into the expressions of scalar multiplication in \eqref{appendix: scalar multiplied trans} and observing that the deformation terms in $(V_\mu\star U_\nu\star\ldots)$ do not change the covariant properties of the classical insertion $(V_\mu \cdot
U_\nu \cdot \ldots)$. Therefore, whatever the transformation law the classical counterpart of an expression had, it will stay the same in the deformed case. In other words, one can say that the isomorphism \eqref{isomorphism map} (applied to generic-rank tensors) leaves the covariant properties given that transformations are executed by the suitably deformed set of diffeomorphisms. 

\subsubsection*{Introducing $\star$-Geometry}

Generalizing the derivative to introduce curvature, we compute the connection coefficients, which are uniquely determined using \eqref{connection formula},
\begin{equation}\label{connection kappa} 
    \nablaST_{\partialST_{\mu}}\partialST_{\nu} = \Gamma_{\mu \nu }^{\sigma}\star\partialST_{\sigma} \neq \Gamma_{\mu \nu }^{\sigma}\partialST_{\sigma},
\end{equation}
holding similarly for the 1-form basis. We will not explicitly expand each $\star$-product in the construction (it can be done in a future perturbative expansion using \eqref{dependent: star product kappa}).  \\

The \textbf{$\star$-Covariant Derivative} is calculated using \eqref{star covariant derivative}, which, for generic vector fields $u$ and $z$, simplifies to (see \eqref{App. dependent: calculation of covariant derivative General}),
\beq\label{dependent: calculation of covariant derivative General}
\nablaST_{z}u
=
[\lie^{\star}_{z}(u^{\nu})+\bar{\text{R}}^{\{\alpha\}}(u^{\sigma})\star\bar{\text{R}}_{\{\alpha\}}^{\ \ \mu'}(z)\star\Gamma_{\mu'\sigma}^{\nu}]\star\partialST_{\nu},
\eeq
we wrote $\bar{\text{R}}_{\{\alpha\}}(z):=\bar{\text{R}}_{\{\alpha\}}(z)^\mu\star\partialST_\mu $, which is always possible since $\bar{\text{R}}_{\{\alpha\}}(z)$ is a vector field. For the case of $z=\partialST_{\mu}$ and only considering the $u=u^{\nu}$ part, we derive \eqref{App. dependent: calculation of covariant derivative}
\beq\label{dependent: calculation of covariant derivative}
\nablaST_{\partialST_{\mu}}(u^{\nu})
    = \partial_{\mu}(u^{\nu})+O^{\mu'}_{\mu}[\lambda', 2](u^{\sigma})\star\Gamma_{\mu'\sigma}^{\nu}
\eeq
Note, in \eqref{dependent: calculation of covariant derivative General}, we expressed a $\star$-vector field as $u = u^\nu \star \partialST_{\nu}$. However, if a vector field $ u = u^\nu \partialST_\nu $ is given in the commutative vector space $\Xi$, its expression in the $\star$-vector space will be $ u = \uST^{\nu} \star \partialST_\nu $.
The components $\uST^\nu$ are related to the components in the commutative space by equating $ u^\nu \partialST_\nu $ to $ \uST^\nu \star \partialST_\nu $ (see \cite{NC_Geometry_Simplified}). This relation can be solved order by order as $ \uST^\nu = \uST^\nu_{(0)} + \lambda' \uST^\nu_{(1)} + \ldots$\\
For example: $ \uST_{(0)}^\nu = u^\nu 
- \uST_{(1)}^\nu = -\lambda' \buphi(u^\rho) \bar{\phi}^\mu_{\{\alpha\}\rho} $.
Here, $ \bar{\phi}^\mu_{\{\alpha\rho\}} \partialST_\mu = \partialST_\rho $.
Nevertheless, we treat general $\star$-vector fields in our calculations, so we can write $u = u^\nu \star \partialST_{\nu} $. We must keep in mind that to relate this quantity to its counterpart in the commutative space; we shall need to use the expression for $ \uST^\nu $ in terms of $ u^\nu $ and $ \lambda' $.

On top of all that, expression \eqref{dependent: calculation of covariant derivative} does not require using $\uST^\nu$, and the $u^\nu$ can also be considered as the component in $u\in\Xi$. This is because we can write \eqref{dependent: calculation of covariant derivative General} with $\uST^\nu$ and use $\uST^\nu\star\partialST_{\nu}=u^\nu\partialST_{\nu}$. 

The \textbf{$\star$-Curvature Tensor} is calculated using \eqref{curvature definition} and \eqref{coefficient frame curvature}, which, after evaluating, takes the form (see \eqref{curvature kappa1}),
\bea\label{curvature kappa}
\textgoth{R}_{\mu\nu\rho}^{\sigma}=\partial_{[\mu}\Gamma_{\nu]\rho}^{\sigma}
+O^{\gamma}_{[\mu}[\lambda',2](\Gamma_{\nu]\rho}^{\tau})\star\Gamma_{\gamma\tau}^{\sigma}
\eea

As anticipated, there are additional deformations in the $\star$-curvature tensor compared to the constant NC case. \\

The \textbf{$\star$-Ricci Tensor} is calculated using \eqref{coefficient frame curvature} while treating the $\star$-curvature as some given tensor-coefficients (see \eqref{Ricci tensor kappaA})   
\bea\label{Ricci tensor kappa}
\textgoth{R}_{\nu \rho } = \langle dx^{\mu}\commaST \textgoth{R}^{\sigma}_{\mu \nu \rho}\star\partialST_{\sigma}\rangle'
=e^{-i\lambda'\partial_{0}}\textgoth{R}_{j\nu\rho}^{j}+\textgoth{R}_{0\nu\rho}^{0}.
\eea

To establish the relation between the metric and the Christoffel symbols, we first evaluate the metric compatibility condition \eqref{metric compatibility} as follows (see \eqref{App. dependent: covariant derivative of the metric}),
\bea\label{dependent: covariant derivative of the metric}
    \nablaST_{\gamma}(g)
    &=&[-\Gamma_{\gamma\text{\tiny(}\mu}^{\sigma}\star g_{\nu\text{\tiny)} \sigma}+g_{\mu\nu,\gamma}\star(dx^{\mu}\otimesST dx^{\nu})] 
    \neweqline
    &\Rightarrow&
g_{\mu\nu;\gamma}=g_{\mu\nu,\gamma}-\Christoffel{\sigma}{\gamma\text{\tiny(}\mu}\star g_{\nu\text{\tiny)} \sigma}.
\eea
Observe that \eqref{dependent: covariant derivative of the metric} shares a similar \textit{form} with its constant NC counterpart \cite{Gravity_Non}; a distinctive property of the metric since the basis 1-forms are $\star$-multiplied from the left. For a general tensor, the covariant derivative will take the form
\bea
\label{dependent: right multiplied tensor covariant derivative}
\nablaST_{\gamma}(\tau_{\mu\nu}\!\star\! dx^{\mu}\!\otimesST\! dx^{\nu})&=&
\left[\tau_{\mu\nu,\gamma}
-O^{\gamma'}_{\gamma}[\lambda',2]( \tau_{\sigma\nu})\!\star\!\Gamma^{\sigma}_{\gamma'\mu} \right.\\
&&\left.-O^{\gamma'}_{\gamma}[\lambda',2](\tau_{\mu\sigma})\!\star\!\Gamma^{\sigma}_{\gamma'\nu}
\right]
\!\star\!(dx^{\mu}\!\otimesST\! dx^{\nu}),
\nonumber
\eea
differing in \textit{form} from the $\star$-covariant derivative in the constant NC scenario \cite{Gravity_Non}. Using the result \eqref{dependent: covariant derivative of the metric}, we derive the $\star$-connection in terms of the metric:
\beq \label{dependent: connection-metric}
\Gamma^{\sigma}_{\nu \gamma} =\invG^{ \sigma\mu}\star\Gamma_{\mu \nu \gamma}=\invG^{ \sigma\mu}\star\frac{1}{2}[g_{\mu (\nu,\gamma)}-g_{\nu \gamma,\mu}].
\eeq
This result does not imply identity between the connection in $\kappa$-Minkowski spacetime with constant NC result \cite{Gravity_Non}; the $\star$-product differs significantly. However, it does signify that the \textbf{form} of the deformation is the same.\\

The \textbf{$\star$-Ricci scalar} is computed using Eqs.~\eqref{star definition of the Ricci scalar} and \eqref{Ricci tensor kappa}:
\beq\label{dependent: Ricci scalar}
\textgoth{R} = \invG^{\mu\nu} \star \textgoth{R}_{\nu\mu} = \invG^{\mu\nu} \star \left(O^{\rho}_{\sigma}[\lambda', -2](\textgoth{R}_{\rho\nu\mu}^{\sigma})\right).
\eeq
We emphasize here the issue of ordering; as indicated in Eq.~\eqref{star definition of the Ricci scalar}, we have \textit{explicitly imposed} the use of \textit{left $A_\star$-linearity}. Consequently, multiplying $\invG$ from the right would yield a different result, meaning that this definition is not unique. Nevertheless, depending on the case considered, there are established ways to handle this ambiguity, as we will discuss shortly.

Note that the $\star$-metric is treated as a generic dynamical variable (as in the constant NC case), not as a $\star$-quantity that needs to be expanded in the non-commutativity parameter.

Finally, we are in a position to write down the $\star$-Einstein tensor \eqref{deformed EE}, along with the corresponding Einstein equation in vacuum, for a NCST locally described by the $\kappa$-Minkowski spacetime. 
To this end, we begin by expanding the $\star$-Ricci tensor \eqref{Ricci tensor kappa}, making use of the operator defined in \eqref{Notation Definition O[l,n]}:
\bea\label{Expansion of Ricci tensor}
\GeomR_{\mu\nu}&=&O^\alpha_\beta[\lambda',-2](\GeomR_{\alpha\mu\nu}^\beta)
\neweqline
&=& O^\alpha_\beta[\lambda',-2]\left(\partial_{[\alpha}\Gamma^\beta_{\mu]\nu}\right)
\neweqline
&&+O^\alpha_\beta[\lambda',-2]\left(O^\gamma_{[\alpha}[\lambda',2](\Gamma_{\mu]\nu}^\tau)\star\Gamma^\beta_{\gamma\tau}\right)
.
\eea
Expanding the third line above is nontrivial: the operator $O^\alpha_\beta[\lambda',-2]$ does not satisfy a simple Leibniz rule, nor is it an algebra homomorphism with respect to $\partial_0$—unlike, for instance, the exponential operator $e^{-\partial_0}$. 

To illustrate the challenge, recall from \eqref{Notation Definition O[l,n]} that
\beq
O^\alpha_\beta[\lambda',-2] = e^{-\hat{D}}\delta_\beta^j\delta_j^\alpha + \delta_\beta^0\delta^\alpha_0,
\quad \text{with} \quad \hat{D} := i\lambda'\partial_0.
\eeq
A typical term appearing in the expansion of the third line thus takes the form:
{\small
\bea\label{Ricci expansion: demo}
&O&^\alpha_\beta[\lambda',-2]\left(A^\tau_{\mu\nu}\star A_{\alpha\tau}^\beta\right)=\left(e^{-\hat{D}}\delta_\beta^j\delta_j^\alpha+\delta_\beta^0\delta_0^\alpha\right)\left(A_{\mu\nu}^\tau\star A_{\alpha\tau}^\beta\right)
\neweqline
&&~~=\left[\left(1+\tsum_{n=1}^{\infty}\tfrac{(-\hat{D})^n}{n!}\right)\delta_\beta^j\delta_j^\alpha+\delta_\beta^0\delta_0^\alpha\right]\left(A_{\mu\nu}^\tau\star A_{\alpha\tau}^\beta\right)
\neweqline
&&~~=\left[\delta_\beta^\alpha+\delta_\beta^j\delta_j^\alpha\tsum_{n=1}^{\infty}\tfrac{(-\hat{D})^n}{n!}\right]\left(A_{\mu\nu}^\tau\star A_{\alpha\tau}^\beta\right)
= A_{\mu\nu}^\tau\star A_{\beta\tau}^\beta
\neweqline
&&+\delta_\beta^j\delta_j^\alpha \sum_{n=1}^{\infty}\sum_{0\leq k\leq n}\tfrac{(-\hat{D})^n}{k!(n-k)!}\partial_0^{(n-k)}(A_{\mu\nu}^\tau)\star \partial_0^k(A_{\beta\tau}^\beta),
\eea
}
for some tensor \( A_{\mu\nu}^\tau \). In this derivation—as in others to follow—we make use of the generalized Leibniz rule for $e^{-\hat{D}}$, as well as the fact that \( O^\alpha_\beta \) is compatible with the twist, i.e., it commutes with the twist element: \( [O^\alpha_\beta, \Phi] = 0 \). 

Clearly, expanding \eqref{Ricci expansion: demo} to all orders is algebraically demanding. For this reason, we first verify that the expression \eqref{Expansion of Ricci tensor} correctly reproduces the classical Ricci tensor at zeroth order in $\lambda'$, denoted $\textgoth{R}^{(0)}_{\mu\nu}$:
\bea\label{ricci: zero order}
\textgoth{R}^{(0)}_{\mu\nu} &=& \delta_\beta^\alpha \partial_{[\alpha}\Gamma^\beta_{\mu]\nu}
+ \delta_\beta^\alpha \left[\delta_\alpha^\gamma \Gamma_{\mu\nu}^\tau \Gamma_{\gamma\tau}^\beta 
- \delta_{\mu}^\gamma \Gamma_{\alpha\nu}^\tau \Gamma_{\gamma\tau}^\beta \right]
\neweqline
&=& \partial_{[\beta}\Gamma^\beta_{\mu]\nu}
+ \Gamma_{\mu\nu}^\tau \Gamma_{\beta\tau}^\beta 
- \Gamma_{\beta\nu}^\tau \Gamma_{\mu\tau}^\beta,
\eea
which matches the standard (commutative) result for the Ricci tensor.

Let us now proceed to compute the \textit{first-order} correction to $\textgoth{R}_{\mu\nu}$, denoted by $\textgoth{R}^{(1)}_{\mu\nu}$. This computation offers insight into the structure of the deformed Ricci tensor and will also be used later on.
Firstly, we define the following notation for brevity:
\beq\label{matrix differential notaion}
O^\alpha_\beta[\lambda', +2] := S^\alpha_\beta \quad \Rightarrow \quad O^\alpha_\beta[\lambda', -2] = S^{*\alpha}_\beta,
\eeq
where $*$ denotes complex conjugation.

Using this, we write:
\bea
\textgoth{R}_{\mu\nu}^{(1)} &=& S^{*\alpha}_\beta\left(\partial_{[\alpha}\Gamma_{\mu]\nu}^\beta\right)
\neweqline
&&+ S^{*\alpha}_\beta\left[
S^{\gamma}_\alpha(\Gamma_{\mu\nu}^\tau)\star\Gamma_{\gamma\tau}^\beta
- S_{\mu}^\gamma(\Gamma_{\alpha\nu}^\tau)\star\Gamma_{\gamma\tau}^\beta
\right]. \nonumber
\eea

Let us now focus on the second line, which we denote $B_{\mu\nu}^{1}$, and expand it to first order in $\lambda'$:
{\small
\bea
B_{\mu\nu}^1&=&\left(e^{-i\lambda'\partial_0}\delta_\beta^j\delta_j^\alpha+\delta_\beta^0\delta^\alpha_0\right)\left[\left(e^{i\lambda'\partial_0}\delta_\alpha^j\delta_j^\gamma+\delta_\alpha^0\delta_0^\gamma \right)(\Gamma_{\mu\nu}^\tau)\star\Gamma_{\gamma\tau}^\beta\right.
\neweqline
&&\left.-\left(e^{i\lambda'\partial_0}\delta_{\mu}^j\delta_j^\gamma+\delta_\mu^0\delta_0^\gamma\right)(\Gamma_{\alpha\nu}^\tau)\star\Gamma^\beta_{\gamma\tau}\right]
\neweqline
&=& \left[\delta_\beta^\alpha-i\lambda'\partial_0\delta_\beta^j\delta_j^\alpha\right]\left[\left(\delta_\alpha^\gamma+i\lambda'\partial_0\delta_\alpha^j\delta_j^\gamma\right)(\Gamma_{\mu\nu}^\tau)\right.
\neweqline
&&\left.-\left(\delta_\mu^\gamma+i\lambda'\partial_0\delta_\mu^j\delta_j^\gamma\right)(\Gamma_{\alpha\nu}^\tau)\right]\star \Gamma^\beta_{\gamma\tau},
\eea
}

Neglecting all zeroth-order terms $\mathcal{O}(1)$ (which are purely classical), and treating all $\star$-products as $\mathcal{O}(\lambda')$, we continue:
{\small
\bea\label{Ricci: First order compx part}
B_{\mu\nu}^1&=& \delta_\beta^\alpha\delta_\alpha^\gamma\Gamma_{\mu\nu}^\tau\star\Gamma_{\gamma\tau}^\beta-\delta_{\beta}^\alpha\delta_\mu^\gamma\Gamma_{\alpha\nu}^\tau\star\Gamma_{\gamma\tau}^\beta
\neweqline
&&+(i\lambda')\left[\delta_\beta^\alpha\delta_\alpha^j\delta_j^\gamma\partial_0(\Gamma_{\mu\nu}^\tau)-\delta_\beta^\alpha\delta_\mu^j\delta_j^\gamma\partial_0(\Gamma_{\alpha\nu}^\tau)\right.
\neweqline
&&\left.-\delta_\gamma^\alpha \delta_\beta^j 
\delta_j^\alpha 
\partial_0(\Gamma_{\mu\nu}^\tau)+\delta_\beta^j\delta_j^\alpha \delta_\mu^\gamma \partial_0(\Gamma_{\alpha\nu}^\tau) 
\right]\Gamma_{\gamma\tau}^\beta
\\
&=&\Gamma_{\mu\nu}^\tau
\star\Gamma_{\tau\gamma}^\gamma-\Gamma_{\beta\nu}^\tau
\star\Gamma_{\mu\tau}^\beta
+i\lambda'\partial_0(\Gamma_{\beta\nu}^\tau)\Gamma_{\tau\mu}^\beta
\mathcal{P}^j_{[\mu,\beta]},\nonumber
\eea
}
where we defined the following \textbf{notation}, which will be also useful later on:
\beq\label{Projector notation}
\mathcal{P}^j_{[\gamma,\rho]} := \delta_\gamma^j \delta^\gamma_j - \delta_\rho^j \delta_j^\rho,
\eeq
which transforms as a scalar under coordinate changes.

We can now express the first-order Ricci tensor as:
\bea\label{Ricci: first order whole}
\textgoth{R}^{(1)}_{\mu\nu} &=& \mathcal{O}(1)
- i\lambda' \delta_{\beta}^j \delta^\beta_j \partial_0\left(\partial_{[\beta}\Gamma_{\mu]\nu}^\beta\right)
+ B_{\mu\nu}^1.
\eea

As expected, the $\kappa$-deformation manifests in specific structural features: the appearance of preferred spatial directions via the projector $\mathcal{P}^j_{[\mu,\beta]}$, and the emergence of time derivatives $\partial_0$, both reflecting the anisotropy of the $\kappa$-Minkowski spacetime. Additionally, as mentioned in \eqref{coefficient frame curvature}, the $\star$-Ricci tensor $\textgoth{R}_{\mu\nu}$ indeed turns out to be \textbf{non-symmetric} under the exchange of $\mu$ and $\nu$.

\subsubsection*{The deformed Equations}

Now, if one considers empty space without any matter source, the deformed Einstein equation \eqref{deformed EE} simplifies to the following form:\footnote{
For a vacuum scenario, equation \eqref{deformed EE} reduces in its form (though not identical since $\GeomR$ is different), as in the classical scenario, to $0=\textgoth{R}_{\mu\nu}-\frac{1}{2}g_{\mu\nu}\star \textgoth{R}$, which implies
\bea
0&=&\invG^{\mu\nu }\star \textgoth{R}_{\mu\nu}-\frac{1}{2}\invG^{\mu\nu }\star g_{\mu\nu}\star (\invG^{\rho\sigma }\star \textgoth{R}_{\rho\sigma})
\nonumber \\
&=&\invG^{\mu\nu }\star(\textgoth{R}_{\mu\nu}-\frac{1}{2}\times4\textgoth{R}_{\mu\nu})
\\
\Rightarrow \textgoth{R}_{\mu\nu}&=&0. \nonumber
\eea
} 
\beq\label{vacuum EE deformed}
\textgoth{R}_{\mu\nu}=0.
\eeq
Importantly, such an equation resolves the ambiguity of the non-uniqueness of the contraction in $\textgoth{R}$. Nevertheless, similar to the constant non-commutativity scenario, the Ricci tensor $\textgoth{R}_{\mu\nu}$ is not real/Hermitian. At the level of the equations of motion, one way to address this is to consider a symmetrized version of $\textgoth{R}_{\mu\nu}$ before solving (see, for example, \cite{Metric_Perturbations_In_NC}). However, we note that at the level of the associated deformed \textit{action}, this ambiguity can be resolved differently, as shown in \eqref{equation from action!}. This alternative approach can also clarify the ambiguity of the non-uniqueness of \eqref{dependent: Ricci scalar}, even when not dealing with vacuum solutions. 

For the deformed Einstein equation in non-vacuum spaces at the level of equations of motion, we need the deformed Einstein tensor \eqref{result: deformed EE} explicitly, keeping in mind that the deformed Ricci scalar \eqref{dependent: Ricci scalar} is just $\GeomR=\invG^{\mu\nu}\star\GeomR_{\nu\mu}$, producing the following equation: 
\bea\label{expanded defomed EE}
0&=& \GeomR_{\mu\nu}-\frac{1}{2}g_{\mu\nu}\star\GeomR
\\
&=&
O^\alpha_\beta[\lambda',-2]\left(\partial_{[\alpha}\Gamma_{\mu]\nu}^\beta\right)+S^\alpha_\beta\left(S^{*\gamma}_{[\alpha}(\Gamma_{\mu]\nu}^\tau)\star\Gamma^\beta_{\gamma\tau}\right)\nonumber
\\
&&+g_{\mu\nu}\star\invG^{\rho\phi}\star\left[S^{*\alpha}_\beta\left(\partial_{[\alpha}\Gamma_{\phi]\rho}^\beta\right)+S^\alpha_\beta\left(S^{*\gamma}_{[\alpha}(\Gamma_{\rho]\phi}^\tau)\star\Gamma^\beta_{\gamma\tau}\right)\right].\nonumber
\eea

Written compactly, the deformed Einstein equation \eqref{expanded defomed EE} is a perturbation expansion in the deformation parameter $\lambda'$:
\bea\label{result: deformed EE}
0&=&\textgoth{R}_{\mu \nu}[\mathbf{g},\lambda'^1,\ldots,\lambda'^n,\ldots]
\\
&\,&-\frac{1}{2}\bar{\phi}^{\{\alpha\}}(\mathbf{g}_{
\mu \nu })\bar{\phi}_{\{\alpha\}}(\textgoth{R}[\mathbf{g},\lambda'^1,\ldots,\lambda'^n,\ldots]).
\nonumber
\eea
The brackets in $\textgoth{R}_{\mu\nu}[\ldots]$ and $\textgoth{R}[\ldots]$ denote a functional dependence on the deformation parameter and the metric.
In principle, all terms in \eqref{expanded defomed EE} can be expanded to a chosen order in the non-commutativity parameter $\lambda'$. The zeroth order of \eqref{expanded defomed EE} is equivalent to the classical expression. This is due to the nature of the operator \eqref{Notation Definition O[l,n]} and the $\star$-product \eqref{dependent: star product kappa}. When combined with the fact that the underlying twisted $\text{IGL}'(3,1)$ symmetry (see \eqref{generators of igl'}) approaches classical Poincaré symmetry in the commutative limit ($\lambda'=0$), equation \eqref{expanded defomed EE} exhibits a meaningful classical limit and obeys the correspondence principle.

Given a scheme to address the non-reality and non-uniqueness of terms (such as adding $c.c$ in the action or $\textgoth{R}$-symmetrization), Equation \eqref{expanded defomed EE} or \eqref{vacuum EE deformed} can theoretically be solved to a given order in $\lambda'$. This provides valuable constraints on the perturbation (non-commutativity) parameter. These constraints can then serve as a learning tool for various theories involving the $\kp$-Minkowski spacetime \eqref{Kappa minkowsky} (see Sec.\ref{section: Introduction}).

Note that \eqref{expanded defomed EE} and \eqref{vacuum EE deformed} represent the deformed version of Einstein's equation \textit{in vacuum}, and thus describe a deformed Einstein manifold. An important aspect of such manifolds is that they may admit the same solutions as the undeformed case, provided that the spacetime possesses Killing vector fields which coincide with the twist vector fields \eqref{twist vector fields}; see, e.g., \cite{Symmetry_Reduction, NC_Geometry3}.
This property, namely, the appearance of trivial (classical) solutions, will generally no longer hold in the presence of a source that lacks the relevant Killing symmetry. To use \eqref{expanded defomed EE} with a source, we must evaluate the source’s energy-momentum tensor (EMT) $T^{\mu\nu}$ within the NC framework, obtaining $\hat{T}^{\mu\nu}$ in the sense of \eqref{algebra isomorphism}. This aspect will be discussed after we address the action formulation.

However, it is important to note that even in the absence of a source (i.e., in the vacuum case described by \eqref{expanded defomed EE}), non-trivial solutions can still arise---specifically when the twist vector fields \eqref{twist vector fields} do \textbf{not} generate symmetries of the spacetime \cite{NC_Geometry3}. This situation occurs, for example, in perturbations of a Schwarzschild black hole \cite{Metric_Perturbations_In_NC}, or in any perturbation that breaks time or scaling symmetries. For this reason, it is worthwhile to study \eqref{expanded defomed EE} even in the vacuum case.

\subsection*{The $\star$--Einstein–Hilbert Action}

For completeness and as an alternative derivation of the deformed Einstein
equations \eqref{expanded defomed EE} we now construct the
deformed Einstein–Hilbert (EH) action.  
A rigorous treatment requires a precise notion of integration over an
$m$-form with (graded) cyclicity.  
Here we exploit a key advantage of using an Abelian twist
\eqref{abelian twist1}:  
for any $\star$–$m$-form,
$w\overset{\star}{\wedge} w' := \buphi(w)\wedge\bdphi(w')$,
the integral is inherently cyclic
\cite{Twistgeneral1, Integral_cyclic}.  
Inside the integral one may drop the explicit “$\star$’’ in
$\overset{\star}{\wedge}$, since the deformation terms contribute only total
derivatives. We note that this property is true \textit{only} for Abelian twist theories and not in general, see Sec.\ref{section: Building Blocks For Construction} for more details.

\vspace{0.5em}
\noindent
\textbf{Ansatz for the action.}
Guided by the correspondence with the classical limit, we \textit{assume} 
\footnote{The square root of the $\star$‐determinant and its
transformation properties can be read from what follows, specifically see \eqref{vielbein trans 2},\eqref{identity 2}.}
the deformed EH action to be
\begin{equation}
\label{deformed Einstein Hilbert}
\overset{\star}{S}_{\text{EH}}
  = \!\! \int\!  \sqrt{-\overset{\star}{\det}[g]}\,\star\textgoth{R}\;+\;c.c.
  = \!\! \int\!  \sqrt{-\overset{\star}{\det}[g]}\,\textgoth{R}\;+\;c.c.,
\end{equation}
where the $\star$‐determinant of an $n\times n$ matrix $\mathbf A$ is
defined by
\begin{equation}\label{star determinant}
\overset{\star}{\det}[\mathbf A]
   = \frac{\epsilon^{\mu_1\ldots\mu_n\nu_1\ldots\nu_n}}{n!}\;
     \mathbf A_{\mu_1\nu_1}\star\mathbf A_{\mu_2\nu_2}
     \star\cdots\star\mathbf A_{\mu_n\nu_n},
\end{equation}
with $(\mu,\nu)\in[0,n]$.

\subsubsection*{Invariance under deformed GCTs.}
We must verify that \eqref{deformed Einstein Hilbert} is invariant under
deformed general coordinate transformations (GCTs) and that its variation
with respect to $\invG^{\mu\nu}$ reproduces
\eqref{expanded defomed EE}. 
To do so, we follow the approach of \cite{Gravity_Non} (noting that our derivation of invariance closely parallels that in \cite{Gravity_Non}, albeit with a few differences), and rewrite the metric field in terms of vielbeins as:
\bea\label{vielblein metric}
g_{\mu\nu}:=\frac{1}{2}(e_\mu^A\star e_\nu^B+e_\nu^B\star e_\mu^A)\Tilde{\eta}_{AB},
\eea
where we have introduced a set of \textit{vielbeins} $\{e_\alpha^\Gamma\}$, with $\alpha$ denoting a spacetime covariant index and $\Gamma$ a local symmetry group transformation index. The symbol $\Tilde{\eta}_{AB}$ represents the corresponding local (flat and constant) metric. 
Here, the index $\Gamma$ is associated with the transformations governed by those discussed in Sec.~\ref{subsec: consistency transl}, and accordingly, so is the metric $\Tilde{\eta}_{AB}$. For the purposes of the current discussion, we focus solely on the covariant spacetime indices $\alpha$, and therefore defer the full specification of $\Tilde{\eta}_{AB}$ (beyond noting its constancy) and the transformation rules of the $\Gamma$ indices to future work.

Using the linearity of \eqref{star determinant} and the constancy of
$\Tilde{\eta}_{AB}$, we can factor out the determinant $\det(\Tilde{\eta}_{AB})$, and write:
\bea
\sqrt{-\overset{\star}{\det}[g]}\star\textgoth{R}&=&\sqrt{\bigl(\overset{\star}{\det}[e_\mu^{A}]\bigr)^{2}\star\overset{\star}{\det}[\tilde{\eta}_{AB}]}\star\textgoth{R}
\\
&=& \det[\tilde{\eta}_{AB}]\overset{\star}{\det}[e_\mu^{A}]\star\textgoth{R}:= \tilde{\eta}e_\star\star\textgoth{R},\nonumber
\eea
where we denoted $\tilde{\eta} := \det(\tilde{\eta}_{AB})$ (a constant), and $e_\star := \overset{\star}{\det}(e_\mu^A)$.  
Since constant factors do not affect invariance, we can drop $\tilde{\eta}$ in the action. Thus, to show the invariance of \eqref{deformed Einstein Hilbert}, it suffices to analyze the transformation of $e_\star\star\textgoth{R}$.

To begin with, from the definition of $\textgoth{R}$ in \eqref{star definition of the Ricci scalar} we know it behaves as a scalar by construction (and its non-reality merely
requires adding the complex conjugate in the action; it will not change invariance properties).  
For $e_\star$ we know that each vielbein separately transforms as a vector as in \eqref{trans rule vectors}; therefore, by the observation above (see discussion after \eqref{trans rule vectors})  the transformation rules of $\star$-multiplication of quantities of known transformations rules is the same as that the rules the regular multiplication has.
Regarding the transformation of $\overset{\star}{\det}(e_\mu^A)=e_\star$, it is known that the same expression with usual products transforms as $\delta_\xi(\det(e_\mu^A))=-\lie_\xi(\det(e_\mu^A))-\lie_{(\partial_\mu\xi^\mu)}(\det(e_\mu^A))$. Therefore, we conclude that 
\begin{align}
\delta_\xi^\star(e_\star)
  = -\lie^\star_\xi \!\left[
     \frac{\epsilon^{\mu_1\!\ldots}}{n!}
     e_{\mu_1}^{A_1}\!\star\! e_{\mu_2}^{A_2}\!\star\!\cdots\right]
  -\lie^\star_\xi(e_\star) - \lie^\star_{\partial_\mu\xi^\mu}(e_\star),
\label{vielbein trans 2}
\end{align}
Once we know how each term in $e_\star\star
\textgoth{R}$ transforms, and keeping in mind the observation that \textit{its} transformation rule will be same as the one for $e_\star\textgoth{R}$, we conclude that 
\bea
\delta^\star_\xi(e_\star\star\textgoth{R})&=&-\lie^\star_\xi(e_\star\star\textgoth{R})-\lie^\star_{\partial_\mu\xi^\mu}(e_\star\star\textgoth{R})
\neweqline
&=&
-(\partial_\mu\xi^\mu)(e_\star\star\textgoth{R})-\xi^\mu(\partial_\mu(e_\star\star\textgoth{R}))
\neweqline
&=&-\partial_\mu[\xi^\mu(e_\star\star\textgoth{R})],
\eea
with the action of the derivative and vector field to be taken as a $\star$-Lie derivative. The last term is what we aimed at showing: that $e_\star\star \textgoth{R}$ transforms to a total derivative, hence validating that $\delta_\xi^\star(\overset{\star}{S_{EH}})=0$.

\subsubsection*{Variation with respect to $\invG^{\mu\nu}$.}
To get the equations of motion, we follow the first formulation of gravity in the context of NC geometry as in \cite{Gravity_Non} and vary the integrand in \eqref{deformed Einstein Hilbert}:
\bea\label{variation of EH}
\frac{\delta}{\delta\invG^{\mu\nu}}
 \bigl(\sqrt{-g_\star}\star\invG^{\rho\sigma}\star\textgoth{R}_{\rho\sigma}\bigr)
 &=&
\frac{\delta\sqrt{-g_\star}}{\delta\invG^{\mu\nu}}\star\textgoth{R}
\\
&+&
\sqrt{-g_\star}\star
\frac{\delta\invG^{\rho\sigma}}{\delta\invG^{\mu\nu}}\star\textgoth{R}_{\rho\sigma}
\nonumber\\
&+&\sqrt{-g_\star}\star\invG^{\rho\sigma}\star\frac{\delta\textgoth{R}_{\rho\sigma}}{\delta\invG^{\mu\nu}},\nonumber
\eea
where for brevity $g_\star:=\overset{\star}{\det{g}}_{\mu\nu}$.
Note that since twist deformations act on space–time symmetries, not on field
variations, the usual (undeformed) Leibniz rule for
$\delta(\cdots)$ was applied, see \cite{Gravity_Non}.  
Consequently, we must evaluate the following $\star$ expressions
\beq\label{3 evaluations}
\frac{\delta\invG^{\rho\sigma}}{\delta g_{\mu\nu}},\quad
\frac{\delta\sqrt{-g_\star}}{\delta\invG^{\mu\nu}},\quad
\frac{\delta\textgoth{R}_{\rho\sigma}}{\delta\invG^{\mu\nu}}.
\eeq

\medskip\noindent
\textbf{First identity.}
Following the classical treatment, from
$\delta(g_{\mu\nu}\star\invG^{\nu\sigma}) = 0$, and applying the classical Leibniz rule for $\delta$ we obtain
\bea\label{identity 1}
0&=&\delta(g_{\mu\nu})\star\invG^{\nu\sigma}+g_{\mu\nu}\star \delta(\invG^{\nu\sigma})
\neweqline
 \Rightarrow 
\delta(\invG^{\rho\sigma})
&=& -\invG^{\mu\rho}\!\!\!\star\!\delta(g_{\mu\nu})\!\star\!\invG^{\nu\sigma}=-\invG^{\mu\rho}\!\!\!\star\!\invG^{\nu\sigma}\!\!\!\!\star\!\delta(g_{\mu\nu}).~~~~~
\eea
using cyclicity to move the variation rightward.
\\

\textbf{Second identity.}
Following the classical derivation, we first consider the variation $\delta(g_\star)$ by using \eqref{star determinant} and absorbing the overall constants 
\bea\label{det variation}
&&\frac{\delta}{\delta g_{\mu\nu}}\left(\epsilon^{\mu_1\mu_2...\nu_1\nu_2...}g_{\mu_1\nu_1}\!\star\! g_{\mu_2\nu_2}\!\star\!\ldots\right)=
\neweqline
&&~~~~~~~~~=\epsilon^{\mu_1...\nu_1...}\left(\delta(g_{\mu\nu})\star g_{\mu_2\nu_2}\star\ldots\right.
\nonumber\\
&&~~~~~~~~~~~~~~~~~~~~~~~~\left.+g_{\mu_1\nu_1}\star\delta(g_{\mu\nu})\star g_{\mu_3\nu_3}\star\ldots+\ldots\right)
\nonumber\\
&&~~~~~~~~~=g_\star\star\invG^{\mu\nu}\star\delta(g_{\mu\nu}).
\eea
Where the classical Leibniz rule for $\delta$ and cyclicity properties for each of $g_{\mu\nu}$ and for moving $\delta(g) $ to the right were used. By use of \eqref{identity 1}, we derive 
\bea\label{det variation result}
\delta(g_{\mu\nu})&=&-g_{\mu\rho}\star\delta(\invG^{\rho\sigma})\star g_{\nu\sigma}
=-g_{\mu\rho}\star g_{\nu\sigma}\star\delta(\invG^{\rho\sigma}) 
\neweqline \Rightarrow \delta(g_\star)&=&-g_\star\star g_{\rho\sigma}\delta(\invG^{\rho\sigma}). 
\eea

By all of these, we can now compute $\delta(\sqrt{-g_\star})$: 
\bea\label{identity 2}
\frac{\delta(\sqrt{-g_\star})}{\delta\invG^{\mu\nu}}=\frac{1}{2\sqrt{-g_\star}}\frac{\delta g_\star}{\delta\invG^{\mu\nu}} 
= -\frac{1}{2}\sqrt{-g_\star}\star g_{\mu\nu}
.
\eea
\\
\textbf{Third identity (generalized Palatini relations).}

finally, we turn to the third term in \eqref{3 evaluations}, namely the variation $\delta \textgoth{R}_{\mu\nu}$. In classical GR, the well-known \textit{Palatini identity} provides the variation of the Ricci tensor:
\begin{equation}\label{classical palatini}
\delta(R_{\mu\nu}) = \nabla_\lambda(\delta\Gamma_{\mu\nu}^{\lambda}) - \nabla_\nu(\delta\Gamma_{\mu\lambda}^{\lambda}).
\end{equation}
This identity holds under the standard assumption (typically proven explicitly) that $\delta\Gamma_{\mu\nu}^\lambda$ transforms as a rank-(1,2) tensor under spacetime diffeomorphisms. The utility of this identity is that, under metric compatibility, the last term in \eqref{variation of EH} can be cast schematically as $\sqrt{-g} \nabla(g^{\mu\nu} \delta\Gamma)$, which, when combined with the other terms in \eqref{variation of EH}, becomes a total derivative and thus vanishes under suitable boundary conditions.

We aim to adopt an analogous strategy in our deformed setting by expressing $\delta \textgoth{R}_{\mu\nu}$ in terms of $\star$-covariant derivatives. Beginning from the expanded expression of the deformed Ricci tensor in \eqref{Expansion of Ricci tensor} and \eqref{Ricci: first order whole}, and using the shorthand notation \eqref{matrix differential notaion}, we find the variation:
{\small
\bea\label{variation of Ricci}
\delta(\textgoth{R}_{\mu\nu}) &=& \partial_\alpha\left(\delta S^{*\alpha}_\beta(\Gamma^\beta_{\mu\nu})\right) - \partial_\mu\left(\delta S^{*\alpha}_\beta(\Gamma_{\alpha\nu}^\beta)\right) 
\neweqline
&&+ \left[\delta\Gamma_{\mu\nu}^\tau \star \Gamma_{\tau\gamma}^{\gamma} + \Gamma_{\mu\nu}^\tau \star \delta\Gamma_{\tau\gamma}^{\gamma}\right] 
\neweqline
&&-\left[\delta\Gamma_{\beta\nu}^\tau \star \Gamma_{\mu\tau}^\beta + \Gamma_{\beta\nu}^\tau \star \delta\Gamma_{\mu\tau}^\beta\right] 
\\
&&+i\lambda'\left(\delta_\beta^j\delta_j^\beta-\delta_\mu^j\delta_j^\mu\right)\left[\partial_0(\delta\Gamma_{\beta\nu}^\tau)\Gamma_{\tau\mu}^\beta+\partial_0(\Gamma_{\beta\nu}^\tau)\delta\Gamma_{\tau\mu}^\beta\right], \nonumber 
\eea
}

Here we used that $\delta$ commutes with both derivatives and twist elements. We now define the shorthand notation:
\begin{equation}\label{Tigam definition}
S_\beta^{*\alpha}(\Gamma_{\mu\nu}^\beta):=\Tigam^\alpha_{\mu\nu},\quad S_\beta^{*\alpha}(\Gamma_{\alpha\nu}^\beta):=\Tigam^\alpha_{\alpha\nu},
\end{equation}
with the index placement implied by the contraction with $\delta^\alpha_\beta$ in $S_\beta^{*\alpha}$; see also \eqref{Notation Definition O[l,n]}. Using this, we rewrite \eqref{variation of Ricci} as:
{\small
\bea\label{variation of Ricci Tigam}
\delta(\textgoth{R}_{\mu\nu}) &=&\partial_\alpha(\delta\Tigam_{\mu\nu}^\alpha) - \partial_\mu(\delta\Tigam_{\nu\alpha}^\alpha) 
+ \left[\delta\Gamma_{\mu\nu}^\tau \star \Gamma_{\tau\gamma}^{\gamma} + \Gamma_{\mu\nu}^\tau \star \delta\Gamma_{\tau\gamma}^{\gamma}\right] 
\neweqline
&&-\left[\delta\Gamma_{\beta\nu}^\tau \star \Gamma_{\mu\tau}^\beta + \Gamma_{\beta\nu}^\tau \star \delta\Gamma_{\mu\tau}^\beta\right] 
\\
&&+i\lambda'\mathcal{P}^j_{[\beta,\mu]}\left[\partial_0(\delta\Gamma_{\beta\nu}^\tau)\Gamma_{\tau\mu}^\beta+\partial_0(\Gamma_{\beta\nu}^\tau)\delta\Gamma_{\tau\mu}^\beta\right],\nonumber
\eea
}
using the projector notation $\mathcal{P}^j_{[\beta,\mu]}$ as defined in \eqref{Projector notation}. 

To determine whether $\delta(\textgoth{R}_{\mu\nu})$ admits a representation in terms of $\nablaST(\cdot)$, we recall the general structure of the $\star$-covariant derivative acting on arbitrary tensors (cf. \eqref{dependent: right multiplied tensor covariant derivative}):
\bea\label{general tensor}
\nablaST_\beta(T_{\mu_1 \mu_2 \ldots}^{\beta \nu_1 \nu_2 \ldots}) &=& \partial_\beta(T^{\beta\nu\ldots}_{\mu\ldots}) 
\neweqline
&&+ S^{\beta'}_\beta(T^{\tau\nu_1\ldots}_{\mu_1\ldots}) \star \Gamma^\beta_{\tau\beta} - S^{\beta'}_{\beta}(T^{\beta\nu_1\ldots}_{\tau\mu_2\ldots}) \star \Gamma^\tau_{\mu_1\beta'} 
\neweqline
&&+ S^{\beta'}_\beta(T^{\beta\tau\nu_2\ldots}_{\mu_1\mu_2\ldots}) \star \Gamma^{\nu_1}_{\tau\beta'} - \ldots
\eea

At zeroth order in $\lambda$, all terms in \eqref{variation of Ricci} reduce to their classical expressions, yielding:
\bea\label{Ricci zero}
\delta(\textgoth{R}_{\mu\nu}^{(0)}) &=& \partial_\alpha(\delta\Gamma_{\mu\nu}^\alpha) - \partial_\mu(\delta\Gamma_{\nu\alpha}^\alpha) 
+ \delta\Gamma_{\mu\nu}^\tau \Gamma_{\tau\gamma}^\gamma + \Gamma_{\mu\nu}^\tau \delta\Gamma_{\tau\gamma}^\gamma 
\neweqline
&&\quad-\, \delta\Gamma_{\beta\nu}^\tau \Gamma_{\mu\tau}^\beta -\, \Gamma_{\beta\nu}^\tau \delta\Gamma_{\mu\tau}^\beta,
\eea
which coincides with the classical Palatini identity \eqref{classical palatini}, where $\delta\Gamma_{\mu\nu}^{\alpha}$ is treated as a rank-(1,2) tensor and $\delta\Gamma_{\nu\alpha}^{\alpha}$ as a rank-(0,1) tensor.

For the first-order-in-$\lambda$ identity, let us isolate the following term, denoted $\mathbb{A} \subset \delta\textgoth{R}_{\mu\nu}^{(1)}$, from \eqref{variation of Ricci Tigam}:
\bea\label{Ricci variation first term}
\mathbb{A}&:=&\partial_\alpha\delta\Tigam_{\mu\nu}^\alpha+\delta\Gamma_{\mu\nu}^\tau\star\Gamma_{\tau\gamma}^\gamma-\delta\Gamma_{\beta\nu}^\tau\star\Gamma_{\mu\tau}^\beta 
\neweqline
&&+i\lambda'\partial_0(\delta\Gamma_{\beta\nu}^\tau)\Gamma_{\tau\mu}^\beta\left(\delta_\beta^j\delta_j^\beta-\delta_\mu^j\delta^\mu_j\right),
\eea
where we consider all the $\star$-products to be evaluated at first order in $\lambda'$ as prescribed in \eqref{dependent: star product kappa}, neglecting $\mathcal{O}(1,\lambda^{n\geq 2})$ terms.

Our goal is to cast $\delta\textgoth{R}_{\mu\nu}^{(1)}$ in the form of a $\star$-covariant derivative. The structure of the first term in $\mathbb{A}$ suggests examining $\nablaST_{\alpha}(\delta\Tigam_{\mu\nu}^\alpha)$. Using \eqref{general tensor}, we write:
\bea\label{Ricci variation: 1st cov deriv term}
\nablaST_{\alpha}(\delta\Tigam_{\mu\nu}^\alpha)&=& \partial_\alpha(\delta\Tigam_{\mu\nu}^\alpha)+S^\sigma_\alpha(\delta\Tigam_{\mu\nu}^\tau)\star\Gamma_{\tau\alpha}^\alpha
\neweqline
&&
-S^\sigma_\alpha(\delta\Tigam_{\tau\nu}^\alpha)\star\Gamma_{\mu\sigma}^\tau
-S^\sigma_\alpha(\delta\Tigam_{\tau\mu}^\alpha)\star\Gamma_{\nu\sigma}^\tau
\neweqline
&=&\partial_\alpha(\delta\Tigam_{\mu\nu}^\alpha)+S^\sigma_\alpha(\delta\Tigam_{\mu\nu}^\tau)\star\Gamma_{\tau\alpha}^\alpha\neweqline
&&- \delta\Gamma_{\nu\tau}^\alpha\star\Gamma_{\mu\alpha}^\tau- \delta\Gamma_{\mu\tau}^\alpha\star\Gamma_{\nu\alpha}^\tau, 
\eea
where we used the contraction identity:
\beq\label{operator matrix contraction}
S^\sigma_\alpha(S^{*\alpha}_\gamma\delta\Gamma^\gamma_{\tau\nu}) = \delta^\sigma_\gamma \delta\Gamma_{\tau\nu}^\gamma = \delta\Gamma_{\tau\nu}^\sigma.
\eeq

To simplify \eqref{Ricci variation: 1st cov deriv term} explicitly to $\mathcal{O}(\lambda')$, we compute the matrix contribution 
$\mathbb{S}_{\mu\nu} := S^\sigma_\alpha(\delta\Tigam_{\mu\nu}^\tau)\star\Gamma_{\tau\alpha}^\alpha$ to first order in $\lambda'$, we find:
\bea
\mathbb{S}_{\mu\nu}&=&\left(\delta_\alpha^\sigma+i\delta_\alpha^j\delta_j^\sigma\lambda'\partial_0\right)(\delta\Tigam_{\mu\nu}^\tau)\star\Gamma_{\tau\sigma}^{\sigma} 
\neweqline
&=& \delta_{\alpha}^\sigma(\delta\Tigam_{\mu\nu}^\tau)\star\Gamma_{\tau\alpha}^\alpha +i\lambda'\partial_0\delta_\alpha^j\delta_j^\sigma(\delta\Tigam_{\mu\nu}^\tau)\star\Gamma_{\tau\sigma}^\sigma
\neweqline\
&=&\left(\delta^\tau_\rho-i\lambda'\partial_0\delta_\rho^j\delta_j^\tau\right)(\delta\Gamma_{\mu\nu}^\rho)\star\Gamma_{\tau\sigma}^\sigma+i\lambda'\delta_j^\sigma\delta^j_\sigma\partial_0(\delta\Gamma_{\mu\nu}^\tau)\Gamma_{\tau\sigma}^\sigma
\neweqline
&=& \delta_{\mu\nu}^\tau\star\Gamma_{\tau\sigma}^\sigma+i\lambda'\partial_0(\delta\Gamma_{\mu\nu}^\tau)\Gamma_{\tau\sigma}^\sigma\times\left[\delta_\tau^j\delta_j^\tau-\delta_j^\sigma\delta^j_\sigma\right]. 
\eea
Substituting into \eqref{Ricci variation: 1st cov deriv term}, we find:
\bea\label{Ricci variation: 1st cov deriv explict}
\nablaST_\alpha(\delta\Tigam_{\mu\nu}^\alpha)&=&\partial_\alpha(\delta\Tigam_{\mu\nu}^\alpha)+\delta\Gamma_{\mu\nu}^\tau\star\Gamma_{\tau\sigma}^\sigma - \delta\Gamma_{\nu\tau}^\sigma\star\Gamma_{\mu\sigma}^\tau
\neweqline
&&-\delta\Gamma_{\mu\tau}^\sigma\star\Gamma_{\nu\sigma}^\tau+i\lambda'\mathcal{P}^j_{[\tau,\sigma]}\partial_0(\delta\Gamma_{\mu\nu}^\tau)\Gamma_{\tau\sigma}^\sigma.~~~~~~
\eea

Therefore, the quantity $\mathbb{A}$ can be written as:
\beq\label{Ricci variation: cov to A}
\mathbb{A} = \nablaST_\alpha(\delta\Tigam_{\mu\nu}^\alpha) + \delta\Gamma_{\mu\tau}^\alpha\star\Gamma_{\nu\alpha}^\tau.
\eeq

We now turn to the remaining terms in \eqref{variation of Ricci Tigam}, denoting them as $\mathbb{B} \subset \delta\textgoth{R}_{\mu\nu}^{(1)}$, such that $\mathbb{A} + \mathbb{B} = \delta\textgoth{R}_{\mu\nu}^{(1)}$. To first order in $\lambda'$:
\bea
\mathbb{B}&=&-\partial_\mu(\delta\Tigam_{\nu\alpha}^\alpha)+\Gamma_{\beta\nu}^\tau\star\delta\Gamma_{\mu\tau}^\beta-\Gamma_{\mu\nu}^\tau\star\delta\Gamma_{\tau\gamma}^\gamma 
\neweqline
&&+ i\lambda'\partial_0(\Gamma_{\beta\nu}^\tau)\delta\Gamma_{\tau\mu}^\beta\left(\delta_\beta^j\delta^\beta_j-\delta_\mu^j\delta^\mu_j\right)
\eea
Applying the Leibniz property $\delta(a \star b) = \delta(a)\star b + a\star\delta(b)$, we obtain:
\bea
\mathbb{B}&=&-\delta\bigg[\partial_{\mu}(\Tigam_{\nu\alpha}^\alpha)+\Gamma_{\beta\nu}^\tau\star\Gamma_{\mu\tau}^\beta-\Gamma_{\mu\nu}^\tau\star\Gamma_{\tau\gamma}^\gamma
\neweqline
&&~~~~
+i\lambda'\partial_0(\Gamma^\tau_{\beta\nu})\Gamma_{\tau\mu}^\beta\mathcal{P}_{[\beta,\mu]}^j\bigg]
\\
&&-\left[\delta\Gamma_{\beta\nu}^\tau\star\Gamma_{\mu\tau}^\beta-\delta\Gamma_{\mu\nu}^\tau\star\Gamma_{\tau\gamma}^\gamma+i\lambda'\partial_0(\delta\Gamma^\tau_{\beta\nu})\Gamma_{\tau\mu}^\beta\mathcal{P}_{[\beta,\mu]}^j\right].\nonumber
\eea
Using \eqref{Ricci variation: 1st cov deriv explict}, we can write:
\bea\label{Ricci variation: B term simplified}
\mathbb{B}&=&-\partial_\mu(\delta\Tigam_{\nu\alpha}^\alpha)-\delta\left[-\nablaST_\alpha(\Gamma_{\mu\nu}^\alpha)+\partial_\alpha(\Gamma_{\mu\nu}^\alpha)-\Gamma_{\mu\tau}^\alpha\star\Gamma_{\nu\alpha}^\tau\right]
\neweqline
&&+\left[-\nablaST_\alpha(\delta\Gamma_{\mu\nu}^\alpha)+\partial_\alpha(\delta\Gamma_{\mu\nu}^\alpha)-\delta\Gamma_{\mu\tau}^\alpha\star\Gamma_{\nu\alpha}^\tau\right]
\neweqline
&=&-\partial_\mu(\delta\Tigam_{\nu\alpha}^\alpha)=-\nablaST_\mu(\delta\Tigam_{\alpha\nu}^\alpha)+S_\alpha^\sigma(\delta\Gamma_{\tau\mu}^\alpha)\star\Gamma_{\nu\sigma}^\tau
\neweqline
&=&-\nablaST_\mu(\delta\Tigam_{\alpha\nu}^\alpha)+\delta\Gamma_{\mu\tau}^\alpha\star\Gamma_{\nu\alpha}^\tau,
\eea
where we again used the contraction identity \eqref{operator matrix contraction}.

Finally, combining \eqref{Ricci variation: cov to A} and \eqref{Ricci variation: B term simplified}, we obtain the first-order deformed Palatini identity:
\bea\label{def Palatini 1st order local}
\delta\textgoth{R}_{\mu\nu}^{(1)}&=&\mathbb{A}+\mathbb{B}
\neweqline
&=&\nablaST_\alpha(\delta\Tigam_{\mu\nu}^\alpha)+\delta\Gamma_{\mu\tau}^\alpha\!\star\!\Gamma_{\nu\alpha}^\tau\! -\!\left(\nablaST_\mu(\delta\Tigam_{\alpha\nu}^\alpha)
\!+\!\delta\Gamma_{\mu\tau}^\alpha\!\star\!\Gamma_{\nu\alpha}^\tau\right)
\neweqline
&=&\boxed{\nablaST_\alpha(\delta\Tigam_{\mu\nu}^\alpha)-\nablaST_\mu(\delta\Tigam_{\nu\alpha}^\alpha)}
\eea

This result aligns with our goal: just as in the classical Palatini identity \eqref{classical palatini}, we find that the deformed variation $\delta\textgoth{R}_{\mu\nu}$—up to first order in $\lambda'$—can be expressed as a difference of $\star$-covariant derivatives acting on appropriately defined tensors.
At this stage, it remains an open question whether such a compact $\star$-covariant form persists at higher perturbative orders in $\lambda'$, and further investigation is required to address this point.

Nevertheless, since our interest lies in evaluating $\invG \star \delta\textgoth{R}$ \textit{within an integral}, we can exploit the cyclicity of the integral under an Abelian twist to commute $\star$-factors. At zeroth order, the integral is trivial. At higher order—specifically, at order $\lambda'^s$ with $s \geq 1$—we write the Ricci tensor \eqref{Expansion of Ricci tensor} as:
{\small
\bea\label{ricci tensor global to order s}
\textgoth{R}_{\mu\nu}^{(s)}&:=&S^{*\alpha}_\beta(\partial_{[\alpha}\Gamma_{\mu]\nu}^\beta)+\left[\delta_\beta^\alpha+\left(\sum_{n=1}^{s}\tfrac{(-i\lambda')^n}{n!}\partial_0^n\right)\delta_\beta^j\delta^\beta_j\right]\times
\neweqline
&&\left[\delta_\mu^\gamma
+ \left(\sum_{r=1}^{s}\tfrac{(i\lambda')^r}{r!}\partial_0^r\right)\delta_\mu^j\delta^\gamma_j\right](\Gamma^\tau_{\alpha\nu})\Gamma_{\gamma\tau}^\beta 
\neweqline
&=& S^{*\alpha}_\beta(\partial_{[\alpha}\Gamma_{\mu]\nu}^\beta)+\delta_\beta^\alpha\delta_\mu^j\delta^\gamma_j \tfrac{(i\lambda')^s}{s!}\partial_0^s(\Gamma_{\alpha\nu}^\tau)\Gamma_{\gamma\tau}^\beta 
\neweqline
&&+ \delta_\mu^\gamma\delta_\beta^j\delta_j^\alpha \tfrac{(-i\lambda')^s}{s!}\partial_0^s(\Gamma^\tau_{\alpha\nu})\Gamma^\beta_{\gamma\tau} 
\neweqline
&&+ \delta_\mu^j\delta_j^\gamma\delta_\beta^j\delta^\alpha_j(i\lambda')^s\partial_0^s\times \sum_{n=0}^{s}\tfrac{(-1)^n}{n!(s-n)!}
\\
&=&S^{*\alpha}_\beta(\partial_{[\alpha}\Gamma_{\mu]\nu}^\beta)+\tfrac{(i\lambda')^s}{s!}\partial_0^s(\Gamma^\tau_{\beta\nu})\Gamma_{\tau\mu}^\beta\mathcal{P}_{[\beta,\mu]}^j(s),\nonumber
\eea
}
where we defined the following \textbf{notation}, generalizing \eqref{Projector notation}:
\beq
\mathcal{P}_{[\beta,\mu]}^j(s):=\left[\delta_\beta^j\delta_j^\beta + (-1)^s\delta_\mu^j\delta^\mu_j\right].
\eeq

Note that in \eqref{ricci tensor global to order s} we have discarded all terms of the form $\Gamma\star\Gamma$, which reduce to $\Gamma\Gamma$ inside the integral and are thus of zeroth order in $\lambda'$. We now express the variation of $\textgoth{R}_{\mu\nu}$ at order $s$ under the integral as:
\bea\label{ricci variation all orders int}
\delta\textgoth{R}_{\mu\nu}^{(s)} &=& \partial_\alpha(\delta\Tigam_{\mu\nu}^\alpha) - \partial_\mu(\delta\Tigam_{\nu\alpha}^\alpha)
\\
&&+ \tfrac{(i\lambda')^s}{s!} \mathcal{P}_{[\beta,\mu]}^j(s) \left[\partial_0^s(\delta\Gamma_{\beta\nu}^\tau)\Gamma_{\tau\mu}^\beta + \partial_0^s(\Gamma_{\beta\nu}^\tau)\delta\Gamma_{\tau\mu}^\beta\right]. \nonumber
\eea

At order $s$, the analog of \eqref{Ricci variation: 1st cov deriv explict} simplifies considerably, as all $\Gamma\star\Gamma$ terms again become zeroth-order under integration. This gives a summation structure identical to the one in \eqref{ricci tensor global to order s}, allowing us to write:
\bea\label{ricci variation: global A to all orders}
\nablaST_\alpha^{(s)}(\delta\Tigam_{\mu\nu}^\alpha) &=& \partial_\alpha(\delta\Tigam_{\mu\nu}^\alpha)
\\
&&+ \frac{(i\lambda')^s}{s!}\partial_0^s(\delta\Gamma_{\mu\nu}^\tau)\Gamma_{\tau\sigma}^\sigma\mathcal{P}_{[\tau,\sigma]}^j(s). \nonumber
\eea

Following the same steps as in \eqref{Ricci variation: B term simplified}, and identifying $\nablaST_\mu(\delta\Tigam_{\alpha\nu}^\alpha) = \partial_\mu(\delta\Tigam_{\alpha\nu}^\alpha)$ in this context (again because $\Gamma\star\Gamma$-terms are zeroth-order within the integral), we conclude that, analogously to \eqref{def Palatini 1st order local}:
\bea\label{Palatini global all orders}
\delta\textgoth{R}_{\mu\nu}^{(s)} = \nablaST_\alpha(\delta\Tigam_{\mu\nu}^\alpha) - \nablaST_{\mu}(\delta\Tigam_{\alpha\nu}^\alpha) + \mathcal{O}(\star),
\eea
where $\mathcal{O}(\star)$ denotes terms from \eqref{variation of Ricci} that either belong to lower orders or vanish under integration due to the cyclicity of the $\star$-product (e.g., $\Gamma \star \Gamma \rightarrow \Gamma \Gamma$ inside the integral).

An important outcome of this analysis is worth highlighting: in classical GR, the Palatini identity is essential for establishing the \textit{local} conservation of the Einstein tensor, $\nabla G_{\mu\nu} = 0$, which in turn ensures the local conservation of the energy-momentum tensor. In the NC setting, we have shown that a \textit{global} version of the deformed Palatini identity, as given in \eqref{Palatini global all orders}, holds to all orders in $\lambda'$. However, for the \textit{local} version, we have only established its validity at first order in $\lambda'$, as given in \eqref{def Palatini 1st order local}, while its status at higher orders remains unresolved.
It would be valuable to investigate the local deformed Palatini identity further, both to determine whether it persists beyond first order and to study the implications this may have for local conservation laws. We leave this for future work.

\subsubsection*{Equations from Action}
Returning to \eqref{variation of EH} and inserting
\eqref{identity 1}–\eqref{identity 2} together with the Palatini‐type
results, while freely cycling factors under the integral, we have
\bea\label{variation EH pieces}
\sqrt{-g_\star}\star\delta\invG^{\mu\nu}\star\textgoth{R}_{\mu\nu}
 &=& \sqrt{-g_\star}\star\textgoth{R}_{\mu\nu}\star\delta\invG^{\mu\nu},
\\
\sqrt{-g_\star}\star\invG^{\mu\nu}\star\delta\textgoth{R}_{\mu\nu}
 &\sim& \sqrt{-g_\star}\star\nabla(g\partial_0\delta\Gamma),
\\
\delta\bigl(\sqrt{-g_\star}\bigr)\star\textgoth{R}
 &=& \sqrt{-g_\star}\star\bigl(-\tfrac12 g_{\mu\nu}\star\textgoth{R}\bigr)
    \star\delta\invG^{\mu\nu}.\quad\quad
\eea
All total $\star$‐(co)derivative terms drop under suitable boundary
conditions.  Hence
\bea\label{equation from action!}
\delta\overset{\star}{S}_{EH}&=&\int \sqrt{-g_\star}\star\left[\textgoth{R}_{\mu\nu}-1/2g_{\mu\nu}\textgoth{R}\right]\star\delta\invG^{\mu\nu}+\sqrt{-g_\star}\nabla=0
\neweqline
&&\Longrightarrow\qquad
 \boxed{\textgoth{R}_{\mu\nu}-\tfrac12\,g_{\mu\nu}\textgoth{R}=0},
\eea
reproducing \eqref{expanded defomed EE}. Note that, as discussed in \cite{Gravity_Non}, one can derive a real action by adding the complex conjugate (c.c.) part as in \eqref{deformed Einstein Hilbert}. Thus, the final action to be used is:
\beq\label{EH with CC}
\overset{\star}{S}_{EH}=  \int{ \sqrt{-g_\star}\star\left(\textgoth{R}+\textgoth{R}^{*}\right)},
\eeq
with $\textgoth{R}^{*}$ as the complex conjugate of $\textgoth{R}$. Note the assumption that $\sqrt{-g_\star}$ is real, following \cite{Gravity_Non}, and the fact that the operator ordering ambiguity in $\textgoth{R}$ as in \eqref{dependent: Ricci scalar} is absent due to the cyclic integral. The action \eqref{EH with CC} can be expanded to a chosen order and solved. As was noted, it is expected that all the odd $\mathcal{O}(\lambda^{n_{odd}})$ will vanish, and one will be left with a real action. Explicitly demonstrating this behavior would be interesting, and we leave it for future work.

\vspace{0.5em}
\noindent
In this subsection, we stated the deformed EH action \eqref{deformed Einstein Hilbert} in the NC of $\kp$-Minkowski flat spacetime. We have assessed the transformation properties of each of the ingredients in the deformed action. Using the crucial fact that the action constructed preserves deformed diffeomorphism invariance we could extract the deformed Einstein equations \eqref{expanded defomed EE}.  
This was established by the variations of each term in the action, focusing on the variation of the NC Ricci tensor $ \textgoth{R}_{\mu\nu} $, expanded in powers of the deformation parameter $ \lambda $. At first order, we showed that $ \delta \textgoth{R}_{\mu\nu} $ obeys a generalized Palatini identity \eqref{def Palatini 1st order local}, while higher-order variations reduce to total covariant derivatives under the integral due to cyclicity (Eq.~\eqref{Palatini global all orders}).

The final derivation of the equation of motion \eqref{equation from action!} adds to the internal consistency of the deformed EH action and the applicability of classical variational techniques in the NC setting, while pointing to further work needed to fully understand the structure of higher-order corrections.

\subsection*{The Energy--Momentum Tensor}

With the deformed Einstein–Hilbert action at hand, the natural next step is to determine the corresponding deformed energy–momentum tensor (EMT) for a generic matter source.  A substantial body of work already addresses this question; we shall discuss them after reviewing the relevant general expression that follows from our present framework.

\paragraph{General derivation.}
Consider a generic deformed matter action on the non-commutative (NC) space–time
\bea\label{deformed matter action}
\overset{\star}{S}_M
 \;=\;\int d^4x\,\sqrt{-g_\star}\star\overset{\star}{\mathrm{L}}_{M},
\eea
where $\overset{\star}{\mathrm{L}}_M$ is a suitably constructed deformed Lagrangian density.  
Varying $\overset{\star}{S}_M$ with respect to $\invG^{\mu\nu}$, and using both the identity~\eqref{identity 2} and the cyclic property of the Abelian-twist $\star$-product, we obtain
\bea\label{variation of star matter}
\delta\overset{\star}{S}_M
   &=&\!\int \!\!\! d^4x\!\left[\,\delta(\sqrt{-g_\star})\star\overset{\star}{\mathrm{L}}_M
        +\sqrt{-g_\star}\star\delta\bigl(\overset{\star}{\mathrm{L}}_M\bigr)\right]
\nonumber\\
   &=&\!\int \!\!\! d^4x\!\left[-\tfrac12\sqrt{-g_\star}\!\star\! g_{\mu\nu}\!\star\!
        \delta(\invG^{\mu\nu})\!\star\!\overset{\star}{\mathrm{L}}_M
        +\sqrt{-g_\star}\!\star\!\delta\bigl(\overset{\star}{\mathrm{L}}_M\bigr)\right]
\nonumber\\
   &=&\!\int \!\!\! d^4x\,\sqrt{-g_\star}\!
        \left[\frac{\delta\overset{\star}{\mathrm{L}}_M}{\delta\invG^{\mu\nu}}
              -\tfrac12 g_{\mu\nu}\star\overset{\star}{\mathrm{L}}_M\right]
        \star\delta\invG^{\mu\nu}.
\eea
Comparing with~\eqref{equation from action!} yields the deformed EMT
\bea\label{derived EM general}
T^{\star}_{\mu\nu}
 \;=\; \frac{\delta\overset{\star}{\mathrm{L}}_M}{\delta\invG^{\mu\nu}}
       -\tfrac12 g_{\mu\nu}\star\overset{\star}{\mathrm{L}}_M.
\eea

\paragraph{Example: a deformed scalar field.}
For a $\star$-scalar field $\psi$, take (See, e.g.\ \cite{Gravity_Non})
\bea\label{star psi lagr}
\mathrm{L}_{\psi}^\star
   &=&\tfrac12(\partialST_\mu\psi)\star(\partialST^\mu\psi)
       -\tfrac{m^2}{2}\,\psi\star\psi                                   \\
   &=&\tfrac12\invG^{\mu\nu}(\partialST_\mu\psi)\star(\partialST_\nu\psi)
       -\tfrac{m^2}{2}\,\psi\star\psi.                                  \nonumber
\eea
Applying~\eqref{derived EM general} one finds, in the usual convention,
\bea\label{psi EM}
2\,\overset{\star}{T}_{\mu\nu}
   &=&(\partialST_\mu\psi)\!\!\star\!\!(\partialST_\nu\psi)-\tfrac12\,g_{\mu\nu}\!\star\!\left[(\partialST^\rho\!\!\psi)\!\star\!(\partialST_\rho\psi)
         -m^2\psi\!\star\!\psi\right].        \nonumber\\
\eea
This expression reproduces the familiar flat-space result (cf.\ \cite{EMT_1,EMT_4} taking $g\sim g_{flat}$), modified only by the appearance of the deformed metric.

\paragraph{Symmetrisation issue.}
As in the flat treatment of NC EMT, the tensor~\eqref{psi EM} is not manifestly symmetric.  One convenient symmetrised form is
\bea\label{symmetrization of EM}
2\,\overset{\star}{T}_{\mu\nu}\bigl|_{\text{SYM}}
  &=&\tfrac12\!\left[(\partialST_\mu\psi)\!\!\star\!\!(\partialST_\nu\psi)
                     +(\partialST_\nu\psi)\!\!\star\!\!(\partialST_\mu\psi)\right] 
  -\,g_{\mu\nu}\!\!\star\!\!\mathrm{L^\star_\psi}. 
\neweqline
\eea
Whether the straightforward generalization of local conservation,
$\nablaST^\mu \overset{\star}{T}_{\mu\nu}|_{\text{SYM}} = 0$,
remains valid, and how the associated Noether charges behave, are open questions that require a dedicated analysis; see the discussion below.

\paragraph{Conservation and Noether charges.}
Extensive studies of deformed EMTs in scalar $\phi^4$ and $U(1)$ gauge theories (e.g.\ \cite{EMT_1,EMT_2,EMT_3,EMT_4,EMT_Fundamental_scalar_Ch2_2})
employ the standard prescription: replace ordinary products in the Lagrangian by $\star$-products, then apply the Noether theorem to obtain $\hat{T}^{\mu\nu}$.  
The resulting tensors are symmetric and can be rendered traceless by a divergence-free improvement term; for $U(1)$, they are also gauge-covariant.  
In all cases, however, local conservation fails: $\partialST_\mu \overset{\star}{T}^{\mu\nu}\neq0$.

For NCSTs generated by Abelian twists, including the Moyal one and the $\kp$-deformation we employ, the integral of $\star$-products is cyclic.  Consequently, the spatial integral of $\overset{\star}{T}^{0\nu}$ is still conserved, ensuring global four-momentum conservation \cite{EMT_1,EMT_3,EMT_Fundamental_scalar_Ch2_2}.  This statement applies equally to the $\star$-product~\eqref{dependent: star product kappa} used in $\kp$-GR.

Alternative reformulations that restore local conservation do exist (see \cite{EMT_3,EMT_4}), but they generally do so at the cost of symmetry and tracelessness.  A systematic investigation of these issues---including the fate of the Bianchi identity and the explicit construction of Noether currents using the deformed symmetries of Sec.~\ref{subsec: consistency igl}---is left for future work.

\section{Summary \& Future Work}\label{section: Final Words}

In this work, we have constructed and analyzed a deformation of General Relativity to account for a local Non-Commutative Spacetime (NCST) of the $\kappa$-Minkowski type \eqref{kappa minkowski}. The motivation for introducing such an NCST arises from one of the central approaches in Quantum Gravity Phenomenology, as discussed in Sec.~\ref{section: Introduction}. Studying the physical implications of such a spacetime begins with investigating its \textit{flat sector} properties, its deformed symmetry transformations, the associated invariant quantities, and so on. A key insight in understanding such an NCST lies in a formulation of its relativistic structure, or the frame independence of physical laws. This initially originated doubt on its validity (as the relation \eqref{kappa minkowski} manifestly breaks Lorentz invariance), but eventually led to its formulation through symmetries of the $\kappa$-Poincar\'e group and to its physical realization via the notion of relative locality (RL), as outlined in Sec.~\ref{section: Introduction}.

Our construction was based on two main observations: First, there exists a well-motivated physical and phenomenological need to generalize the \textit{flat} NCST \eqref{kappa minkowski} to incorporate curvature induced by mass distribution; a deformed gravitational theory. Second, there now exists a solid and well-established formalism of \textit{twist deformations} in noncommutative geometry \cite{QuantumGroup_GaugeTheory, NC_Geometry2, NC_Geometry_Simplified, Gravity_Non}, which can be used to introduce gravity on a chosen local NCST. This formalism has already been applied to cases involving \textit{constant} NCSTs. Essentially, this is exactly what we have pursued here: we evaluated a gravitational theory on the local NCST \eqref{kappa minkowski} using the twist formalism, which we refer to as $\kappa$-GR. However, while this may appear to be a straightforward generalization of the constant NCST case, the actual situation turned out to be more subtle.

To employ the twist formalism in constructing geometry on an NCST, the twist must belong to the symmetry group of the underlying commutative space. Yet, as discussed in Secs.~\ref{section: Introduction} and \ref{section: Building Blocks For Construction}, there exists no twist element within the Poincar\'e group that can meaningfully deform spacetime into the $\kappa$-Minkowski structure. Consequently, we could not use the twist formalism to incorporate gravity with locally deformed Poincar\'e symmetries.

To overcome this obstacle, we followed approaches such as \cite{IGL_TWIST_1, IGL_TWIST_2}, which address the issue by enlarging the symmetry group to include dilatations. From this extended group, a twist can be constructed that deforms commutative spacetime into the $\kappa$-Minkowski form. Accordingly, in Sec.~\ref{section: Building Blocks For Construction}, we adopted the classical symmetry group of \textit{local} spacetime to be IGL(3,1) (with the minimal alternative being the $\mathcal{WP}$-group), and we reviewed key properties of the associated twist that yields \eqref{kappa minkowski}.

However, before proceeding with the gravitational construction, we had to reassess the physical viability of this framework. From the perspective of \textit{flat} spacetime, several questions needed to be considered: What is the structure of the deformed symmetry group? Does it possess the same degree of \textit{relativistic consistency} as the $\kappa$-Poincar\'e group? Can we still regard the NCST \eqref{kappa minkowski} as physically meaningful? These questions were the focus of Sec.~\ref{section: Consistency of The NC-Relations}, where, by following reasoning similar to previous studies on the $\kappa$-Poincar\'e case, we analyzed the scenario of twisted IGL(3,1) symmetry.

Specifically, we validated that the twisted symmetry, when applied to the NCST \eqref{kappa minkowski}, satisfies several consistency conditions: (1) the differential one-forms are well-defined and compatible with the deformed transformations; (2) the NCST \eqref{kappa minkowski} remains closed under the action of the deformed transformations, i.e., a coordinate chart acted upon by these transformations remains within the same NCST; and (3) the spacetime \eqref{kappa minkowski}, together with the deformed symmetry transformations, obeys a deformed Leibniz rule. To verify these points, we also derived explicit expressions for the full set of deformed transformations in the twisted IGL(3,1) group and their associated $\star$-coproducts.

The next concern we addressed was regarding the classical limit of the would-be constructed deformed GR with a \textit{local} twisted IGL(3,1) symmetry group. In general, a twist deformation reduces to the associated classical symmetry when the deformation parameter ($\lambda$) vanishes. Therefore, the \textit{form} of the equations is expected to reduce to classical GR. However, using IGL(3,1) as the pre-twisting symmetry implies a residual dilatation enlargement that persists even in the classical limit $\kp\rightarrow\infty$. This issue was discussed in Sec.~\ref{Contraction of the Symmetry}, and specifically in Sec.~\ref{ssection: IW-contraction our case}, where we proposed the Inönü–Wigner (IW) contraction as a suitable tool to deal with this residual symmetry.

The basic idea comes from analyzing 
the vector representation of the dilatation operator and its action on a free wave, which introduces a scaling dependence on the \textit{local} frequency of the wave \eqref{Dilatation action: free wave}. Inspecting the uncertainty relation \eqref{Uncertainty Relation}, we noted that it naturally suggests a dimensionless small parameter $\epsilon_{RL} = \frac{\omega_{\text{local}}}{\omega_p}$, capturing the physical deviation from classicality. A possible additional scaling factor of the form $\left(\frac{x}{l_{\text{sys}}}\right)^n$ may arise from RL considerations, but its form and origin are left for future work. The property of this parameter is that, based on RL, it approaches zero in the classical limit of low energies and is bounded by unity. This led us to identify the contraction parameter with $\alpha_{\text{fund}} = \omega_p = c/l_p$, analogous to the role of the speed of light in SR.

Given this, we utilized the IW contraction and performed a contraction of the dilatation enlargement controlled by the scale $\alpha_{\text{fund}}$ (see \eqref{Dilatation Rescaled} and \eqref{contraction Demonestrated}). This ensured that when the twist deformation is active, the usual deformation persists, and when it is turned off, the enlargement also vanishes. Technically, this involved rescaling the deformation parameter \textbf{in the twist} by $\alpha$ (see \eqref{rescaling lambda deformation}), noting that the relations \eqref{kappa minkowski} and especially \eqref{Uncertainty Relation} remain unaffected: they served as the motivation for introducing $\alpha_{\text{fund}}$ in the first place.

Having established a well-defined local deformed symmetry with a relativistic characterization of its flat sector, which smoothly reduces to local Poincar\'e symmetry in the classical limit $(\kappa, \alpha \to \infty)$, we proceeded to construct a gravitational theory based on this framework. This was the subject of Sec.~\ref{section: The Construction of Gravity}, where we evaluated the deformation of GR using the twist formalism on the NCST geometry with the constructed twisted symmetry as the local symmetry.

To derive a \textit{bona fide} theory of gravitation, we first constructed the deformed diffeomorphisms in terms of deformed Lie derivatives. Using standard techniques from NCST gravity, we then evaluated the explicit form of the deformed Einstein equations in vacuum (see \eqref{Expansion of Ricci tensor}, \eqref{vacuum EE deformed}--\eqref{result: deformed EE}) and discussed associated ambiguities related to reality, and the $\star$-product ordering.

However, our goal was to show that this construction is well-defined at the level of an \textit{action formulation}, which offers manifest symmetry properties. To that end, we introduced an ansatz for the deformed Einstein-Hilbert action \eqref{deformed Einstein Hilbert}, explicitly verified its global invariance under the deformed diffeomorphisms, and detailed the transformation rules for each term. We then applied the variational principle to this deformed action, generalizing classical identities to the NCST framework. This allowed us to explicitly show that the variation of \eqref{deformed Einstein Hilbert} yields the deformed field equations (see \eqref{equation from action!}).

While constructing this variational formalism, we identified a special role for the classical \textit{Palatini identity}, which is responsible for the local and global (covariant) conservation of the Einstein tensor in classical GR. In the NCST context, we verified a global generalization of this identity but stressed the difficulty in defining its local counterpart. This issue is not just computational and is related to the well-known ambiguities in \textit{local conservation laws} in noncommutative geometries. Finally,  we derive the general form of the deformed energy-momentum tensor using the developed deformed variational identities and discuss the related ambiguities in symmetrization and conservation.

\vspace{0.5em}
Collecting all the pieces, we have constructed a deformation of GR with a local $\kappa$-Minkowski NCST. The local symmetry was developed and analyzed with its deformed transformation structure. The theory was built to admit a proper classical limit to GR and equipped with deformed notions of general covariance (diffeomorphisms), a deformed variational principle and action formulation, and the associated treatment of conservation laws. A summary of some of the key results is presented in Table~\ref{summary-table}. We hope these developments will advance the understanding of deformed symmetries on $\kappa$-Minkowski spacetime and support the formulation of well-behaved deformations of gravity, leading to the possibility of computing explicit predictions in the future.
\vspace{0.5em}
Further directions for future research include:

\textbf{1.} Studying the new invariant quantities that arise from the derived deformed symmetries.
\textbf{2.} Formalizing the contraction parametrization in a more rigorous manner, with a concrete definition and physical role of the parameter $\eps_{RL}$.
\textbf{3.} Constructing explicit physical predictions, such as gravitational wave generation, through analyzing the deformation of the linearized gravitational theory.
\textbf{4.} Better understanding the challenges in formulating a local deformed Palatini identity, and the relation to ambiguities in defining local conservation laws in noncommutative spacetime geometries.

\section*{Acknowledgements}
We acknowledge support from the US-Israel Binational Science Fund (BSF) grant No. 2020245 and the Israel Science Fund (ISF) grant No. 1698/22.
We want to thank Eyal Subag, Yarden Shani, and Yogesh Dandekar for helpful discussions. We thank the anonymous referees for helpful criticism and suggestions. 

\appendix
\section{Detailed Calculations for Sec.\ref{section: Building Blocks For Construction}}\label{Appendix: Building Blocks}
Here, we shall detail the calculations beyond the results stated in Sec.\ref{section: Building Blocks For Construction}. 
\\ \, \\
\textbf{The action on basis 1-forms \eqref{action on 1forms}:}\\
Using the definition of a Lie derivative of a one-form along a vector field, we derive
\bea
\label{App. action on 1forms}
X_{0}(dx^{\mu})&=&\mathcal{L}_{X_{0}}(dx^{\mu})
=\left[X_{0}^{\mu}\partial_{\mu}(1)+\partial_{\nu}(X_{0}^{\mu})1\right]dx^{\nu}
\nonumber
\neweqline
&=&\left[0+\partial_{\nu}(1)\right]dx^{\nu}=0,
\\
X_{1}(dx^{\mu})&=&\mathcal{L}_{x^{j}\partial_{j}}(dx^{\mu})
=(\partial_{\mu}x^{j})dx^{\mu}=\delta_{\mu}^{j}dx^{\mu}\neq 0.
\nonumber
\eea
For the basis vectors \eqref{action on basis vectors assumed}, we use the Lie bracket,
\bea\label{App. action on basis vectors}
X_{0}(\partial_{\mu})&=& \left[1\partial_{\mu}(1)-1\partial_{\mu}(1)\right]\partial_{\nu}=0,  
\neweqline
X_{1}(\partial_{\mu})&=& -\partial_{\mu}(x^{j})\partial_{j}= -\delta^{j}_{\mu}\partial_{j}.
\eea
\textbf{The $\star$-pairing of the classical basis vectors and 1-forms \eqref{basis proof}:} \\
 
Explicitly calculating, we get, 
\bea
\langle\partial_{0}\commaST dx^{\nu}\rangle&=&\langle \bar{\phi}^{\{\alpha\}}(\partial_{0}),\bar{\phi}_{\{\alpha\}}(dx^{\nu})\rangle 
\nonumber \\
&=&\langle O^{\nu}_{\nu'}[\lambda,1] \partial_{0},dx^{\nu'}\rangle
=\delta_{0}^{\nu},~~~~~~~~~
\eea
\bea
\langle\partial_{j}\commaST dx^{\nu}\rangle&=&\langle \bar{\phi}^{\{\alpha\}}(\partial_{j}),\bar{\phi}_{\{\alpha\}}(dx^{\nu})\rangle
\nonumber \\
&=& \langle e^{i\frac{\lambda}{2}\partial_{0}}\delta_{j'}^{\nu}\partial_{j},dx^{j'}\rangle
\neweqline
&=&e^{i\frac{\lambda}{2}\partial_{0}}\delta_{j'}^{\nu}\langle\partial_{j},dx^{j'}\rangle=e^{i\frac{\lambda}{2}\partial_{0}}\delta_{j}^{j'}\delta_{j'}^{\nu},
\eea
\bea
\Rightarrow\langle\partial_{\mu}\commaST dx^{\nu}\rangle&=&\boxed{\delta_{\mu}^{0}\delta_{0}^{\nu}+\delta_{\mu}^{j}\delta_{j}^{\nu}e^{i\frac{\lambda}{2}\partial_{0}}}\ .
\label{App. basis proof}
\eea
 
\textbf{The non-commutativity of $\star$-1-forms:}\\
 
Since the relevant (non-zero) part of $\OprR$ here is 
\bea\label{App. Deriving 1-form non-commutativity_1}
\bar{\OprR}_{\text{relevant}} =\exp\left[-i\lambda(-X_{1}\otimes X_{0})\right], 
\eea
we use \eqref{partial twist action on classical derivatives} and \eqref{Deriving 1-form non-commutativity_1} to get
\bea
\bar{\text{R}}^{\{\alpha\}}(dx^{\mu})\star\bar{\text{R}}_{\{\alpha\}}(f) &=&dx^{\mu'}\star (e^{+i\lambda\partial_{0}}\delta^{j}_{\mu'}\delta_{j}^{\mu}+\delta^{0}_{\mu'}\delta^{\mu}_{0})(f)
\neweqline
&:=&
\boxed{dx^{\mu'}\star O^{\mu}_{\mu'}[\lambda,2](f)}~.
\eea

\textbf{The known calculation of the deformed Leibniz rule of $\star$-basis-vectors (see \cite{General_notion_of_Twist_2}):}\\
Calculating the left-hand side in the second line of \eqref{1-form differential},
\bea
        \mathbf{d}(f\star g) &=& (\partialST_{\mu}f)\star dx^{\mu}\star g + f\star(\partialST_{\mu}g)\star dx^{\mu} 
        \\
        &=&((\partialST_{\mu}f)\!\star\! O^{\mu}_{\mu'}[\lambda,-2 ](g))\!\star\! dx^{\mu'} + f\!\star\!(\partialST_{\mu}g)\!\star\! dx^{\mu},
        \nonumber
\eea
and equating with the right-hand side, we derive 
\bea
        (\partial_{j}^{\star}(f\star g))\star dx^{j} &=& ((\partial_{j}^{\star}f)\star e^{-i\lambda\partial_{0}}g+f\star(\partial_{j}^{\star}g))\star dx^{j}, \nonumber\\
        (\partial_{0}^{\star}(f\star g))\star dx^{0} &=& f\star(\partial_{0}^{\star}g)+(\partial_{0}^{\star}f)\star g .
\eea
 
concluding the deformed Leibniz rule for $\star$-derivatives,
\bea
\label{1-form differential Appendix}
        \partial_{j}^{\star}(f\star g) & =& (\partial_{j}^{\star}f)\star e^{-i\lambda\partial_{0}}g + f\star \partial_{j}^{\star} g, \nonumber\\
        \partial_{0}^{\star}(f\star g) & =& \partialST_{0}f\star g+ f\star\partial_{0}^{\star}g.
\eea
We have used,
\begin{equation}\label{dependent: side 2} \bar{\text{R}}^{\{\alpha\}}(h)\star\bar{\text{R}}_{\{\alpha\}}(dx^{\mu})=O^{\mu}_{\mu'}[\lambda,-2](h)\star dx^{\mu'}.
\end{equation}
 
\textbf{The new calculation of the Leibniz rule \\ of $\star$-basis-vector:} \\
 
First, we evaluate the twisted co-product of $\partialST_{\mu}$:
\bea\label{App. Leibniz rule new}
\Delta_{\Phi}(\partialST_{\mu})&=&\Phi\circ(\partialST_{\mu}\otimes 1+1\otimes \partialST_{\mu})\circ \Phi^{-1}
\neweqline
&=&\phi^{\{\alpha\}}(\partialST_{\mu})\bar{\phi}^{\{\beta\}}\otimes \phi_{\{\alpha\}}\bar{\phi}_{\{\beta\}}
\neweqline
&&+\phi^{\{\alpha\}}\bar{\phi}^{\{\beta\}}\otimes \phi_{\{\alpha\}}(\partialST_{\mu})\bar{\phi}_{\{\beta\}}
\\
&=& \partial_{\mu'}^{\star}\otimes O^{\mu'}_{\mu}[\lambda,-1 ]+O^{\mu'}_{\mu}[\lambda, 1]\otimes \partial_{\mu'}^{\star}\nonumber,
\eea
so that the $\star$-co-product is then,
\begin{widetext}
\bea\label{App. Star co-product} 
\Delta_{\star}(\partialST_{\mu})
&=&
(\phi^{\{\alpha\}}\phi_{\{\alpha\}}\otimes \phi^{\{\alpha'\}}\phi_{\{\alpha'\}})\circ \left[\partial_{\mu'}^{\star}\otimes O^{\mu'}_{\mu}[\lambda,-1 ]
+O^{\mu'}_{\mu}[\lambda,1 ]\otimes \partial_{\mu'}^{\star}\right]
\circ \bar{\phi}^{\{\beta\}}\bar{\phi}_{\{\beta\}} 
\nonumber\\
&=& \left[\phi^{\{\alpha\}}\phi_{\{\alpha\}}(\partial_{\mu'}^{\star})\otimes \phi^{\{\alpha'\}}\phi_{\{\alpha'\}}(O^{\mu'}_{\mu}[\lambda,-1 ])\right.
+\left.\phi^{\{\alpha\}}\phi_{\{\alpha\}}(O^{\mu'}_{\mu}[\lambda, 1])\otimes \phi^{\{\alpha'\}}\phi_{\{\alpha'\}}(\partial_{\mu'}^{\star})\right]
\circ \bar{\phi}^{\{\beta\}}\bar{\phi}_{\{\beta\}}
\nonumber\\
&=& \left[\phi_{\{\alpha\}}\partial_{\mu''}^{\star}\otimes \phi^{\{\alpha'\}}O^{\mu''}_{\mu}[\lambda, -2]\right. 
+\left.\phi_{\{\alpha\}}\otimes \phi^{\{\alpha'\}}\partialST_{\mu'}\right]\circ \bar{\phi}^{\{\beta\}}\bar{\phi}_{\{\beta\}}
= \partial_{\mu'}^{\star}\otimes O^{\mu'}_{\mu}[\lambda,-2 ]+1\otimes \partialST_{\mu},
\eea
\end{widetext}
This leads to the deformed Leibniz rule,
\bea\label{App. Leibniz rule final}
\partialST_{\mu}(f\star g)&=&\mu_{\star}\circ \{\Delta_{\star}(\partialST_{\mu})(f\otimes g) \}
\\
&=&\partialST_{\mu'}(f)\star O^{\mu'}_{\mu}[\lambda, -2](g)+f\star(\partialST_{\mu}g).\nonumber
\eea

\section{Calculations for Sec.\ref{subsec: consistency transl}}
We begin by evaluating:
\bea\label{appendix: epsilon non-commutativity}
(x^0)^n\star\epsilon^{j} &=& (x^0)^{n-1}\star (\epsilon^{j}\star x^0 - [\epsilon^{j}, x^0]_{\star})
\nonumber\\
&=& (x^0)^{n-1}\star \epsilon^{j}\star x^0 - A (x^0)^{n-1} \star\epsilon^{j} \nonumber\\
&=& (x^0)^{n-1} \star\epsilon^{j}\star (x^0 - A)
\nonumber\\
&=& (x^0)^{n-2}\star (\epsilon^{j} \star x^0 - [\epsilon^{j}, x^0]_{\star}) \star(x^0 - A)
\nonumber\\
&=& (x^0)^{n-2}\star \epsilon^{j} \star(x^0 - A)^2 ~,
\eea
hence
\bea
(x^0)^n \star\epsilon^{j}|_{\{\beta \in \mathbf{R}|\ \beta \leq n\}}  &=& (x^0)^{n-\beta}\star \epsilon^{j}\star (x^0 - A)^{\beta}|_{\beta=n} 
\nonumber\\
&=& \boxed{\epsilon^{j}\star (x^0 - A)^n}.
\eea
This enables us to compute:
\bea
[\epsilon^{j}, (x^0)^n]_{\star} &=& \epsilon^{j}\star(x^0)^n - (x^0)^n \star\epsilon^{j}
\nonumber\\
&=&\epsilon^{j} \star((x^0)^n - (x^0 - A)^n)~,
\eea
which can be used to derive our end goal \eqref{exponential commutator result}:
\bea 
[\epsilon^{j}, e^{-ik_{0}x^0}]_{\star} &=& \sum_{n=0}^{\infty} \frac{1}{n!} (-ik_{0})^{n} \left[\epsilon^{j}, (x^0)^n\right]_{\star
} 
\nonumber\\
&=& \epsilon^{j} \star
    \left(e^{-ik_{0}x^0} - e^{-ik_{0}(x^0 - A)}\right)
\nonumber\\
&=& \epsilon^{j}\star e^{-ik_{0}x^0} \left(1-e^{-\lambda k_{0}}\right).
\eea

\section{Calculations for Sec.\ref{subsec: consistency igl}}
\textbf{The $\star$-Lie derivative:}\\ For a general vector field $\xi\in\Xi_{\star}$ we have 
\bea
\label{appendix: Lie derivative non non constant}
\lie^{\star}_{\xi}
&=&
\lie_{\bphi^{\{\alpha\}}(\xi)\bphi_{\{\alpha\}}}\equiv\bphi^{\{\alpha\}}(\xi)\bphi_{\{\alpha\}}
\\
&=&
\mu\!\circ\!\left(\exp\left[\frac{i\lambda}{2}\left(\partial_{0}\!\otimes\! x^j\partial_{j}-x^j\partial_{j}\!\otimes\!\partial_{0}\right)\right]
\left(\xi^{\mu}\partialST_{\mu}\!\otimes\!1\right)\right)
\nonumber \\
&=&\mu\!\circ\!\left[\exp\left(\frac{i\lambda}{2}\partial_{0}\!\otimes\! x^j\partial_{j}\right)\exp\left(-\frac{i\lambda}{2}x^j\partial_{j}\!\otimes\!\partial_{0}\right)\left(\xi^{\mu}\partialST_{\mu}\!\otimes\!1\right)\right]
\nonumber \\
&=&
\!\!\sum_{\substack{n=0\\n'=0}}^{\infty}\!\!\!
\frac{\left(i\lambda/2\right)^{n+n'}}{n!n'!}
\left(-x^j\partial_{j}\right)^{n'}\left[\partial_{0}^{n}\left(\xi^{\mu}\right)\partialST_{\mu}\right]\partial_{0}^{n'}\left(x^{j}\partial_{j}\right)^{n}.\nonumber
\eea
We used that $[x^j\partial_{j},\partial_{0}]=0$ and that $\partial_{0}(\partial_{\mu})=0$. 
To get a taste for the first few orders, first note that 
\bea 
\partial_{0}(\xi)=\lie_{\partial_{0}}(\xi)
&=&
(\partial_{0}(\xi^{\nu})-\xi^{\mu}\partial_{\mu}(1))\partial_{\nu}=\partial_{0}(\xi^{\mu})\partial_{\mu},\quad
\nonumber \\
x^j\partial_{j}(\xi)
&=&
[x^j\partial_{j},\xi]=(x^j\partial_{j}(\xi^{\mu})-\xi^{\mu}\delta_{j}^{\mu})\partial_{\mu}. 
\eea
 
Then, up to the second order, we can write \eqref{appendix: Lie derivative non non constant} explicitly;
\bea
\label{appendix: Lie derivative non non constant expansion}
\lie^{\star}_{\xi}
&=&
\xi^{\mu}\partialST_{\mu}
+\frac{i\lambda}{2}\left[x^j\left(\partial_0(\xi^\mu)\partial_j-\partial_j(\xi^\mu)\partial_0\right)+\xi^\mu\delta_j^\mu\partial_0\right]\partialST_{\mu}\nonumber
\\
&&+(\frac{i\lambda}{2})^2x^j\left[(\partial_0(\xi^\mu)-x^j\partial_j\partial_0(\xi^\mu))\partial_j-\partial_j(\xi^\mu)\partial_0\delta_j^\mu \right.\nonumber
\\
&&\left. +1/2\left(\partial_{j}(\xi^\mu)+x^j\partial_j^2(\xi^\mu)\right)\partial_0\right]\partialST_\mu\partial_0\nonumber
\\
&&+\frac{1}{2}(\frac{i\lambda}{2})^2\xi^\mu\delta^{j}_{\mu}\delta_j^{\mu'}\partialST_{\mu'}\partial_0^2. 
\eea
Note that if $\xi^{\mu}$ is constant, we can calculate the entire sum,
\bea
\lie^{\star}_{\xi}&=&\xi+\frac{i\lambda}{2}\xi^{\mu}\partial_{\mu}\delta^{\mu}_{j}\partial_{0}
+\frac{1}{2}\left[\left(\frac{i\lambda}{2}\right)^2\xi^{\mu}\partialST_{\mu'}\right]\delta_{j}^{\mu'}\delta_{\mu}^{j}\partial_{0}^2
+\cdots
\neweqline
&=&
e^{+(i\lambda/2)\partial_{0}\delta^{\mu}_{j}}\xi.
\eea
\\

\section{Detailed Calculations for Sec.\ref{section: The Construction of Gravity}}\label{Appendix. the construction}
Here, we shall detail the calculations beyond the results stated in Sec.\ref{section: The Construction of Gravity}.\\
 \, \\ 
 
\textbf{Transformation properties of $\delta^\star_\xi(\cdot)$:}
Considering the transformation of scalar multiplication $\lie^\star(f\star g)$, we can use \eqref{deformed Leibniz rule} and derive the following first non trivial order (in $\lambda$: the zeroth order trivially equals the classical answer, which we know to produce a scalar quantity):
\begin{widetext}
\bea\label{appendix: scalar multiplied trans}
\lie^\star_\xi(f\star g) &=& \bar{\phi}^{\{\alpha\}}(\lie^\star_\xi(h))\bar{\phi}_{\{\alpha\}}(g) 
+\bar{\phi}^{\{\beta\}}(\bar{R}^{\{\alpha\}}(h))\bar{\phi}_{\{\beta\}}(\lie^\star_{\bar{R}_{\{\alpha\}}(\xi)}(g))
\neweqline
&=&\mathcal{O}(\lambda^0,\lambda^2) 
+\bphi^{1}(\lie^{\star\ (0)}(h))\bphi_{1}(g)
+\bphi^{0}(\lie_\xi^{\star\ (1)}(h))\bphi_{0}(g)+\bphi^{1}(\bar{R}^{0}(h))\bphi_{1}(\lie^{\star\ (0)}_{\bar{R}_{0}(\xi)} (g))
\neweqline
&&+\bphi^{0}(\bar{R}^{1}(h))\bphi_{0}(\lie^{\star\ (0)}_{\bar{R}_{1}(\xi)} (g))
+\bphi^{0}(\bar{R}^{0}(h))\bphi_{0}(\lie^{\star\ (1)}_{\bar{R}_{0}(\xi)} (g))
\neweqline
&=&\bphi^{1}(\xi^\mu\partial_\mu(h))\bphi_{1}(g)+\bphi^{1}(h)\bphi_{1}(\xi^\mu\partial_\mu(g))+\lie^{\star\ (1)}_{\xi^\mu\partial_\mu}(h)g  
+h\lie^{\star\ (1)}_{\xi^\mu\partial_\mu}(g)+\bar{R}^{1}(h)(\bar{R}_{1}(\xi^\mu\partial_\mu)g),
\eea
where $\lie^{\star \ (1)}(\cdot)$ is defined as the first order in $\lambda$ of \eqref{Lie derivative non non constant}. Indeed, from the last equality 
it is evident that the transformation of the product $f\star g$ just transforms as another scalar.
\end{widetext}
\textbf{The $\star$-Covariant Derivative:}\\ 
Directly calculating, we get for the general vector fields case \eqref{dependent: calculation of covariant derivative General}, 
\bea\label{App. dependent: calculation of covariant derivative General}
\nablaST_{z}u
&=&\mathcal{L}^{\star}_{\zST^{\mu}\star\partialST_{\mu}}(\uST^{\nu})\star\partialST_{\nu}+\bar{\text{R}}^{\{\alpha\}}(\uST^{\nu})\star\nablaST_{\bar{\text{R}}_{\{\alpha\}}(\zST^{\mu}\star\partialST_{\mu})}(\partialST_{\nu})
\neweqline
&=&\mathcal{L}^{\star}_{z}(\uST^{\nu})\star\partialST_{\nu} + \bar{\text{R}}^{\{\alpha\}}(\uST^\nu)\star \bar{R}_{\{\alpha\}}(z) \star  \Gamma _{\mu\nu}^\sigma\partialST_\sigma
\neweqline
&=&
\left[\lie^{\star}_{z}(\uST^{\nu})+\bar{\text{R}}^{\{\alpha\}}(\uST^{\sigma})\star\bar{\text{R}}_{\{\alpha\}}^{\ \ \mu'}(z)\star\Gamma_{\mu'\sigma}^{\nu}\right]\star\partialST_{\nu} 
\eea
 
This is the farthest way we can simplify the expression. From here one need to use the $\star$-Lie derivative and the composite action of $\bar{\text{R}}^{\{\alpha\}}$ on $z$. If $z=\partialST_{\mu}$ we can get 
\bea\label{App. dependent: calculation of covariant derivative}
\nablaST_{\partialST_{\mu}}(u)&=&
[\partial_{\mu}(\uST^{\nu})+O^{\mu'}_{\mu}[\lambda', 2](\uST^{\sigma})\star\Gamma_{\mu'\sigma}^{\nu}]\star\partialST_{\nu} 
\neweqline
&=&
[\partial_{\mu}(u^{\nu})+O^{\mu'}_{\mu}[\lambda', 2](u^{\sigma})\star\Gamma_{\mu'\sigma}^{\nu}]\partialST_{\nu}.
\eea
 
Moreover, for the case when we are only interested in the  $u=u^{\nu}$ part, it is immediate to derive \eqref{dependent: calculation of covariant derivative} from \eqref{App. dependent: calculation of covariant derivative General}. 
\\
\, \\ 
\textbf{The $\star$-curvature tensor:}\\ 
The result \eqref{curvature kappa} is derived through the following calculation,
\bea\label{curvature kappa1}
\textgoth{R}(\partialST_{\mu},\partialST_{\nu},\partialST_{\rho})
    &=&
    \nablaST_{\partialST_{\mu}}(\nablaST_{\partialST_{\nu}}\partialST_{\rho})
    -\nablaST_{[\partialST_{\mu}\commaST\partialST_{\nu}]}(\partialST_{\rho})
    \nonumber\\
    &&
    -\nablaST_{\bar{\text{R}}^{\{\alpha\}}(\partialST_{\nu})}(\nablaST_{\bar{\text{R}}_{\{\alpha\}}(\partialST_{\mu})}\partialST_{\rho})
    \nonumber\\
    && \!\!\!\!\!\!\!\!\!\!\!\!\!\!\!\!\!\!\!\!\!\!\!\!= \nablaST_{\partialST_{\mu}}
    \!(\Gamma_{\nu\rho}^{\sigma}\!\star\!\partialST_{\sigma})
    -\bar{\text{R}}^{\{\alpha\}\nu'}_{\ \ \nu}\!\star\!\nablaST_{\partialST_{\nu'}
    }\!(\bar{\text{R}}^{ \ \mu'}_{\{\alpha\} \mu}\!\star\!\nablaST_{\partialST_{\mu'}}(\partialST_{\rho})) \neweqline
    && \!\!\!\!\!\!\!\!\!\!\!\!\!\!\!\!\!\!\!\!\!\!\!\!=
    \nablaST_{\partialST_{\mu}}
    \!(\Gamma_{\nu\rho}^{\sigma}\!\star\!\partialST_{\sigma})
    -\bar{\text{R}}^{\{\alpha\}\nu'}_{\ \ \nu}\!\star\!\nablaST_{\partialST_{\nu'}}
    \!(\bar{\text{R}}^{ \ \mu'}_{\{\alpha\} \mu}\!\star\!\Gamma_{\mu'\rho}^{\sigma}\!\star\!\partialST_{\sigma})
    \nonumber\\
    :&=&(\mathbf{C})+(\mathbf{D})~,
\eea
where we have used $[\partialST_{\mu}\commaST\partialST_{\nu}]=0$
to go from the first line to the second. We calculate the first term as follows:
\bea
\label{curvature first term}
    \mathbf{C}&:=&\nablaST_{\partialST_{\mu}}
    \!\!(\Gamma_{\nu\rho}^{\sigma}\!\star\!\partialST_{\sigma})
    \nonumber\\
    &=& 
    \mathcal{L}^{\star}_{\partialST_{\mu}}
    \!\!(\Gamma^{\sigma}_{\nu\rho})\star\partialST_{\sigma}+\bar{\text{R}}^{\{\alpha\}}(\Gamma^{\sigma}_{\nu\rho})\star\nablaST_{\bar{\text{R}}_{\{\alpha\}}(\partial_{\mu}^{\star
})}(\partialST_{\sigma}) 
    \nonumber\\
    &=& 
     \!\!\!\!\!\!\!\!\!\!\!\!\!\!\!\!\!\!\!\!\!\!\!\!
    [\partial_{\mu}(\Gamma^{\sigma}_{\nu\rho})+\bar{\text{R}}^{\{\alpha\}}(\Gamma^{\tau}_{\nu\rho})\star\bar{\text{R}}_{\{\alpha\} \mu}^{\mu'}\star\Gamma^{\sigma}_{\mu'\tau}]\star\partialST_{\sigma}\, ,~~~~
\eea
and the second term as
\bea\label{curvature second term}
\mathbf{D}&:=&
\bar{\text{R}}^{\{\alpha\}\nu'}_{\ \ \nu}\star\nablaST_{\partialST_{\nu'}}(\bar{\text{R}}^{ \ \mu'}_{\{\alpha\}\ \ \mu}\star\Gamma_{\mu'\rho}^{\sigma}\star\partialST_{\sigma})
\\
&=&\bar{\text{R}}^{\{\alpha\}\nu'}_{\ \ \nu}\star(\bar{\text{R}}^{\{\beta\}}(\bar{\text{R}}^{\mu'}_{\{\alpha\} \ \mu})\star\nablaST_{\bar{\text{R}}_{\{\beta\}}(\partialST_{\nu})}(\Gamma_{\mu'\rho}^{\sigma}\star\partialST_{\sigma}))
\neweqline
&=&\bar{\text{R}}^{\{\alpha\}\nu'}_{\ \ \nu}\!\!\star\![(\bar{\text{R}}^{\{\beta\}}(\bar{\text{R}}^{\mu'}_{\{\alpha\} \mu})\!\star\!\bar{\text{R}}_{\{\beta\} \ \nu'}^{\nu''}\star
\nablaST_{\partial_{\nu''}^{\star}}(\Gamma_{\mu'\rho}^{\sigma}\!\star\!\partialST_{\sigma}))].
\nonumber
\eea
We also observe that,
\bea\label{curvature simplifying relations}
    \bar{\text{R}}^{\{\beta\}}(\bar{\text{R}}_{\{\alpha\}\ \ \mu}^{\ \mu'})\star\bar{\text{R}}_{\{\beta\}\ \ \nu'}^{\ \nu''}
&=&
    \bar{\text{R}}_{\{\alpha\} \ \mu}^{\ \mu'}\delta_{\nu'}^{\nu''} ~~ ,\\
    \bar{\text{R}}^{\{\alpha\}\ \nu'}_{\ \ \nu}\star\bar{\text{R}}_{\{\alpha\}\ \ \mu}^{\ \mu'}
&=&
    \delta_{\mu}^{\mu'}\delta_{\nu}^{\nu'}.
\eea
thus, we can simplify \eqref{curvature second term}:
\bea\label{curature second term simplified}
\bar{\text{R}}^{\{\alpha\}\nu'}_{\ \ \nu}\!&&\star\nablaST_{\partialST_{\nu'}}(\bar{\text{R}}^{ \ \mu'}_{\{\alpha\}\ \mu}\star\Gamma_{\mu'\rho}^{\sigma}\star\partialST_{\sigma})
    =
    \nablaST_{\partialST_{\nu}}(\Gamma_{\mu\rho}^{\sigma}\!\star\!\partialST_{\sigma})
    \nonumber \\
&&
\!\!\!\!
\!\!\!\!
    = [\partial_{\nu}(\Gamma^{\sigma}_{\mu\rho})+\bar{\text{R}}^{\{\alpha\}}(\Gamma^{\tau}_{\mu\rho})\star\bar{\text{R}}_{\{\alpha\} \nu}^{\ \nu'}\star\Gamma^{\sigma}_{\nu'\tau}]\star\partialST_{\sigma}.~~~~~~~
\eea
Finally, \eqref{curvature kappa1} can be now written directly in its components using \eqref{coefficient frame curvature} and the two terms (\eqref{curvature first term},\eqref{curvature second term}),
\bea
\textgoth{R}_{\mu\nu\rho}^{\sigma}&=&[\partial_{\mu}(\Gamma^{\sigma}_{\nu\rho})+\bar{\text{R}}^{\{\alpha\}}(\Gamma^{\tau}_{\nu\rho})\star\bar{\text{R}}_{\{\alpha\}\ \mu}^{\mu'}\star\Gamma^{\sigma}_{\mu'\tau}]
\neweqline
&&
    -[\partial_{\nu}(\Gamma^{\sigma}_{\mu\rho})+\bar{\text{R}}^{\{\alpha\}}(\Gamma^{\tau}_{\mu\rho})\star\bar{\text{R}}_{\{\alpha\}\ \nu}^{\nu'}\star\Gamma^{\sigma}_{\nu'\tau}]
\neweqline
&=&
\boxed{\partial_{[\mu}\Gamma_{\nu]\rho}^{\sigma}
+O^{\gamma}_{[\mu}[\lambda',2](\Gamma_{\nu]\rho}^{\tau})\star\Gamma_{\gamma\tau}^{\sigma}}\ ,
\eea
where we have used the by-now familiar result,
\beq 
\bar{\text{R}}^{\{\alpha\}}(\Gamma^{\tau}_{\nu\rho})\star\bar{\text{R}}_{\{\alpha\}\ \mu}^{\mu'}= O^{\mu'}_{\mu}[\lambda', 2] 
(\Gamma_{\nu\rho}^{\tau}).\nonumber
\eeq
\textbf{The $\star$-Ricci tensor:}\\
Equation \eqref{Ricci tensor kappa} is derived through, 
\bea\label{Ricci tensor kappaA}
\textgoth{R}_{\nu \rho } &=& \langle dx^{\mu}\commaST \textgoth{R}^{\sigma}_{\mu \nu \rho}\star\partialST_{\sigma}\rangle'
\neweqline
&=&\langle dx^{\mu}\commaST \bar{\text{R}}^{\{\alpha\}}(\partialST_{\sigma})\star\bar{\text{R}}_{\{\alpha\}}(\textgoth{R}^{\sigma}_{\mu\nu\rho})\rangle'\ ,
\eea
noting that, 
\bea
\bar{\text{R}}^{\{\alpha\}}(\partialST_{\sigma})\!\star\!\bar{\text{R}}_{\{\alpha\}}(\textgoth{R}^{\sigma}_{\mu\nu\rho})
&=&
\partial_{\sigma'}^{\star}\!\star\!(e^{-i\lambda'\partial_{0}}\delta_{\sigma}^{j}\delta_{j}^{\sigma'}+\delta_{\sigma}^{0}\delta_{0}^{\sigma'})\textgoth{R}_{\mu\nu\rho}^{\sigma}
\neweqline
&=&
\partial_{\sigma'}^{\star}\!\star\! O^{\sigma'}_{\sigma}[\lambda', -2](\textgoth{R}^{\sigma}_{\mu\nu\rho}),
\eea\\
and using this in \eqref{Ricci tensor kappaA}, we get
\bea
\textgoth{R}_{\nu \rho}&=&\langle dx^{\mu}\commaST \partial_{\sigma'}^{\star}\star O^{\sigma'}_{\sigma}[\lambda', -2]\textgoth{R}_{\mu\nu\rho}^{\sigma}\rangle'
\neweqline
&=&\delta_{\sigma'}^{\mu}\star O^{\sigma'}_{\sigma}[\lambda',-2 ] (\textgoth{R}_{\mu\nu\rho}^{\sigma})=O^{\mu}_{\sigma}[\lambda',-2] (\textgoth{R}_{\mu\nu\rho}^{\sigma})
\neweqline
&=&\boxed{e^{-i\lambda'\partial_{0}}\textgoth{R}_{j\nu\rho}^{j}+\textgoth{R}_{0\nu\rho}^{0}}~.
\eea

\textbf{The $\star$-metric-compatibility:}\\ 
Equation \eqref{dependent: covariant derivative of the metric} is derived as follows,
\bea
\label{App. dependent: covariant derivative of the metric}
\nablaST_{\gamma}(g)&=&\nablaST_{\partial_{\gamma}^{\star}}((dx^{\mu}\otimesST dx^{\nu})\star g_{\mu\nu})
    \neweqline
    &=&[\nablaST_{\partial_{\gamma}^{\star}}(dx^{\mu}\otimesST dx^{\nu})\star g_{\mu\nu}
    \neweqline
    &&+\bar{\text{R}}^{\{\alpha\}}(dx^{\mu}\otimesST dx^{\nu})\star\nablaST_{\bar{\text{R}}_{\{\alpha\}}(\partial_{\gamma}^{\star})}(g_{\mu\nu})]\neweqline
    &=&[(\nablaST_{\partial_{\gamma}^{\star}}(dx^{\mu})\otimesST dx^{\nu})\star g_{\mu\nu}
    \neweqline
    &&+(\bar{\text{R}}^{\{\alpha\}}\star\bar{\text{R}}_{\{\alpha\}\ \gamma}^{\ \gamma'}\otimesST \nablaST_{\partial_{\gamma'}^{\star}}(dx^{\nu}))\star g_{\mu\nu}
    \neweqline
    &&+\bar{\text{R}}^{\{\alpha\}}(dx^{\mu}\otimesST dx^{\nu})\star \bar{\text{R}}_{\{\alpha\}\ \gamma}^{\ \gamma'}\star\nablaST_{\partial_{\gamma'}^{\star}}(g_{\mu\nu})]
    \nonumber   \\
    &=&[-\Gamma_{\gamma(\mu}^{\sigma}\star g_{\nu)\sigma}+g_{\mu\nu,\gamma}\star(dx^{\mu}\otimesST dx^{\nu})]. 
\eea
Going from the fourth to the fifth line, we used the relations 
\bea\label{dependent: metric derivative aside}
\bar{\text{R}}^{\{\alpha\}}(dx^{\mu}\otimesST dx^{\nu})\star \bar{\text{R}}_{\{\alpha\}\ \gamma}^{\ \gamma'} &=& 
O^{\gamma'}_{\gamma}[\lambda', 2] (dx^{\mu}\otimesST dx^{\nu})
\neweqline
&=&dx^{\mu}\otimesST dx^{\nu},
\eea
and
\bea
\bar{\text{R}}^{\{\alpha\}}(dx^{\mu})\star\bar{\text{R}}_{\{\alpha\}\ \gamma}^{\ \gamma'} =O^{\gamma'}_{\gamma}[\lambda',2] (dx^{\mu})
=dx^{\mu}.
\eea

\bibliographystyle{apsrev4-1}

\bibliography{PaperIV3Arxiv}

\clearpage


\begin{sidewaystable} 
\caption{Your caption.}
\label{summary-table}
\, \\ \, \\ \, \\ \, \\ \, \\ \, \\ \, \\ \, \\ \, \\ \, \\ \, \\ \, \\ \, \\ \, \\ \, \\ \, \\
TABLE I. Differential Geometry \& Gravity Theories in Commutative, Constantly Non-Commutative, and
$\kappa$-Minkowski Non-Commutative Spacetimes \, \\ \, \\
\begin{tabularx}{\textwidth}{ |p{3.7cm}||p{4.3cm}|p{7cm}|X| }
\hline
\rule{0pt}{2.5ex} 
Quantity & Commutative Spacetime & Constant Non-Commutativity & $\kappa$-Minkowski Non-Commutativity: results the $\kappa$-GR \\
\hline
\rule{0pt}{4ex} The Spacetime Metric & 
$g_{\mu\nu}$ & 
$g_{\mu\nu} $\cite{NC_Geometry_Simplified} &
$g_{\mu\nu}$ \\
\hline
\rule{0pt}{4ex} Coor. Commutators &
$[x^\mu,x^\nu]=0$ & $[\hat{x}^{\mu},\hat{x}^{\nu}]=i\lambda\theta^{\mu\nu}$ \cite{Gravity_Non}& $[\hat{x}^\mu,\hat{x}^\nu]=il_p\hat{x}^{j}\delta_{[0}^{\mu}\delta_{j]}^{\nu}$ \\
\hline
\rule{0pt}{4ex} The product $f\star g$ &
$fg$ & 
$\begin{array}{l}
fg + \frac{i}{2}\lambda\theta^{\mu \nu } (\partial_{\mu}f) (\partial_{\nu}g)\\
~~~+\frac{1}{8}(i\lambda)^2\theta^{\mu\nu}\theta^{\rho\sigma}\partial_\mu\partial_\rho(f)\partial_\nu\partial_\sigma(g)
\end{array}$
\cite{Gravity_Non}&  
$\begin{array}{l}
fg +i\frac{\lambda'}{2}x^{j}f_{,[0} g_{,j]} \\
~~~+\frac{1}{8}(i\lambda')^2x^jx^l\left(f_{,0[0}\cdot g_{,l]j} + f_{,j[l}\cdot g_{,0]0}\right)
\end{array}$ \\
\hline
\rule{0pt}{4ex} Basis Vectors &
$\partial_{\mu}$ & 
$\partial_{\mu}$ \cite{Twist_general_2}& $\partialST_{\mu}=e^{-i\frac{\lambda'}{2}\partial_0\delta_\mu^j\delta_j^{\mu'}}\partial_{\mu'}$ \\
\hline
\rule{0pt}{4ex} Lie Derivative & $\lie_\xi\left(\cdot\right)$ & 
$\sum_{n=0}^{\infty}\frac{1}{n!}(\frac{i\lambda}{2})^{n}\prod_{j=1}^n \theta^{\mu_{j}\nu_{j}} \frac{\partial^n\left(\xi^{\gamma}\right)}{\partial^{\mu_{1}}...\partial^{\mu_{n}}}\frac{\partial^n \left(\partial_{\gamma}(\cdot)\right)}{\partial^{\nu_{1}}...\partial^{\nu_{n}}}$ \cite{Twist_general_2}& 
$\sum_{\substack{n=0\\n'=0\\ \,}}^{\infty} 
\frac{\left(i\lambda'/2\right)^{n+n'}}{n!n'!} [(-x^j\partial_{j})^{n'}[\partial_{0}^{n}(\xi^{\mu})] \partialST_{\mu}\partial_{0}^{n'}(x^{j}\partial_{j})^{n}](\cdot)$ \\
\hline
\rule{0pt}{4ex} Leibniz Rule & 
$\partial_\mu(fg)=\partial_\mu(f)g+f\partial(g)$& 
$\partial_\mu(f\star g)=\partial_{\mu}(f)\star g+f\star\partial_{\mu}(g)$ \cite{Gravity_Non}& 
$\partial_{\mu}(f\star g)=\partial_{\mu'}(f)\star O^{\mu'}_{\mu}[\lambda',-2](g)+f\star(\partial_{\mu}g)$ \\
\hline
\rule{0pt}{4ex} GL(3,1) Symmetry Trans' & 
$\lie_{x^\mu\partial_\nu}=x^\mu\partial_\nu$& 
-----& 
$\lie^\star_{x_\nu\partialST_\mu}=x_\nu\partial_\mu+\frac{i\lambda}{2}x^j\partial_\mu\delta_{\nu[0,}\partial_{j]}$ \\
\hline
\rule{0pt}{4ex} Metric Inverse & 
$g^{\mu\nu}$ & 
$\invG^{\mu\nu}=2g^{\mu\nu}-g^{\mu\alpha}\star g_{\alpha\beta}\star g^{\beta\nu}$ \cite{Gravity_Non}&   
$\invG^{\mu\nu}=2g^{\mu\nu}-g^{\mu\alpha}\star g_{\alpha\beta}\star g^{\beta\nu}$ \cite{Gravity_Non} \\
\hline
\rule{0pt}{4ex} Metric Determinant & 
$\frac{1}{4!}\epsilon^{\mu_1\ldots\mu_4\nu_1\ldots\nu_4}g_{\mu_1\nu_1}\ldots g_{\mu_4\nu_4}$ & 
$\frac{1}{4!}\epsilon^{\mu_1\ldots\mu_4\nu_1\ldots\nu_4}g_{\mu_1\nu_1}\star g_{\mu_2\nu_2}\star g_{\mu_3\nu_3}\star g_{\mu_4\nu_4}$ &   
$\frac{1}{4!}\epsilon^{\mu_1\ldots\mu_4\nu_1\ldots\nu_4}g_{\mu_1\nu_1}\star g_{\mu_2\nu_2}\star g_{\mu_3\nu_3}\star g_{\mu_4\nu_4}$  \\
\hline
\rule{0pt}{4ex} Cristoffel & 
$\Gamma^{\sigma}_{\nu \gamma}=g^{ \sigma\mu}\frac{1}{2}[g_{\mu (\nu,\gamma)}-g_{\nu \gamma,\mu}]$ & 
$\Gamma^{\sigma}_{\nu \gamma}=\invG^{ \sigma\mu}\star\frac{1}{2}[g_{\mu (\nu,\gamma)}-g_{\nu \gamma,\mu}]$\cite{Gravity_Non} & 
$\Gamma^{\sigma}_{\nu \gamma}=\invG^{ \sigma\mu}\star\frac{1}{2}[g_{\mu (\nu,\gamma)}-g_{\nu \gamma,\mu}]$ \\
\hline
\rule{0pt}{4ex} Covariant Derivative & 
$\nabla_{\mu}u^{\nu} = \partial_{\mu}u^{\nu}+u^{\sigma}\Gamma_{\sigma\mu}^{\nu}$  & 
$\nablaST_{\mu}u^{\nu} = \partial_{\mu}u^{\nu}+u^{\sigma}\star\Gamma_{\sigma\mu}^{\nu}$ \cite{Twist_general_2}& 
$\nablaST_{\partialST_{\mu}}(u^{\nu}) = \partial_{\mu}(u^{\nu})+O^{\mu'}_{\mu}[\lambda', 2](u^{\sigma})\star\Gamma_{\mu'\sigma}^{\nu}$ \\
\hline
\rule{0pt}{4ex} Riemann Tensor & 
$\textgoth{R}_{\mu \nu \rho}^{\sigma} =\partial_{[\mu}\Gamma_{\nu]\rho}^{\sigma} +\Gamma_{\rho [\nu }^{\tau}\Gamma_{\mu] \tau}^{\sigma}$ & 
$\textgoth{R}_{\mu \nu \rho}^{\sigma} =\partial_{[\mu}\Gamma_{\nu]\rho}^{\sigma} +\Gamma_{\rho [\nu }^{\tau}\star\Gamma_{\mu] \tau}^{\sigma}$\cite{Twist_general_2} & 
$\textgoth{R}_{\mu\nu\rho}^{\sigma} =\partial_{[\mu}\Gamma_{\nu]\rho}^{\sigma} +O^{\gamma}_{[\mu}[\lambda',2](\Gamma_{\nu]\rho}^{\tau})\star\Gamma_{\gamma\tau}^{\sigma}$ \\
\hline
\rule{0pt}{4ex} Ricci Tensor & 
$\textgoth{R}_{\nu \rho }=\textgoth{R}_{\mu \nu \rho}^{\mu}$ & 
$\textgoth{R}_{\nu \rho }=\textgoth{R}_{\mu \nu \rho}^{\mu}$ \cite{Twist_general_2}& 
$\textgoth{R}_{\nu \rho }=e^{-i\lambda'\partial_{0}}\textgoth{R}_{j\nu\rho}^{j}+\textgoth{R}_{0\nu\rho}^{0}$  \\
\hline
\rule{0pt}{4ex} Ricci Scalar & 
$\textgoth{R}=g^{ \mu\nu} \textgoth{R}^{\sigma}_{\sigma\mu\nu}$  & 
$\textgoth{R}=\invG^{ \mu\nu}\star \textgoth{R}^{\sigma}_{\sigma\mu\nu}$ \cite{Twist_general_2} & 
$\textgoth{R}=\invG^{\mu\nu }\star \textgoth{R}_{\nu\mu}$ \\
\hline
\rule{0pt}{4ex} Einstein Tensor &
$\textgoth{R}_{\mu\nu}-\frac{1}{2}g_{\mu\nu} \textgoth{R}$ &
$\textgoth{R}_{\mu\nu}-\frac{1}{2}g_{\mu\nu}\star \textgoth{R}$ \cite{NC_Geometry_Simplified} & 
$\textgoth{R}_{\mu\nu}-\frac{1}{2}g_{\mu\nu}\star \textgoth{R}$ \\
\hline
\rule{0pt}{5ex} Einstein-Hilbert \white{aaaaaa} \white{a}Lagrangian& $\sqrt{-\det [g]}\textgoth{R}$
 & $\sqrt{-\overset{\star}{\det}[g]}\textgoth{R} + c.c$
 & $\sqrt{-\overset{\star}{\det}[g]}\textgoth{R} + c.c$
 \\
\hline
\rule{0pt}{5ex} $\phi^4$ Stress-Energy tensor 
& $T_{\mu\nu}=\partial_\mu\phi\partial_\nu\phi-g_{\mu\nu}\mathcal{L}_{\phi^4}$
 & $\hat{T}_{\mu\nu}=\frac{1}{2}\partial_{\{\mu}\phi\star\partial_{\nu\}}\phi-g_{\mu\nu}\star\mathcal{L}_{\phi^4}^{\star}$ \cite{EMT_1}
 & $\hat{T}_{\mu\nu}=\frac{1}{2}\partial_{\{\mu}\phi\star\partial_{\nu\}}\phi-g_{\mu\nu}\star\mathcal{L}_{\phi^4}^{\star}$
 \\
\hline
\rule{0pt}{5ex} Global/Local Palatini \white{aaa} \white{a}(conservation) 
& $g^{\mu\nu}\delta R_{\mu\nu}\sim\nabla\left(g\delta\Gamma\right)$
 & -----------
 & $\int \sqrt{-g_\star}\star g^{\mu\nu}\star\delta\textgoth{R}_{\mu\nu}\sim \sum_{n\geq 0}\int\sqrt{-g_\star}\star\nablaST(g\star\partial_0^n\delta\Gamma)$
 \\
\hline
\end{tabularx}
\end{sidewaystable}


\clearpage

\end{document}